\documentclass[12pt,preprint]{aastex}
\usepackage{graphicx}

\begin{document}

\shorttitle{Yellow and Red Supergiants in M33}
\shortauthors{Drout et al.}

\title{The Yellow and Red Supergiants of M33\altaffilmark{1}}

\author{Maria R. Drout\altaffilmark{2,3}, Philip Massey\altaffilmark{3}, and Georges Meynet\altaffilmark{4}}

\altaffiltext{1}{Observations reported here were obtained at the MMT Observatory, a joint facility of
the University of Arizona and the Smithsonian Institution.  MMT telescope time was granted by NOAO, through the Telescope System Instrumentation
Program (TSIP).  TSIP is funded by the National Science Foundation (NSF).  This paper uses data products produced by the OIR Telescope Data Center, supported by the Smithsonian Astrophysical Observatory.}
\altaffiltext{2}{Center for Astrophysics, Harvard University, 60 Garden Street, M-S 10, Cambridge, MA 02138; mdrout@cfa.harvard.edu}
\altaffiltext{3}{Lowell Observatory, 1400 W. Mars Hill Road, Flagstaff, AZ 86001; phil.massey@lowell.edu}
\altaffiltext{4}{Geneva University, Geneva Observatory, CH-1290 Versoix, Switzerland; georges.meynet@unige.ch}

\begin{abstract}
Yellow and red supergiants are evolved massive stars whose numbers and locations on the HR diagram can provide a stringent test for models of massive star evolution.  Previous studies have found large discrepancies between the relative number of yellow supergiants observed as a function of mass and those predicted by evolutionary models, while a disagreement between the predicted and observed locations of red supergiants on the HR diagram was only recently resolved.  Here we extend these studies by examining the yellow and red supergiant populations of M33.  Unfortunately, identifying these stars is difficult as this portion of the color-magnitude diagram is heavily contaminated by foreground dwarfs.  We identify the red supergiants through a combination of radial velocities and a two-color surface gravity discriminant and, after re-characterizing the rotation curve of M33 with our newly selected red supergiants, we identify the yellow supergiants through a combination of radial velocities and the strength of the OI $\lambda$7774 triplet.  We examine $\sim$1300 spectra in total and identify 121 yellow supergiants (a sample which is unbiased in luminosity above $\log(L/L_\odot) \sim 4.8$) and 189 red supergiants. After placing these objects on the HR diagram, we find that the latest generation of Geneva evolutionary tracks show excellent agreement with the observed locations of our red and yellow supergiants, the observed relative number of yellow supergiants with mass and the observed red supergiant upper mass limit.  These models therefore represent a drastic improvement over previous generations.
\end{abstract}


\section{\label{Intro}Introduction}

Yellow and red supergiants (F- to G-type supergiants, 
and K- to M-type supergiants, respectively) represent evolved phases in the lives of massive stars whose populations can provide stringent tests of current models of stellar evolution.  The yellow supergiant (YSG) phase (defined as the region where 7500 K $>$ T$_{\rm eff}$ $>$ 4800 K and $\log{L/L_\odot} \gtrsim 4$) is a short-lived, transitionary, phase.  Stars with initial masses between approximately 9 M$_\odot$ and 40 M$_\odot$ briefly pass through this region of the Hertzsprung-Russell (HR) diagram while either transitioning from the main sequence to the red supergiant (RSG) phase, or, in some cases, from the red back to the blue.  The lifetime of the YSG phase is only on the order of tens of thousands of years and, as such, these stars are very rare. The RSG phase, on the other hand, is longer lived, occupying most of the He-burning phase, and representing the final stage in the evolutionary process for stars of an initial mass between approximately 9 M$_\odot$ and 25 M$_\odot$.

\begin{figure}
\plottwo{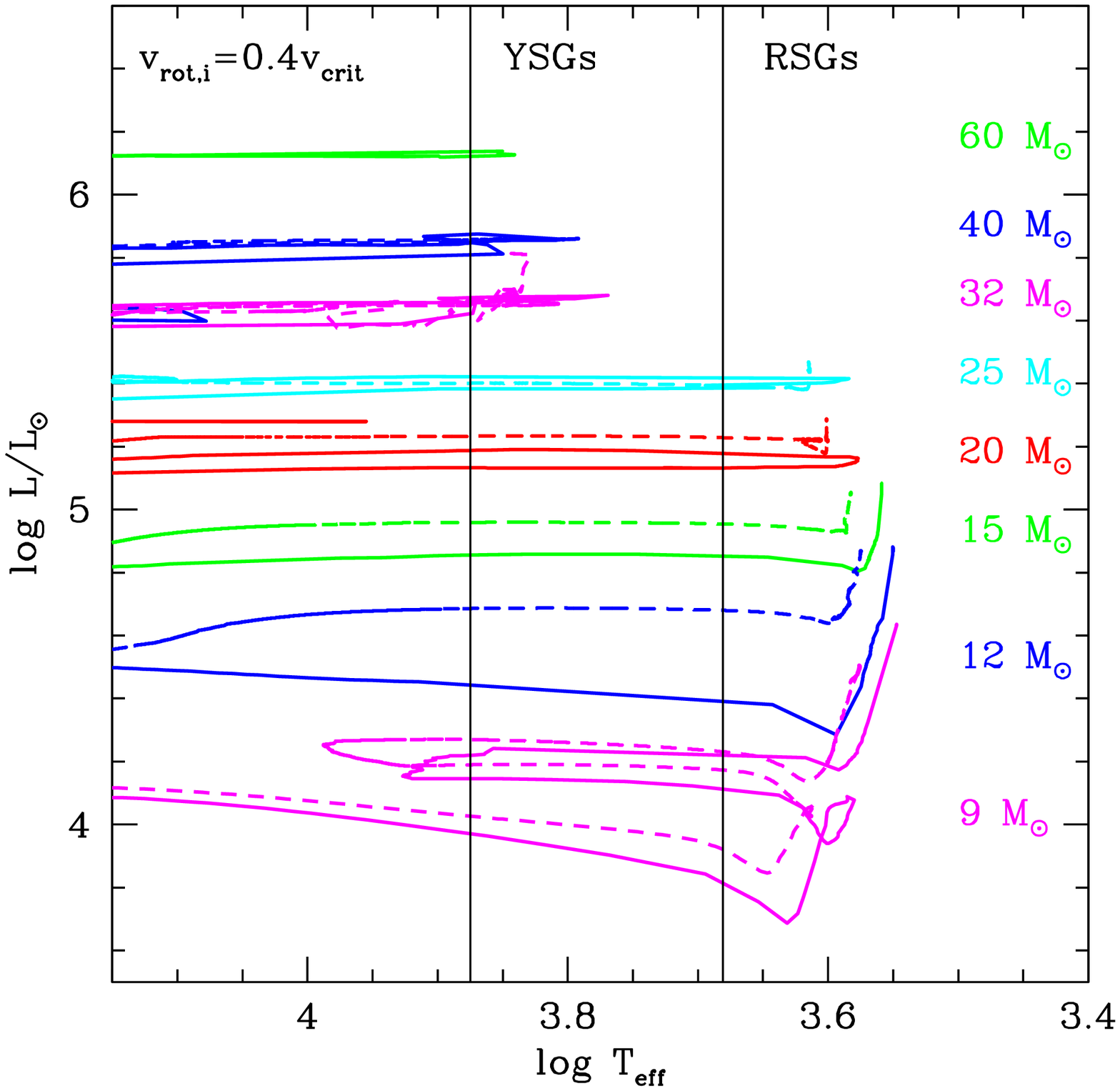}{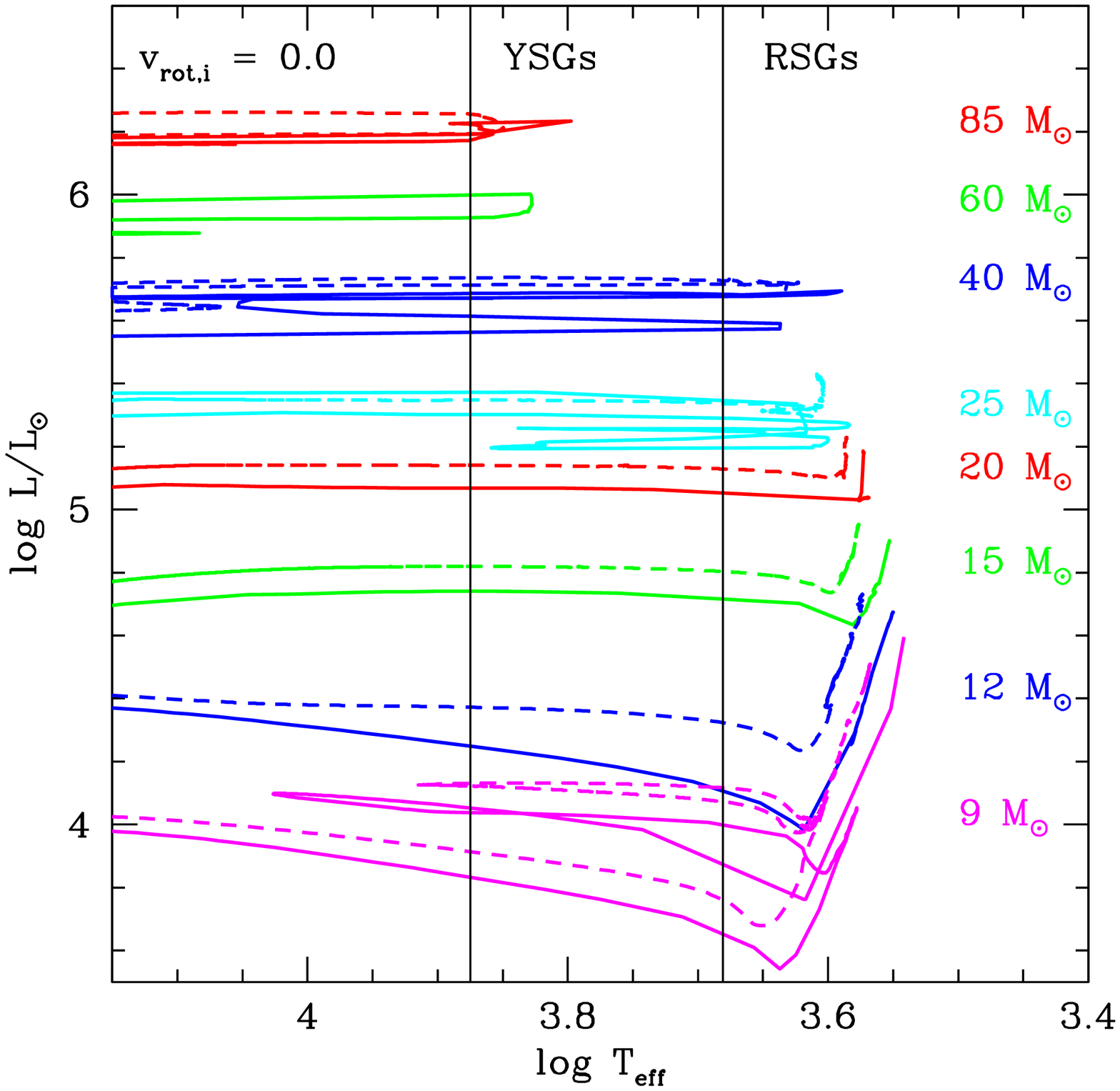}
\caption{\label{fig:toyHR} New Geneva Group $Z=0.014$ (solid lines) and $Z=0.006$ (dashed lines) evolutionary models (Ekstr\"{o}m et al.\ 2012; Chomienne et al.\ \emph{in prep.}). The left panel displays models with an initial rotation of $0.4v_{\rm crit}$, where $v_{\rm crit}$ is critical velocity at which the centrifugal force exactly balances the gravitational acceleration (see Section~\ref{HRdis} for further discussion).  The right panel presents models with no initial rotation, and is shown for comparison.  Vertical lines designate the yellow supergiant region (4800 K $<$ T$_{\rm eff}$ $<$ 7500 K).}
\end{figure}

Details of the evolutionary process for stars in this portion of the HR diagram are far from straightforward.  To emphasize this point, in Figure~\ref{fig:toyHR} we present a set of $Z=0.014$ (solid lines) and $Z=0.006$ (dashed lines) evolutionary tracks from the Geneva group (Ekstr\"{o}m et al.\ 2012; Chomienne et al.\ \emph{in prep.}) zoomed in on this region.  The left 
panel presents models computed with realistic initial rotation speeds, while the right panel presents models with no initial rotation for comparison. The number of loops in these models is striking, with the 9 M$_\odot$ tracks going through a characteristic `blue loop' before terminating in the RSG region, while many of the higher mass tracks only briefly pass through this portion of the HR diagram before looping back to end their lives as Wolf-Rayet (WR) stars or Luminous Blue Variables (LBVs).  The location at which these upper loops occur therefore dictates the divide between massive stars who end their lives as RSGs and those who do not (i.e., they define an upper mass limit for RSGs). This, in turn, has a \emph{significant} effect on the predicted ratios of both supergiant and supernova types (for a discussion see Meynet et al.\ 2011). 

The precise locations at which both these loops and the main RSG branches occur provide one of the key observational tests that can be carried out in this portion of the HR diagram.  Not only are their locations incredibly sensitive to both metallicity and to the treatment of uncertain parameters such as mass-loss, overshooting, and rotationally induced mixing\footnote{This can easily be seen from Figure~\ref{fig:toyHR} where the higher metallicity tracks both possess cooler RSG branches and loop back to the blue portion of the HR diagram at a lower initial mass than the low metallicity tracks, and the non-rotating tracks systematically extend to cooler temperatures before looping back to the blue.}, but they also define forbidden regions on the HR diagram --- a prediction easily tested by obtaining effective temperatures and luminosities for observed RSG and YSG populations.  

Indeed, until recently, there was a discrepancy between the locations of observed RSGs on the HR diagram and those predicted by stellar evolution theory, with the models failing to produce RSGs as cool or luminous as those observed (Massey 2003).  Although a number of assumptions made in the modeling process could cause this discrepancy (e.g., uncertain RSG opacities and simplifications invoked by mixing-length theory), the fault actually lay with the derived locations for the observed samples.  Using state-of-the-art MARCS model atmospheres (Gustafsson et al. 2003, 2008; Plez et al.\ 1992; Plez 2003), Levesque et al.\ (2005) derived new temperatures and luminosities for Galactic RSGs and found excellent agreement with the solar metallicity Geneva evolutionary models (Meynet \& Maeder 2003).  Since then, agreement has been found across a wide range of metallicities from the Magellanic Clouds (Levesque et al.\ 2006) to M31 (Massey et al.\ 2009).

However, the main reason this portion of the HR diagram is so ideal for observational testing is the incredibly short lifetime of the YSG phase.  
With assumptions about the star formation rate and initial mass function (IMF) the YSG lifetimes predicted by the models for stars of various masses can be used to predict the \emph{relative} number of stars that should appear in various luminosity (mass) bins\footnote{We note that a similar test is not possible for the RSGs because the near-vertical nature of the tracks in the region make it very difficult to distinguish stars in different mass bins.}.  However, these predicted lifetimes are also highly sensitive to the uncertain parameters mentioned above (mass-loss, overshooting, rotationally induced mixing), and because of their short duration, even small variations can dramatically change the predicted relative number of stars.   As Kippenhahn \& Weigert (1990,  p.\ 468) put it, ``[The yellow supergiant] phase
is a sort of magnifying glass, revealing relentlessly the faults of calculations of earlier phases."

In previous studies, Drout et al.\ (2009) and Neugent et al.\ (2010) examined the YSG populations of M31 and the Small Magellanic Cloud (SMC), respectively.  They found that the relative number of observed YSGs as a function of luminosity differs dramatically from those predicted by the lifetimes of the evolutionary models.  In both cases, the models \emph{overestimate} the number of high luminosity stars with respect to low luminosity stars by factors of 10 or more.  We note that since this problem has manifested itself in \emph{both} M31 and the SMC (whose metallicities differ by a factor of 10) the cause cannot \emph{simply} be the prescriptions used for the metal line driven (and, hence, metallicity dependent) main sequence mass-loss.  

In this paper we examine the case of M33, whose metallicity, although debated (see Section~\ref{make}), likely lies intermediate between the SMC and solar ($Z=0.006$ and $Z=0.014$, respectively).  This, and a companion paper examining the red and yellow supergiant content of the Large Magellanic Cloud (LMC; Neugent et al.\ 2011), are part of a continued effort to characterize the yellow and red supergiant populations in local group galaxies of various metallicities in order to more fully assess the problems that may be present in the current evolutionary models.  These tests are vital not only for our understanding of stellar physics, but also because spectral synthesis codes (such as Starburst99, Leitherer et al.\ 1999) are only as good as the evolutionary models which they utilize. At the same time, we are employing a newer generation of the Geneva evolutionary models.

\begin{figure}
\epsscale{0.5}
\plotone{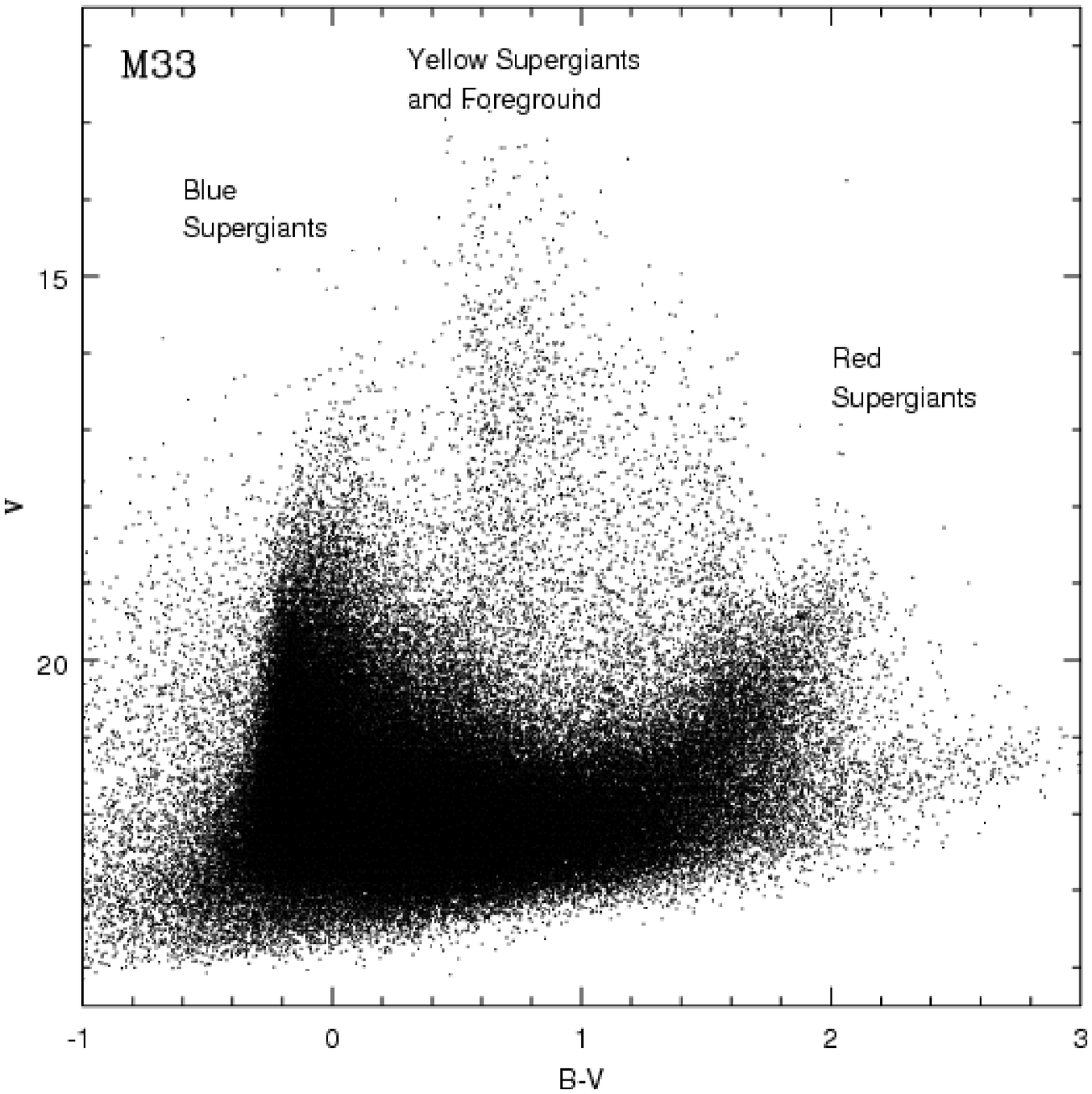}
\plotone{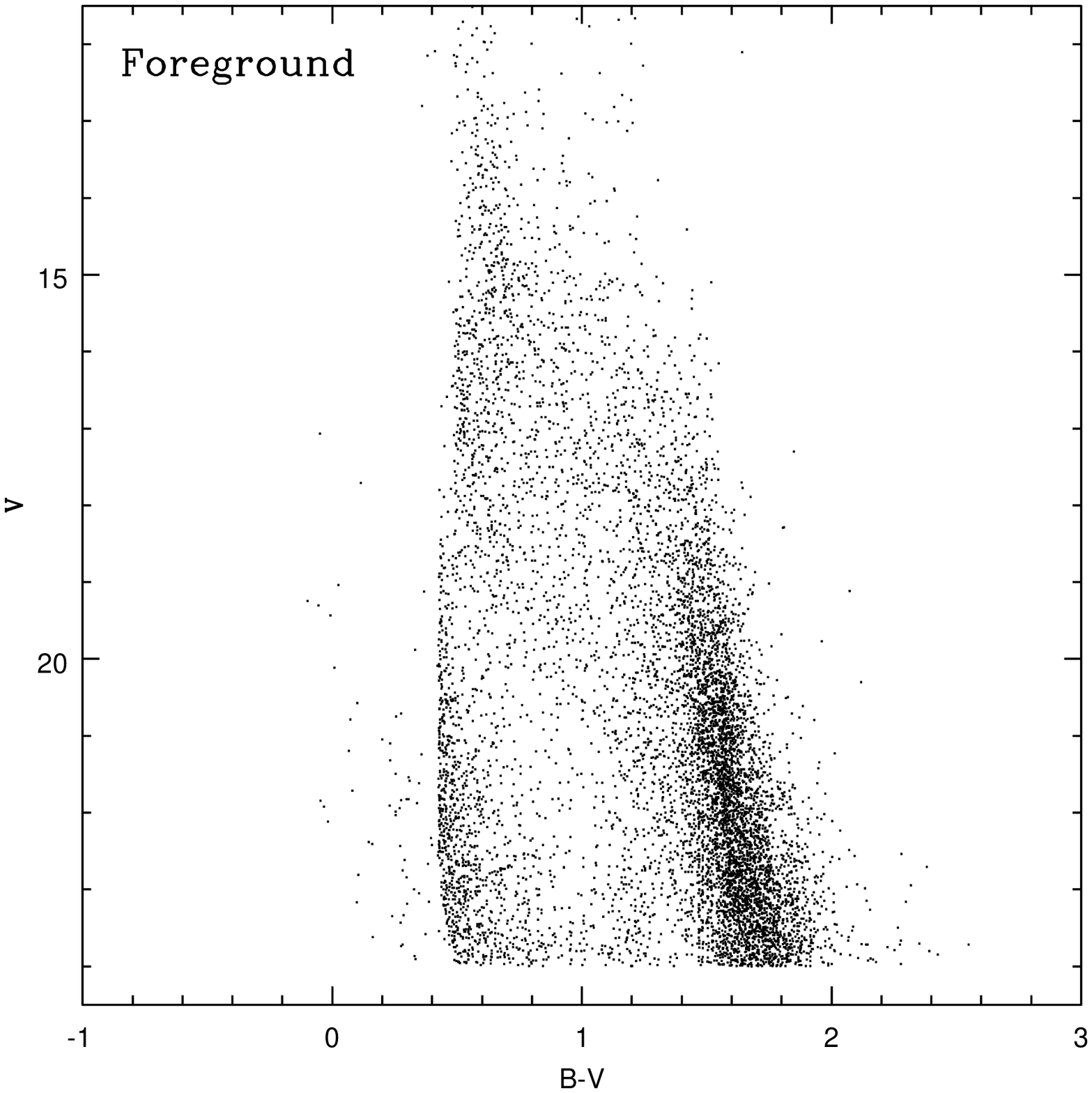}
\caption{\label{CMD} \emph{top:} A color-magnitude diagram for M33 constructed from the LGGS photometry. \emph{bottom:} Location of foreground contamination as predicted by the Besan\c{c}on model for the Milky Way (Robin et al.\ 2003).}
\end{figure}

One large hurdle that must be overcome in order to actually carry out these tests is, ironically, \emph{identifying} the supergiants.  When one looks towards a local galaxy at the colors and magnitudes of yellow supergiants, the bona fide supergiants are masked by a veritable sea of foreground dwarfs.  This is clearly demonstrated in Figure~\ref{CMD} where the upper panel is a color magnitude diagram (CMD) of M33 constructed from the photometry of the Local Group Galaxies Survey (LGGS, Massey et al.\ 2006) and the lower panel is the predicted locations for foreground dwarfs constructed from the Besan\c{c}on Model for the Milky Way (Robin et al.\ 2003).  Although the red and blue portions of the CMD are relatively uncontaminated, the overlap in the YSG regime in striking. Drout et al.\ (2009) and Neugent et al.\ (2010) overcome this problem by using the relatively large systematic velocities of M31 and the SMC to distinguish extragalactic supergiants based on their radial velocities.  In M33, modest systematic ($-180$ km s$^{-1}$) and line-of-sight rotational ($\sim$80 km s$^{-1}$) velocities (Kerr \& Lynden-Bell 1986; Zaritsky et al.\ 1989) should still allow some separation of supergiant candidates via this method.

\begin{figure}
\epsscale{1}
\plotone{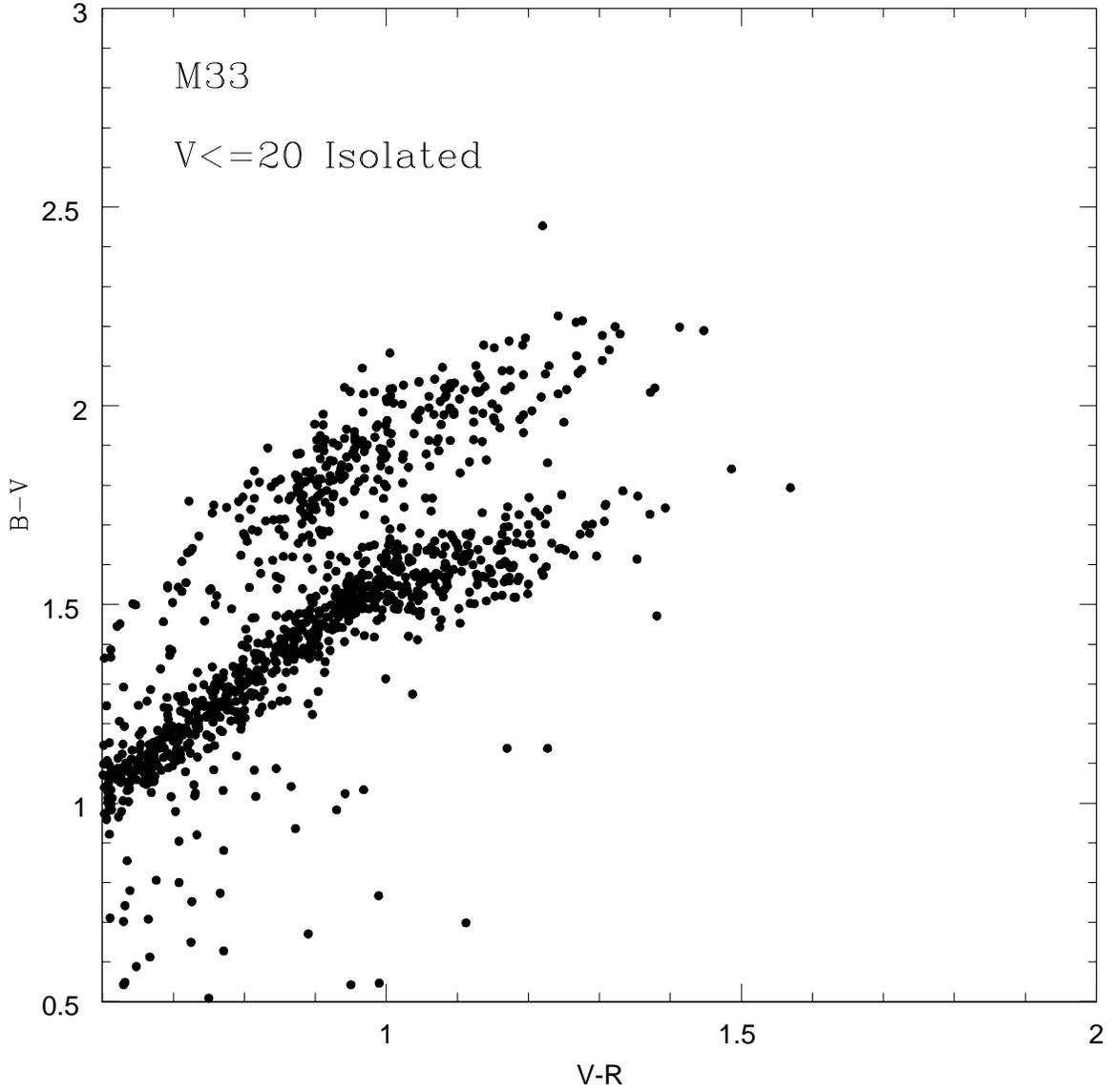}
\caption{\label{Isolated} Photometry of isolated RSG candidates in M33.  Two separate bands are obvious, with the upper band representing likely supergiants and the lower band representing likely dwarfs.} 
\end{figure}

For the case of the red supergiants, the situation is not nearly as dire.  Massey (1998) demonstrates that local group red supergiants may be separated from foreground dwarfs by means of a $B-V$, $V-R$ two-color diagram.  At cool effective temperatures $B-V$ is primarily a surface gravity indicator (due to line-blanketing in the blue), while $V-R$ remains an effective temperature indicator.  Using the improved photometry of the LGGS, Massey et al.\ (2009) examine M31 in more detail and demonstrate that stars selected in this manner show excellent agreement with stars whose radial velocities are consistent with their location in M31.  In Figure~\ref{Isolated} we plot the $B-V$ versus $V-R$ colors from the LGGS for isolated stars toward M33 with $V < 20$ and $V - R > 0.6$.  Two distinct bands are evident in this plot with the upper band representing likely supergiants and the lower band likely dwarfs.  However, the exact selection criteria which should be utilized has yet to be thoroughly tested at sub-solar metallicities.

Thus, our aims in this paper are two-fold: (1) to characterize the red and yellow supergiant content of M33, and (2) further assess any contradictions that may be present between the current evolutionary tracks and our supergiant populations.  In Section~\ref{obsRed} we describe our data and reductions. In Section~\ref{An} we describe the process by which we distinguish M33's red and yellow supergiants from the population of foreground dwarfs, 
and in Section~\ref{HR} we present a comparison of our results to the current set of evolutionary tracks.

\section{Observations and Reductions\label{obsRed}}

In order to distinguish genuine red and yellow supergiants in M33 from foreground Milky Way dwarfs, we obtained radial velocities for $\sim$1300 red and yellow stars toward M33 using the Hectospec multi-fiber spectrograph on the 6.5 m MMT telescope.  The observations were obtained over the course of two years, with our yellow sample observed in the fall of 2009 and our red sample (along with repeat observations for a subset of the yellow sample) in fall of 2010.  In this section, we describe our sample selection, observations, and initial data reduction.

\subsection{Selection Criteria\label{SC}}

Our samples were initially selected photometrically from the contents of the Local Group Galaxies Survey (LGGS) of Massey et al.\ (2006). The yellow supergiant candidates were selected to have $V < 18.7$, $U - B > -0.4$, and $0.0 \leq B - V \leq 1.4$.  We note that these selection criteria are very similar to those used in Drout et al.\ (2009).   The magnitude cutoff (roughly corresponding to $\log{L/L_\odot} \sim 4.5)$ is selected to provide adequate signal-to-noise in our spectroscopy, and the color range \emph{roughly} corresponds to that over which F- and G-type supergiants cannot be photometrically distinguished from foreground dwarfs.  With these definitions we constructed a catalog of 1438 yellow supergiant candidates. 

The sample of red stars was initially selected to have $V < 20$ and $V - R > 0.6$.  These criteria were selected to provide adequate signal to noise for spectroscopy, to help avoid confusion with low mass asymptotic giant branch (AGB) stars, and to restrict the sample to K-type stars and later (see discussion in Massey et al.\ 2009).  We then applied the additional criteria that our targets be `isolated' to ensure, to the best of our ability, that our spectroscopy probes a single star.   Using the $V$-band photometry of the LGGS, we estimated the crowding of each red star which falls within the above criteria. We restricted our sample to stars for which we can expect the contamination from objects other than the star itself in an Hectospec fiber (diameter $\sim$1.5$^{\prime\prime}$) to be $<20\%$.  With this additional requirement, we construct a catalog of 1098 red stars.  This represents $\sim$42\% of the 2602 stars in the LGGS which matched the initial photometric criteria.  We then applied the photometric criteria described in Section~\ref{Intro} to assess the likelihood that each star was a bona fide M33 supergiant.  As in Massey et al.\ (2009), we adopt the line:
$$B-V= - 1.599(V-R)^2 + 4.18(V-R) - 0.83$$
as the divide between supergiant candidates and likely dwarfs (located above and below the line, respectively).

Fiber assignment files for Hectospec are created prior to observations.  Due to fiber configuration constraints, we were unable to assign all of our catalog objects.  For the fall 2009 observations of the yellow targets, priority for assignment was based on the level of crowding of each object, with higher priority going to isolated stars.  For the fall 2010 observations, which contained our red targets as well as repeat observations for a subset of our yellow stars, the priority assignments were more complex.  In the end, highest priority was given to red stars whose $BVR$ photometry place them in the supergiant candidate regime and who also have $K$-band data from the Two Micron All Sky Survey (2MASS, Skrutskie et al.\ 2006). The 2MASS criteria was imposed because we would like to utilize $K$-band photometry and $V-K$ colors to determine the effective temperatures and bolometric luminosities of our red supergiant candidates. After these stars, the order of priority was our repeat yellow targets, the remaining RSG candidates, and finally, the red catalog stars whose $BVR$ photometry indicates they are likely dwarfs.  This final category is included to act as a test of our photometric selection criteria.  During this process the RSGs were given some short shrift as our higher priority was obtaining an unbiased selection of YSGs.  In the end, we were able to assign fibers to $\sim$68\% of our yellow catalog and $\sim$15\% of our red catalog ($\sim$37\% of the isolated stars).

\subsection{\label{Obs}Observations}
  
Hectospec is a 300 optical fiber fed spectrograph (Fabricant et al.\ 2005)  with a $1^{\circ}$ field of view deployed on the MMT 6.5 m telescope.  Hectospec observations are carried out in a queue mode in order to spread the effects of bad weather over all of the selected programs.   As a result, our observations were obtained over three nights in October 2009, one night in November 2009, one night in December 2009, and two nights in November 2010 (NOAO proposals
2009B-0149 and 2010B-0260).   All observations were obtained using the 600 line mm$^{-1}$ grating, yielding a dispersion of 0.55 \AA\  pixel$^{-1}$ and a (5 pixel) spectral resolution of 2.8 \AA.  The grating was centered on 7800 \AA\  such that the wavelength coverage (6500 to 9000 \AA) included both the Ca II triplet ($\lambda\lambda$ 8498, 8543, 8662) for cross-correlation and the OI $\lambda$7774 line (which has been shown to be a luminosity indicator in galactic F-type supergiants, see Section~\ref{yellow}).

The observations obtained in 2009 consisted of one field containing bright catalog stars ($V\sim 12.7-15.5$) and four fields containing faint catalog stars ($V\sim 15.0-18.7$), with overlap to check on the consistency
of radial velocity measurements. Each field was observed in 3 consecutive exposures to help eliminate cosmic rays; the total
integration times for the bright and faint fields were 15 min and 60 min, respectively.

The observations in 2010 were again broken into a bright field ($V\sim 15.3-17.7$ for the yellow stars, and $V\sim 14.2-17.0$ for the red stars) and two fields of fainter stars ($V\sim 17.0-20.0$ for the YSG candidates and $\sim 17.0-18.7$ for the RSG candidates).  The bright field was observed through inclement weather for 36 minutes.  One of the two faint fields was also observed that same night for 75 minutes, but was repeated the next evening under clear conditions  for 80 minutes, along with the second faint field.  Information on our observed fields is summarized in Table~\ref{tab:obs}.  

For our radial velocity standards we utilize ten yellow stars in the direction of M31 whose radial velocities were determined in Drout et al.\ (2009).  The observations were taken in the fall of 2009 with the same grating and over the same wavelength range as the M33 data described here.

In observing our fields, we relied upon the third US Naval Observatory (USNO) CCD Astrograph Catalog (UCAC3) (Zacharias et al.\ 2010) 
to provide guide stars with negligible proper motions.  The LGGS coordinates were tied to an earlier version
of the USNO catalogs.  In order to transform the LGGS coordinates to the UCAC3 ``system" (presumably identical to the International Celestial Reference System, ICRS) we used stars in common to the two surveys to determine
$\alpha_{\rm UCAC3}=\alpha_{\rm LGGS}-0\fs03$ and $\delta_{\rm UCAC3}=\delta_{\rm LGGS}-0\farcs14$. The names of all our targets contain their LGGS coordinates, and these relations may therefore be used to determine their ICRS coordinates.

The reductions followed the same procedure as described in Drout et al.\ (2009) for Hectospec data, except that
the zero-point of the wavelength scale was set by the OH $\lambda 8399$ line.  We are indebted to Susan Tokarz
of the CfA's OIR group for performing these reductions through their standard pipeline.

\section{\label{An}Analysis}

\subsection{\label{Vels}Observed Radial Velocities}

In order to begin assigning membership to our supergiant candidates, we first measured radial velocities for our $\sim$1300 observed spectra via Fourier cross-correlation with a set of radial velocity standards.  Before cross-correlation, each spectra was normalized with a 9th order cubic spline fit and then had 1.0 subtracted to remove the continuum.   All cross-correlations were carried out in the IRAF routine ``fxcor''.

As mentioned in Section~\ref{Obs}, for radial velocity standards we used spectra of stars in the direction of M31 whose radial velocities had been measured by Drout et al.\ (2009).  We initially intended to carry out all of our cross-correlations in the region of the Ca II triplet, and thus choose five stars, J004535.23+413600.5, J004251.90+413745.9, J004532.62+413227.8, J004618.59+414410.9, and Mag-253496, whose spectra contained particularly strong Ca II lines.  When the velocities from Drout et al.\ (2009) were applied to the spectra and the standards were cross-correlated against each other the results were consistent to within $\sim$1.5 km s$^{-1}$ with standard deviations of the mean ranging from 0.8 to 1.3 km s$^{-1}$.  Our $\sim$1300 program spectra were then cross-correlated against all five standards.  All cross-correlations involving these standards were carried out over the range 8465 $-$ 8680 \AA\ to include the Ca II triplet and to exclude featureless continuum and strong atmospheric bands.

\begin{figure}
\plottwo{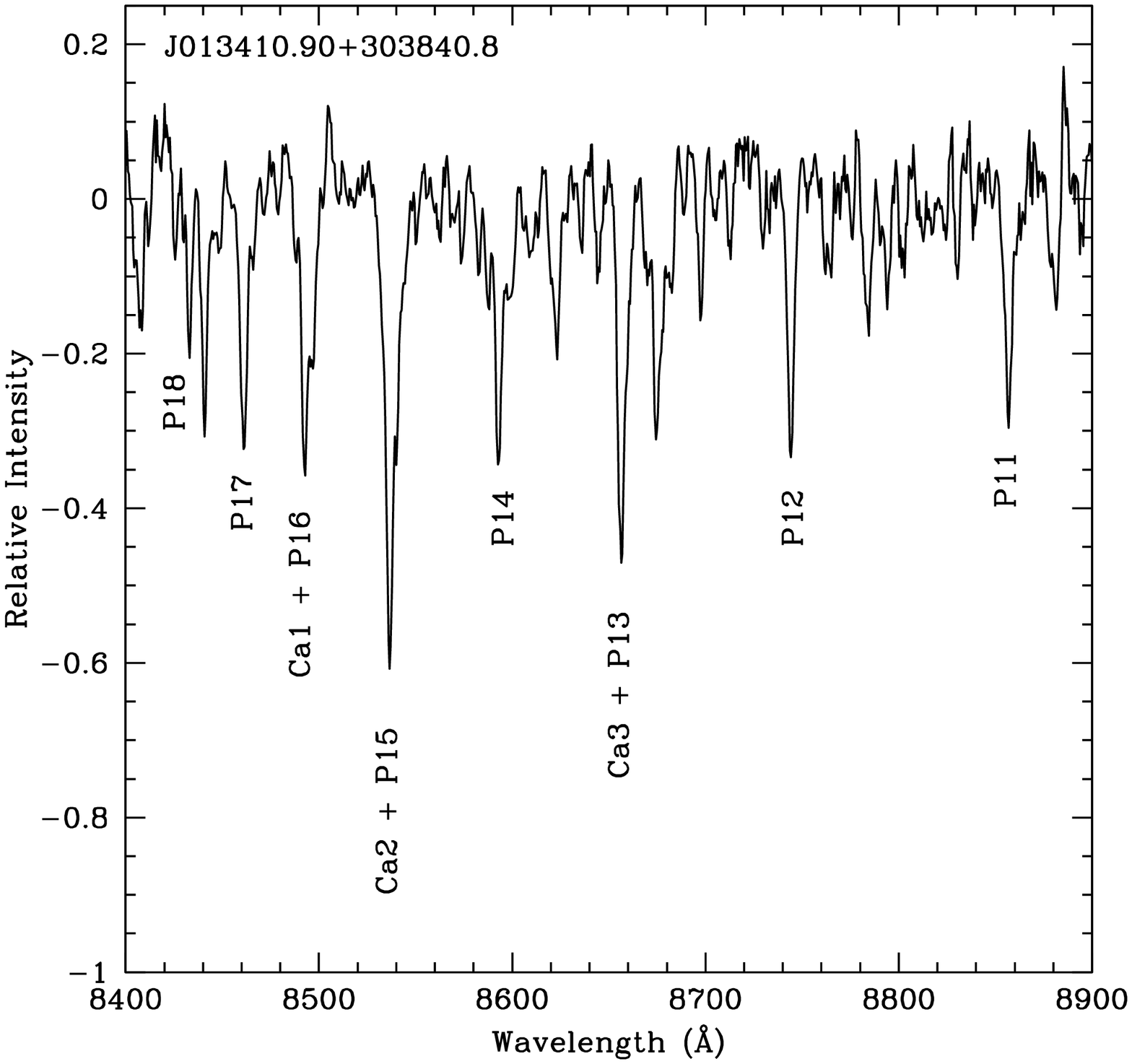}{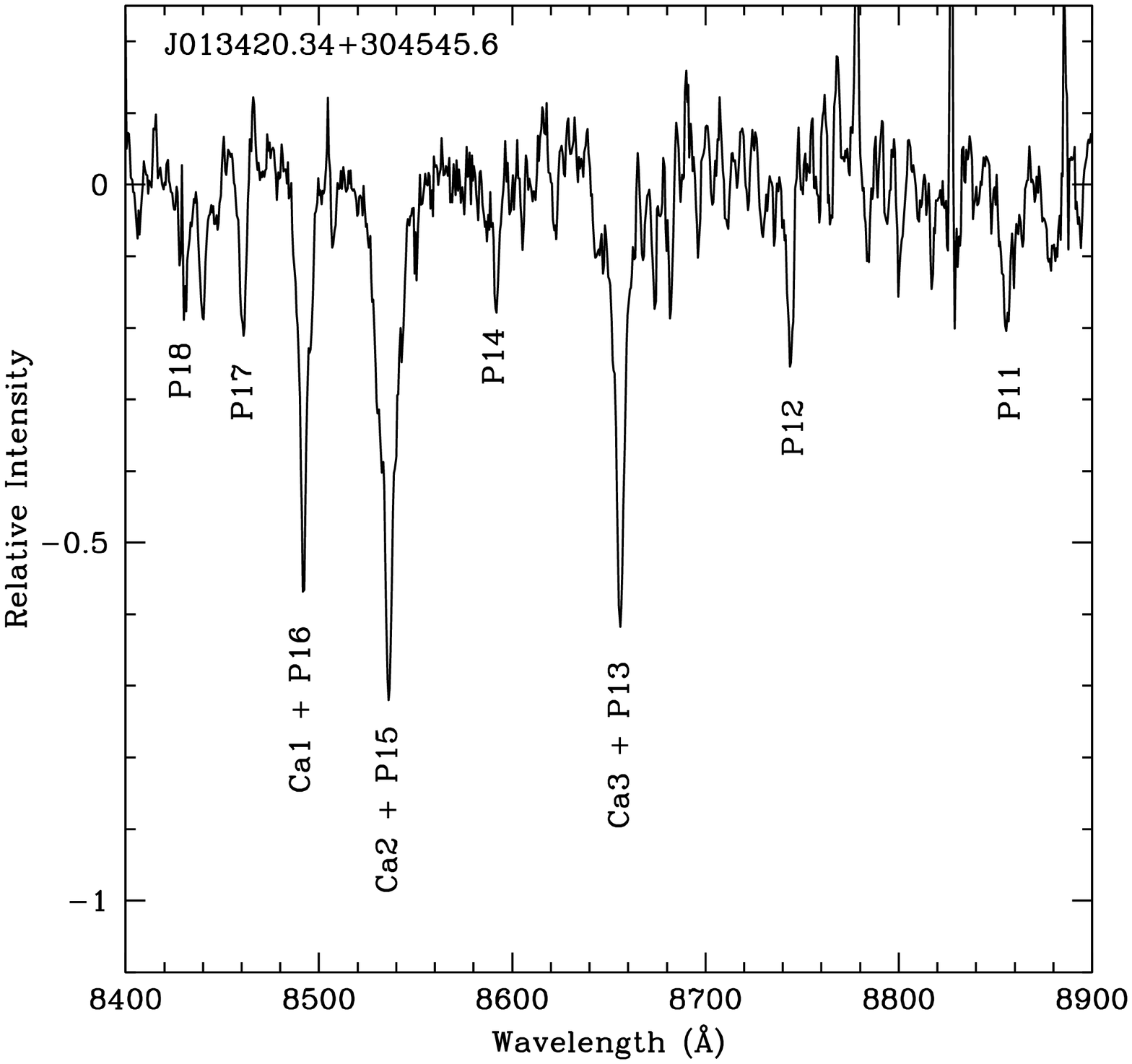}
\caption{\label{Pa} Examples of spectra demonstrating the progression of spectral lines with temperature for a subset of our yellow sample. \emph{left:} A star with $T_{\rm eff} \sim 6000$ K.  Although the three Paschen lines coincident with the Ca II triplet stand out, they are comparable in strength to the other lines in the Paschen series. \emph{right:}.  A cooler star with $T_{\rm eff} \sim 5000$ K.  In this case, all the Paschen lines are still present, but their relative strength has decreased significantly with respect to the three lines of the Ca II triplet.}
\end{figure}

As described in detail by Cenarro et al.\ (2001) at the temperatures of A- to F-type stars the lines of the Ca II triplet ($\lambda\lambda$ 8498, 8543, 8662) can become contaminated by the Paschen lines, P13, P15 and P16 ($\lambda\lambda$8502, 8545, 8665).  Suspecting that this could affect the accuracy of the radial velocity measurements for our warmer stars, we examine our spectra for evidence of the other lines in the Paschen series.  Indeed, we found $\sim$120 spectra in our sample ($\sim$9\% of the total sample) which showed obvious Paschen lines.  As expected, within these stars we observe a general trend such that, as $B-V$ increases (and therefore effective temperature decreases), the strength of the three lines near the Ca II triplet increases relative to the other Paschen lines.  We find that at a $B-V$ of $\sim$0.55 (in the middle of the YSG range) this relative increase is apparent, but not overwhelming, indicating the lines in the spectra
are likely a blend of the Ca II triplet lines and Paschen lines.  Conversely, by a $B-V$ of $\sim$1.4 (approximately the edge of our defined YSG range), either no additional Paschen lines are evident above the noise, or they are so dwarfed by the main three lines in the spectra that it is evident the Ca II triplet completely dominates. See Figure~\ref{Pa} for an illustration of this effect.

For the spectra with Paschen lines we sought new radial velocity standards.  From the spectra of yellow stars in M31 we identify five stars with obvious Paschen lines: J003930.55+403135.2, J003948.85+403844.8, J004009.13+403142.9, J003949.86+405305.8, and J004120.56+403515.4.  When these were cross-correlated against one another, the results were consistent with the `known' values from Drout et al.\ (2009) to within $\sim$1.5 km s$^{-1}$ and the standard deviations from the mean varied between 0.5 and 1.0 km s$^{-1}$.  The $\sim$120 spectra with obvious Paschen lines were then cross-correlated against these standards over the range 8400 $-$ 8900 \AA, to include the P11 to P19 lines of the Paschen series.

These new radial velocity standards all possess intermediate temperatures within the yellow supergiant range.  We found use of these standards improved our cross-correlations for all but our hottest yellow stars (which represent spectra most strongly dominated by the Paschen lines).  For these $\sim$30 spectra we found no suitable standards within our M31 sample.  Instead, we chose six of the highest quality spectra (J013254.33+303603.8, J013244.40+301547.7, J013252.56+303419.6, J013327.81+302106.0, J013350.71+304254.3, J013327.65+303751.2), measured their radial velocities by hand, and used them as standards for the rest of the sample.  Given our means of measuring the standards' velocities, we expect our radial velocities for these $\sim$30 spectra to process higher errors than the rest of our sample.  However, we do not expect this fact to alter which stars we select as M33 members (see Section~\ref{yellow}).  In addition, all these stars actually fall outside our defined yellow supergiant range (4800 $-$ 7500 K) and, thus, will not affect our comparison to the current set of evolutionary tracks.

The error associated with our measured radial velocities was quantified in several independent ways.  First, each measured velocity, $V_{\rm obs}$, is produced by taking the average result of cross-correlating the spectra against five independent standards.  The standard deviation of the mean for this averaging process was typically on the order of 1.5 km s$^{-1}$, although this number may underestimate our true error. For the $\sim$30 hottest stars in our sample, the average standard deviation was slightly larger: on order of 4 km s$^{-1}$.  This is a result of the larger uncertainty in the standards' velocities. Additionally, the IRAF package ``fxcor'' outputs both an internal error, obtained from the uncertainty of the Gaussian fit to the center of the correlation function, and the Tonry \& Davis (1979) r-parameter, the ratio of the peak of the correlation function to its noise.   The average internal error and r-parameter for our sample are 2.4 km s$^{-1}$ and 38.6, respectively. Further examination reveals that stars with low (r $<$ 25 ), medium (25 $<$ r $<$ 40 ), and high (r $>$ 40 ) r-parameters have typical errors of $3.5 - 5.5$ km s$^{-1}$,  $2.0 - 3.5$ km s$^{-1}$, and $1.5 - 2.0$ km s$^{-1}$, respectively.

\begin{figure}
\plottwo{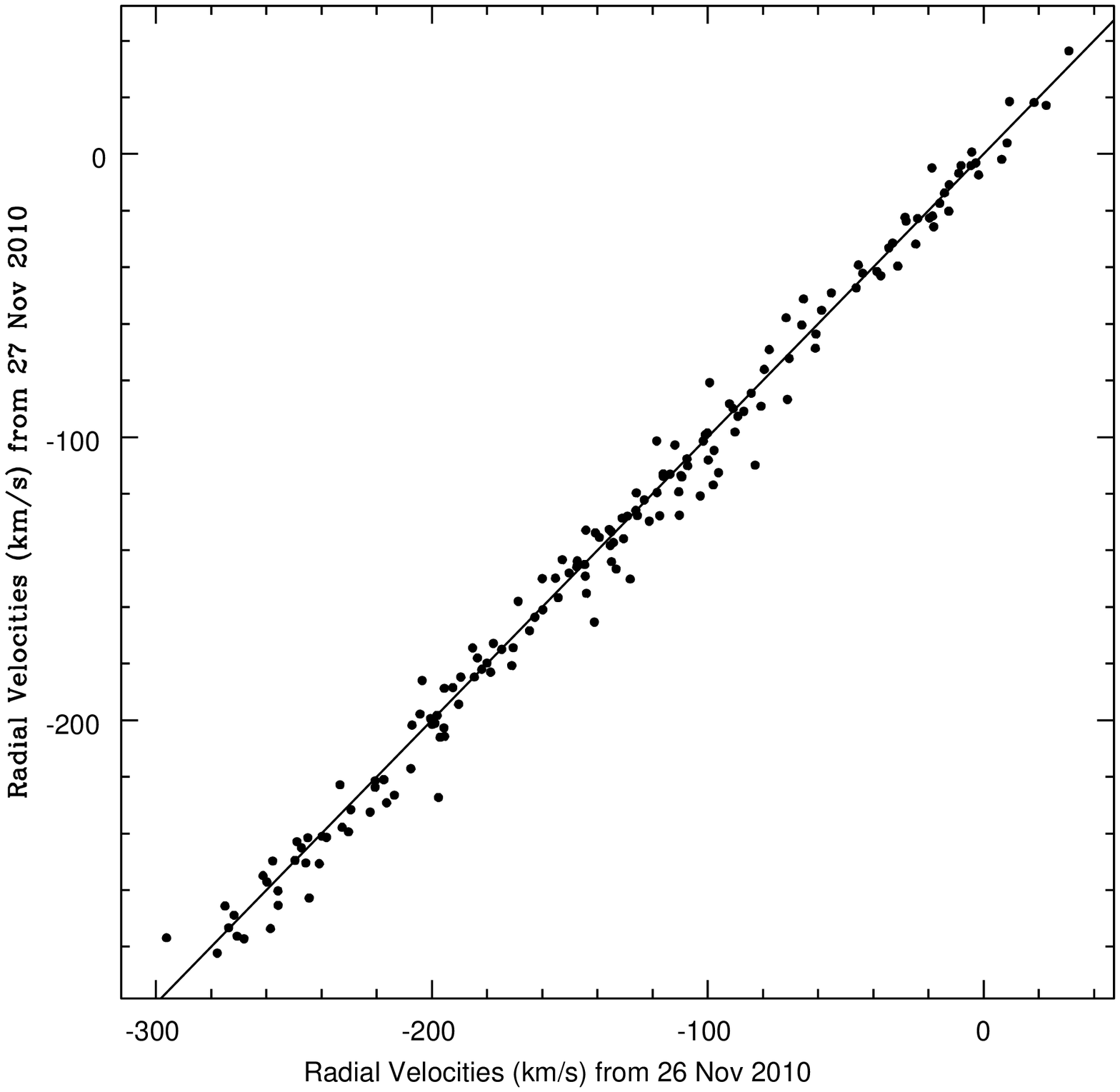}{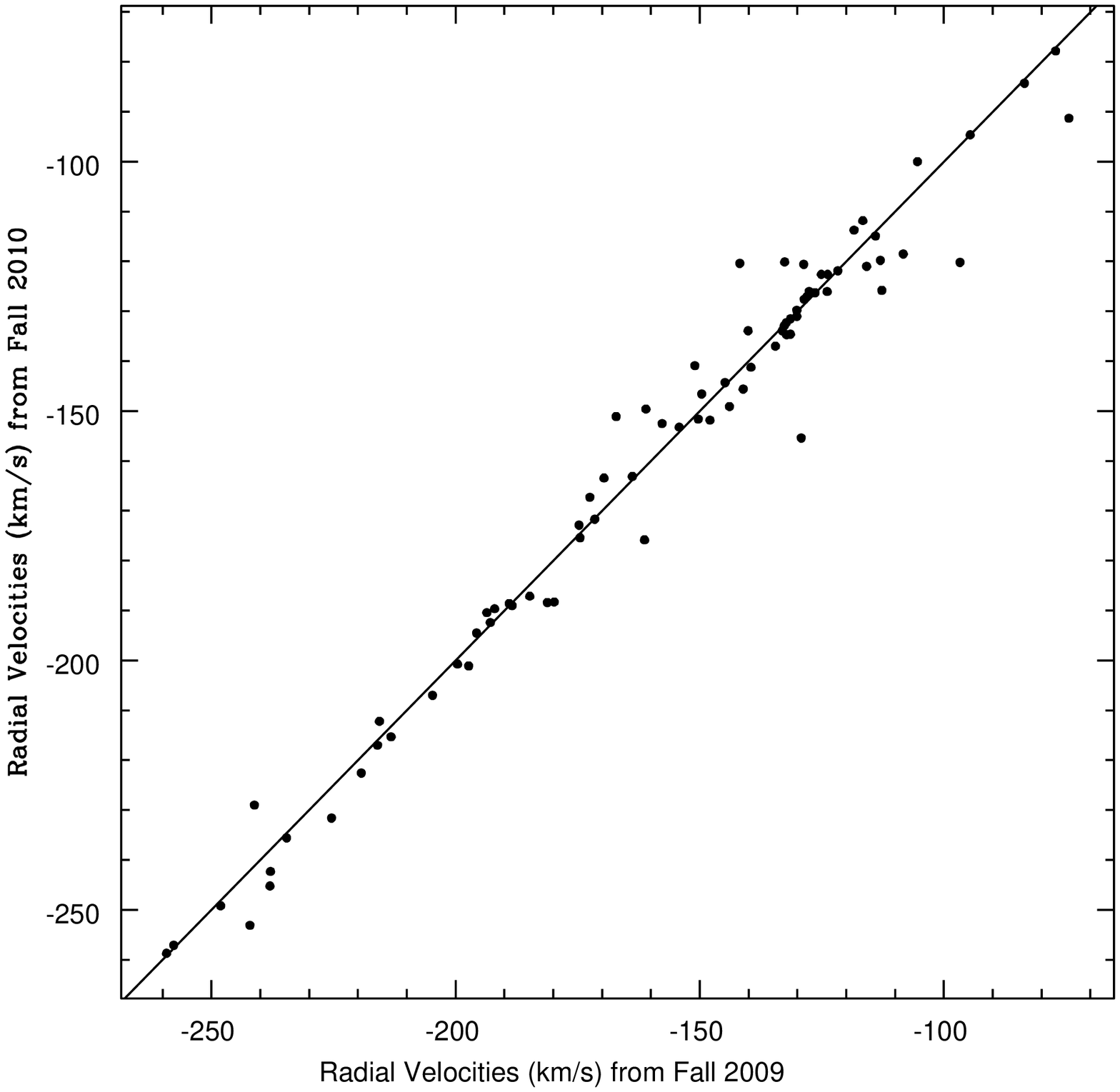}
\caption{\label{fig:fnt1_0910} \emph{left:} A comparison of the radial velocities measured during the two observations of the fnt-1 field in 2010. \emph{right:} A comparison of the radial velocities measured for the 79 stars observed in both 2009 and 2010. The lines show a 1:1 relation.}
\end{figure}

Over the course of our program, we also obtained multiple observations of a number of our targets.   Notably, in the 2010 set of observations, the fnt-1 field was repeated on two consecutive nights.  In Figure~\ref{fig:fnt1_0910} (left) we compare the velocities measured each night.  We see that, in general, the agreement is good.  The average difference between the observations is $-1.3$ km s$^{-1}$ with a standard deviation of 8.0 km s$^{-1}$.  Additionally, 25 and 42 stars were observed in common between the `faint' and `bright' fields in 2010 and 2009, respectively.  A comparison of the two measured velocities for the 25 stars from 2010 yields similar to results to the fnt-1 stars.  However, the 42 stars observed twice in 2009 had an average difference of $-$1.0 km s$^{-1}$ with a standard deviation of only 3.1 km s$^{-1}$.  Thus, we suspect that the poor weather during our second observing run has produced higher uncertainties in the 2010 data.  Finally, 79 of the stars observed in 2010 were repeat observations of yellow stars observed in 2009.  In Figure~\ref{fig:fnt1_0910} (right) we plot the velocities measured each year.  The average difference between the observations is $-$0.7 km s$^{-1}$ with a standard deviation of 7.0 km s$^{-1}$.

\begin{figure}
\plotone{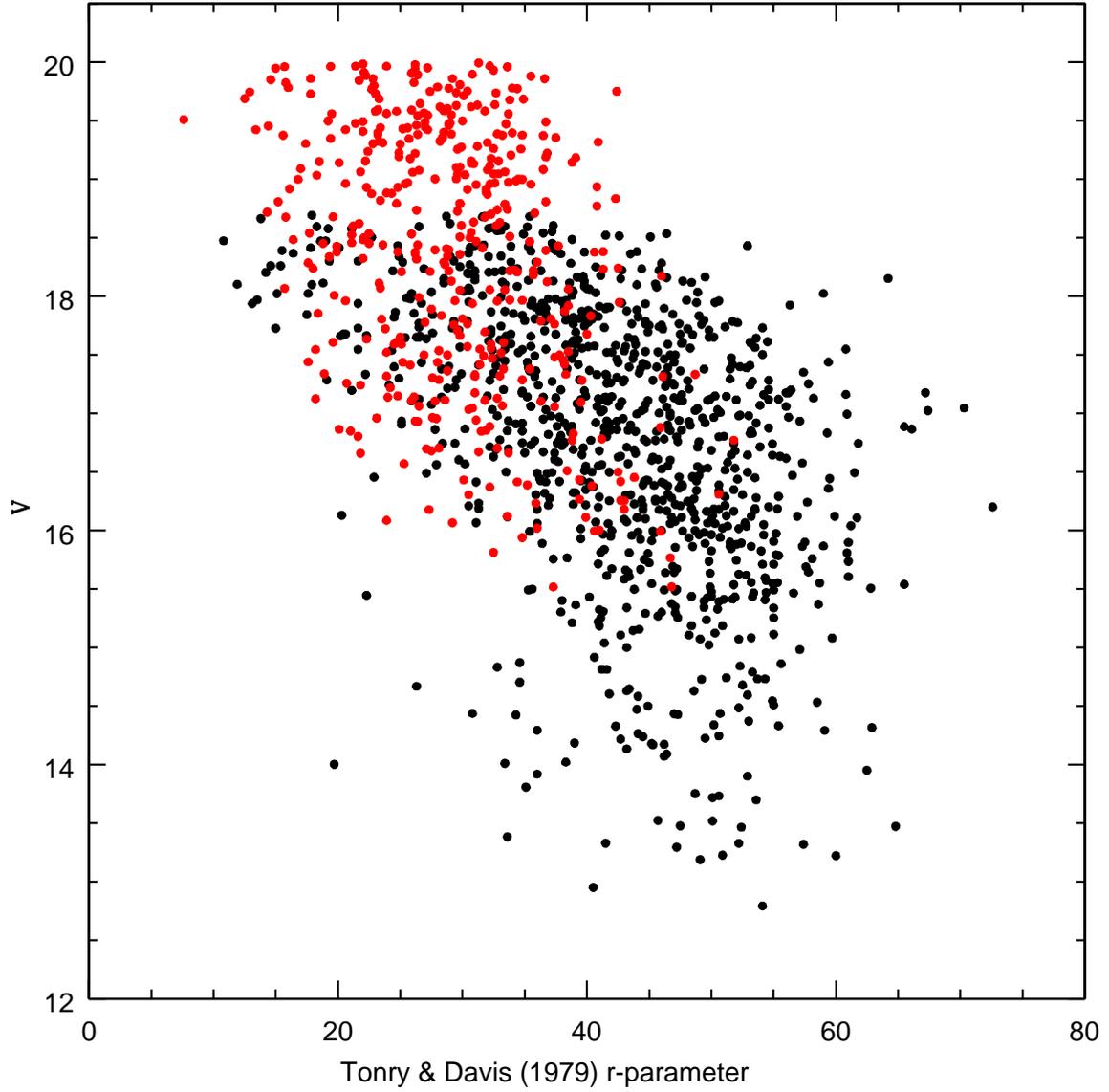}
\caption{\label{fig:Rmag} Tonry \& Davis (1979) r-parameter versus $V$-band magnitude for our sample.  Black points are yellow stars and red points are red stars.  The red stars have a lower average r-parameter for a given magnitude.}
\end{figure}

We do note that, although our sample contained both red and yellow stars, all of our radial velocity standards were yellow stars.  Thus, it is likely that the velocities we measure for our red sample will have a higher associated error.  In Figure~\ref{fig:Rmag} we plot Tonry \& Davis (1979) r-parameter vs.\ $V$-band magnitude for both our red and yellow sample.  Indeed, it can be seen that, for a given magnitude, the red sample has a lower average r-parameter.  However, we do not believe that this effect influences our results.

Recall that our purpose in measuring radial velocities was to attempt to discern membership in M33.  As mentioned in Section~\ref{Intro}, M33 has a systematic velocity of approximately $-180$ km s$^{-1}$ and has a line-of-sight rotation of approximately $80$ km s$^{-1}$.  We therefore expect M33 members to have velocities roughly ranging from $-$100 km s$^{-1}$ to $-$260 km s$^{-1}$ while the foreground dwarfs should be scattered about zero km s$^{-1}$. Thus, we find that, for essentially all of our targets, we obtained radial velocities to suitable precision for our purposes.  We did choose to reject a handful stars whose errors were above $\sim$10 km s$^{-1}$, and examination revealed no obvious spectral features above the noise.  In the end, we obtained radial velocities for 408 individual red stars and 916 individual yellow stars in the direction of M33.  Their velocities, errors, and other identifying information is summarized in Tables \ref{tab:TableMeas:Red} and \ref{tab:TableMeas:Yel}.  For stars observed in multiple fields we adopt the average values.  If the velocities obtained from two different observations of the same star varied by more that 10 km s$^{-1}$ we note this fact in Tables \ref{tab:TableMeas:Red} and \ref{tab:TableMeas:Yel}.

\subsection{\label{reds}Membership Determination: Red Supergiants}

In order to assess the accuracy of our two-color discriminant for red supergiants, we compare the stars we select photometrically as supergiants to those we expect to be M33 members based on their radial velocities.  In general, the radial velocity of a disc galaxy can be approximated as $V_r=V_0+V(R) \sin{\xi} \cos{\theta}$ where $V_0$ is the systemic radial velocity, $\xi$ is the angle between the line-of-sight and the perpendicular to the plane of the galaxy, V(R) is the rotational velocity within the plane at a radial distance R, and $\cos{\theta} = X/R$ where X is the position along the major axis (Rubin \& Ford 1970).  If the rotation curve were flat ($V(R) = {\rm constant}$) this would be a linear relationship between $V_r$ and $X/R$.

\begin{figure}
\plotone{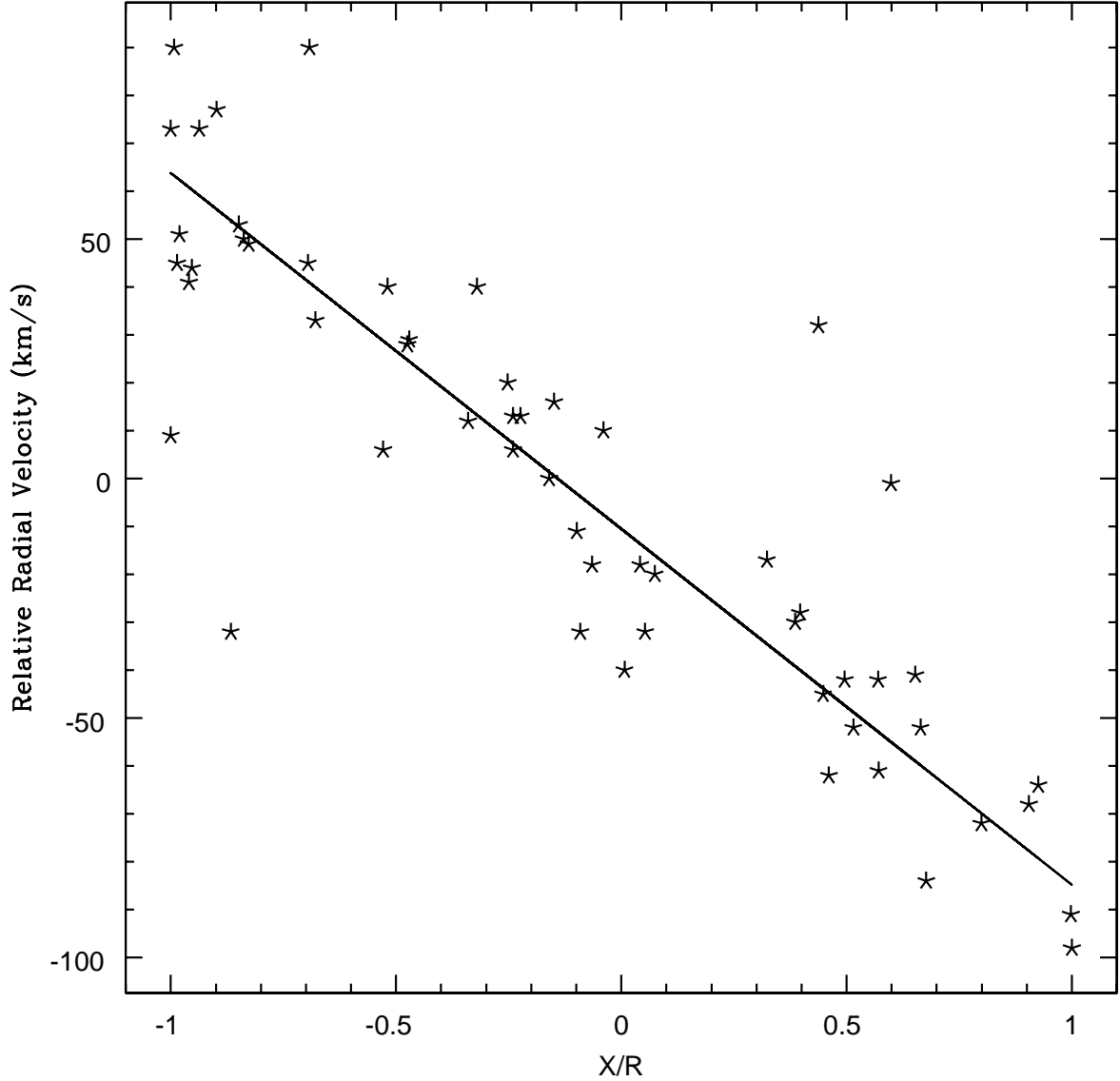}
\caption{\label{fig:HII} Radial velocity versus $X/R$ for 55 HII regions in M33 from Zaritsky et al.\ (1989).  Radial velocities were measured with respect to one of the HII regions near the center of the galaxy.  The relation between the two variables is well approximated by a linear fit (solid black line) indicating that the rotation curve of M33 is nearly flat.  Using this fit we are able to define an expected radial velocity for a bona fid M33 member as a function of position in the galaxy ($X/R$).}
\end{figure}

To assess whether such a linear relationship is a valid approximation for M33, we examine the radial velocities obtained for 55 HII regions in M33 by Zaritsky et al.\ (1989).  We first calculate the parameter $X/R$ for each HII region, assuming a position angle of 23 degrees and an inclination angle of 56 degrees (following Kwitter \& Aller 1981 and Zaritsky et al.\ 1989), and plot the results versus radial velocity in Figure~\ref{fig:HII}. This shows a clear linear trend, indicating the kinematics of M33 are dominated by dark matter. We note that Zaritsky et al.\ (1989) provides velocities relative to one of the HII regions in the central portion of the galaxy, rather than absolute velocities.  Thus, the data plotted in Figure~\ref{fig:HII} will only be able to provide us with the slope of the linear relation above, not the offset (i.e., the systematic velocity of M33).  After removing two outliers, a linear least square fit to the HII data yields a slope of $-74.3$ km s$^{-1}$ with a dispersion of 18 km s$^{-1}$.  Applying a systematic velocity of $-180$ km s$^{-1}$ (Kerr \& Lynden-Bell 1986) yields a relation for the expected radial velocity (in km s$^{-1}$) of an M33 member of:

$$V_{\rm expect} = -180 - 74.3 (X/R)$$ 

\begin{figure}
\plotone{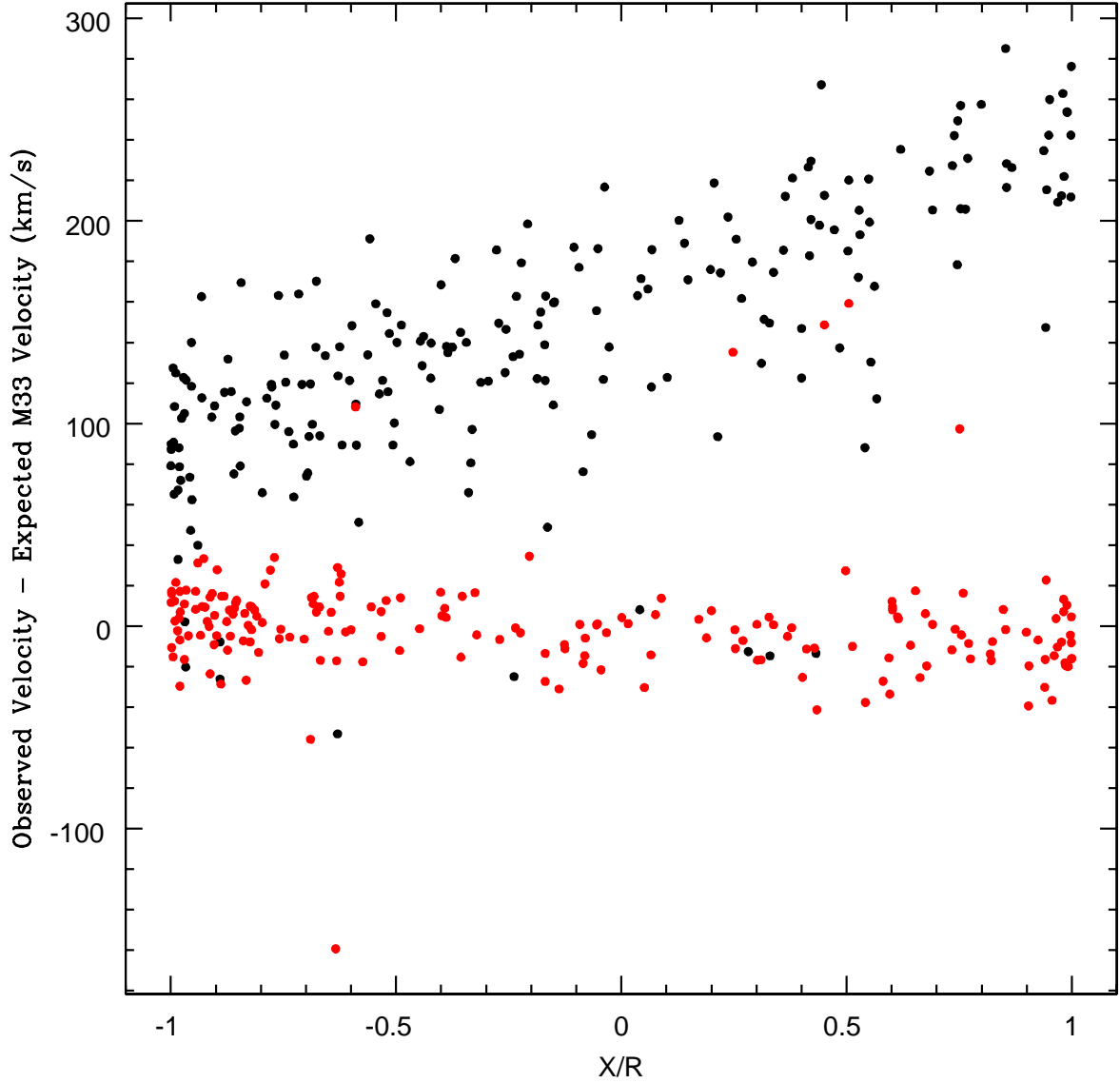}
\caption{\label{fig:XRDif:Red} Observed velocity minus expected M33 velocity versus $X/R$ for our sample of red stars.  Genuine M33 members should be centered about zero difference, while the foreground dwarfs make up the diagonal band.  Stars photometrically classified as supergiants are plotted in red.  One can see there excellent agreement between these two selection criteria.}
\end{figure}

Using this relation, we calculate the expected radial velocity for our 408 red program stars.  In Figure~\ref{fig:XRDif:Red} we plot the difference between our observed radial velocities and expected M33 velocities versus $X/R$ (which serves as a proxy for position within M33).  Within the plot, bona fide M33 members should be centered around zero difference while the foreground dwarfs (whose radial velocities should be approximately zero) make up the diagonal band.  The right-hand side of the plot represents the portion of M33 rotating \emph{towards} the Milky Way, where the velocity separation between foreground dwarfs and extragalactic supergiants is largest.  Stars which were photometrically selected as supergiants (as defined in Section~\ref{SC}) are plotted in red.  We immediately see that there is excellent agreement between the stars photometrically selected as supergiants, and stars kinematically consistent with M33 members.  

\begin{figure}
\plotone{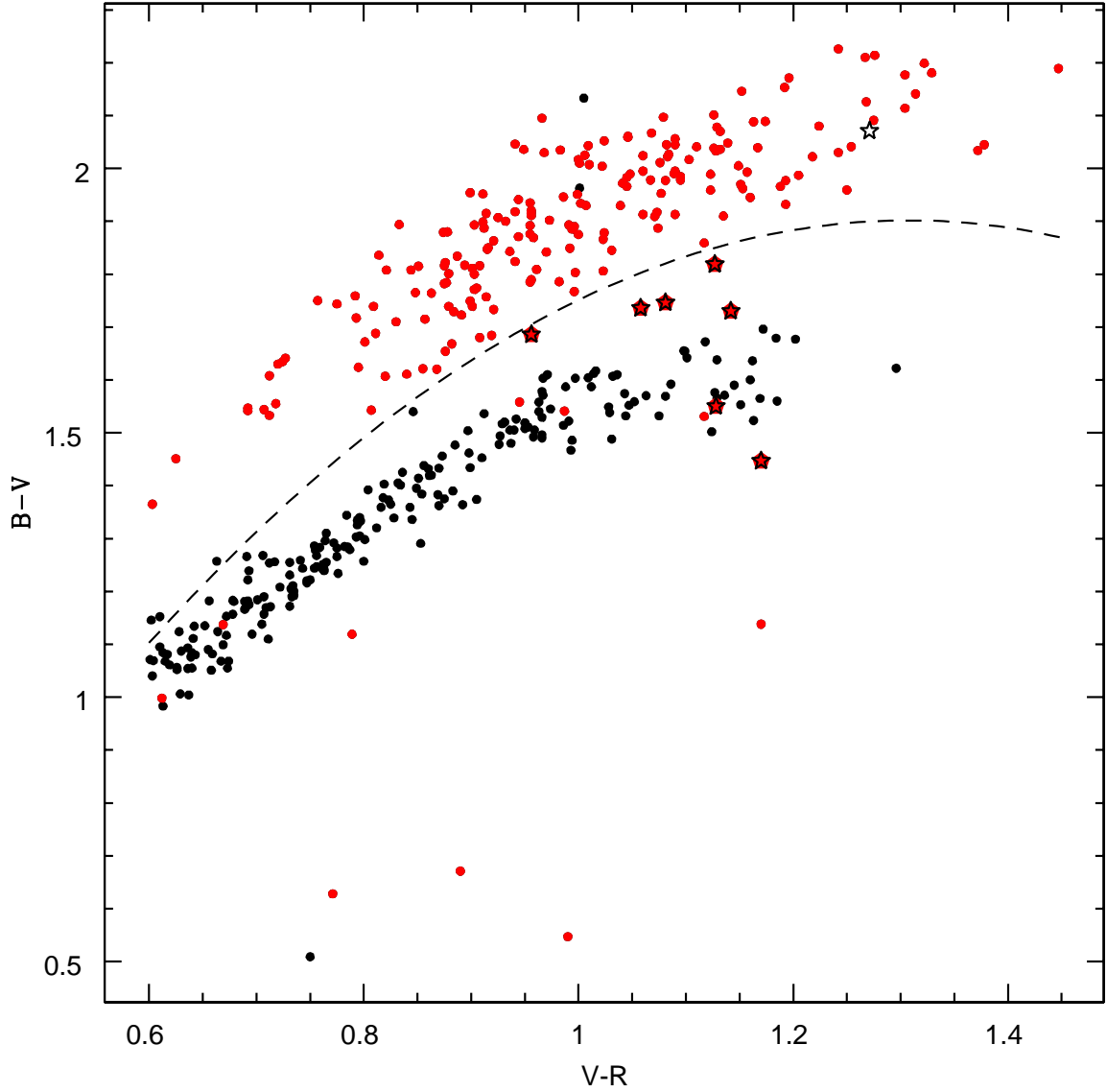}
\caption{\label{fig:BVVRprob} A two-color diagram for our observed sample of red stars (filled points) and the stars from Massy (1998) whose classifications are at odds with those predicted by the LGGS photometry (stars).  The dashed line represents our adopted photometric cutoff between supergiant candidates (above line), and likely foreground dwarfs (below line).  Stars whose radial velocities are consistent with the kinematics of M33 are plotted in red.}
\end{figure}

Although this is very promising, we wish to examine in more detail the handful of stars whose photometry and kinematics appear to offer conflicting results.  In Figure~\ref{fig:BVVRprob} we plot $B-V$ versus $V-R$ for our final sample of observed red stars.  The dashed line represents our adopted photometric cutoff between supergiant candidates and likely dwarfs, with stars above the line representing probable supergiants (i.e., stars above the dashed line are those presented in red in Figure~\ref{fig:XRDif:Red}).  Stars which would have been kinematically selected as supergiants (possessing an observed minus expected velocity less than 20 km s$^{-1}$) are plotted in red.
 
From this we see that of the five of the stars classified photometrically as supergiants (but with radial velocities inconsistent with M33) three lie very close to our photometric cutoff, and are more naturally associated with the band of likely dwarfs.  An examination of the two remaining stars reveals that they are highly contaminated by light from other objects in the $B$-band.  Had our initial criteria that the stars be `isolated' been based on $B$-band photometry (rather than $V$-band) these stars would not have been included in our sample.

To explain the stars which appear photometrically as dwarfs but kinematically as supergiants, we note that falling near the zero point of the y-axis in Figure~\ref{fig:XRDif:Red} does not necessarily \emph{prove} membership in M33.  A portion of M33's $-$180 km s$^{-1}$ systematic velocity is actually reflex motion of the sun (Courteau \& van den Bergh 1999) and we thus expect some halo dwarfs and giants to posses negative radial velocities.  An examination of the Atlas of Galactic Neutral Hydrogen by Hartmann \& Burton (1997) reveals that M33 only stands out from the Galactic clutter at radial velocities less (more negative) than $-$150 km s$^{-1}$. This suggests that a star with $V_{\rm obs} > - 150$ km s$^{-1}$ should not be selected as a member of M33 based on radial velocity \emph{alone}.  

\begin{figure}
\plotone{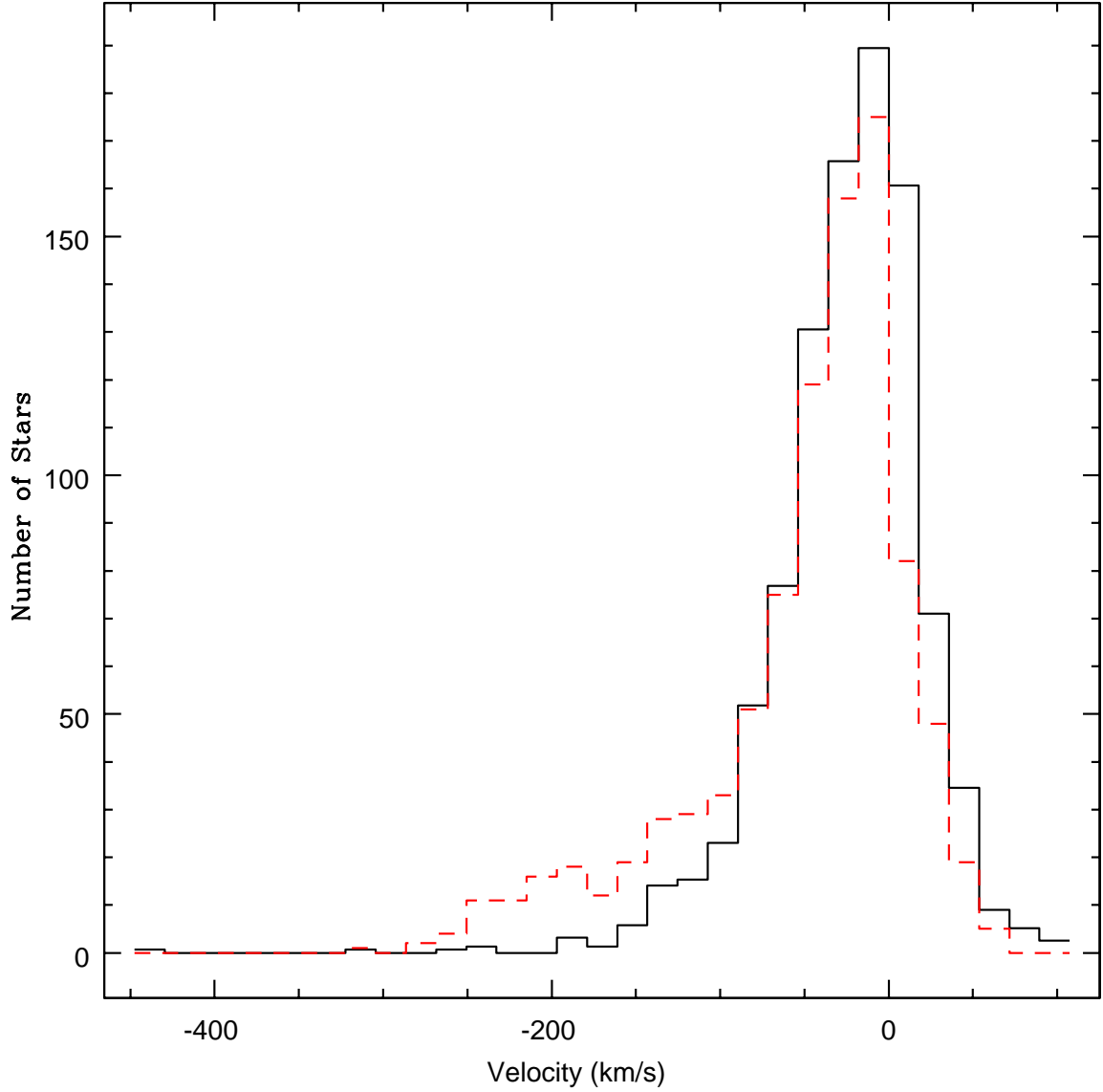}
\caption{\label{fig:Bescan} A histogram of the velocities predicted for red foreground dwarfs by the Besan\c{c}on model (black, solid) and the observed velocities of our red stars (red, dashed).}
\end{figure}

We additionally examine the velocities of red foreground stars (covering the same area as our observations) predicted by the Besan\c{c}on model for Galactic dwarfs (Robin et al.\ 2003).  In Figure~\ref{fig:Bescan} we plot a histogram of the velocities predicted by the model (solid, black) and the velocities of our observed red sample (dashed, red).  From this we see an excess of stars at high radial velocities in our observed sample, representing the genuine M33 RSGs.  We note that the peak of stars centered about 0 km s$^{-1}$ (the foreground dwarfs) is much smaller in our observed sample because we only observed a fraction of the red stars in our initial catalog.  We choose not to scale the output of the Besan\c{c}on model by this factor because during our sample selection we favored stars we believed to be supergiants based on our photometric criteria.  Thus, the fraction of total stars we observe is likely not representative of the fraction of foreground dwarfs we observe.  From this histogram we expect to find a handful of foreground dwarfs scattering out to approximately $-260$ km s$^{-1}$.  A foreground star with such a velocity would fall within 20 km s$^{-1}$ or below its expected M33 velocity over the entire range of $X/R$ in Figure~\ref{fig:XRDif:Red}.

As a final check, we re-examine the 65 red stars toward M33 which were classified as RSGs
by Massey (1998). Sixteen of these stars were also observed as part of our sample, and we find complete agreement between our classifications and those of Massey (1998).  In addition, 58 of the stars now have improved photometry as a result of the Local Group Galaxies Survey.   When we place these stars on the $B-V$ versus $V-R$ diagram (with the improved LGGS photometry) we find that for eight of the stars the classifications given by Massey (1998) would not be supported by our photometric cut.  These are plotted as stars in Figure~\ref{fig:BVVRprob}.  An examination reveals that all eight stars are located in very crowded regions of M33, thus validating our decision to include only isolated stars in our sample. 

Thus, we conclude that it is likely possible for (1) some foreground star to posses radial velocities as (or more) negative than genuine M33 members over the entire surface of the galaxy and (2) stars which are sufficiently crowded to appear to possess colors in disagreement with our photometric selection criteria.  Despite these facts, when considering only ($V$-band) isolated stars, our two selection criteria agree for 96\% of our sample. We therefore believe that the two-color discriminant described in Massey (1998), does, in fact, act well at discerning RSGs at M33 metallicities.  For the purposes of comparing our results to the current evolutionary models we divide our stars into three ranks: rank one stars were selected as supergiants both photometrically and kinematically, rank 2 stars were selected by one method, but not the other, and rank 3 stars were selected by neither method and are likely foreground dwarfs.  After checking our rank 1 and rank 2 samples for known stellar clusters (finding none) and assigning the three stars initially selected photometrically as supergiants, but which fall much closer to the band of foreground dwarfs as rank 3 stars, we are left with 189 rank 1 stars, 12 rank 2 stars, and 207 rank 3 stars. 
 
Prompted by the two stars described above which were included in our sample, despite being crowded in the $B$-band, we also examine the level of crowding of our rank 1 stars in the $B$- and $R$-bands.  Only two stars showed moderate crowding in the $R$-band, but 67 stars ($\sim$35 \% of our rank 1 stars) showed moderate to high crowding in the $B$-band.  Given that all of these objects have observed minus expected radial velocities less than 20 km s$^{-1}$, we do not believe that this crowding invalidates our classification of these stars as very probable M33 RSGs.  However, we do indicated rank 1 and rank 2 stars which possess moderate levels of crowding in the $B$-band in Table~\ref{tab:TableMeas:Red}.

\subsection{\label{yellow}Membership Determination: Yellow Supergiants}

\subsubsection{Radial Velocities}

We now turn to our yellow sample and begin by examining their radial velocities.  In Section~\ref{reds}, we characterized the rotation curve of M33 based on the radial velocities of 55 HII regions from Zaritsky et al.\ (1989).  However, these radial velocities were obtained in a relative, as opposed to absolute (heliocentric), sense and often had high associated errors (10 to 15 km s$^{-1}$).  Thus, to calculate the expected radial velocities for our yellow sample, we opt to use a characterization of M33's rotation curve based on our probable red supergiants.

\begin{figure}
\plotone{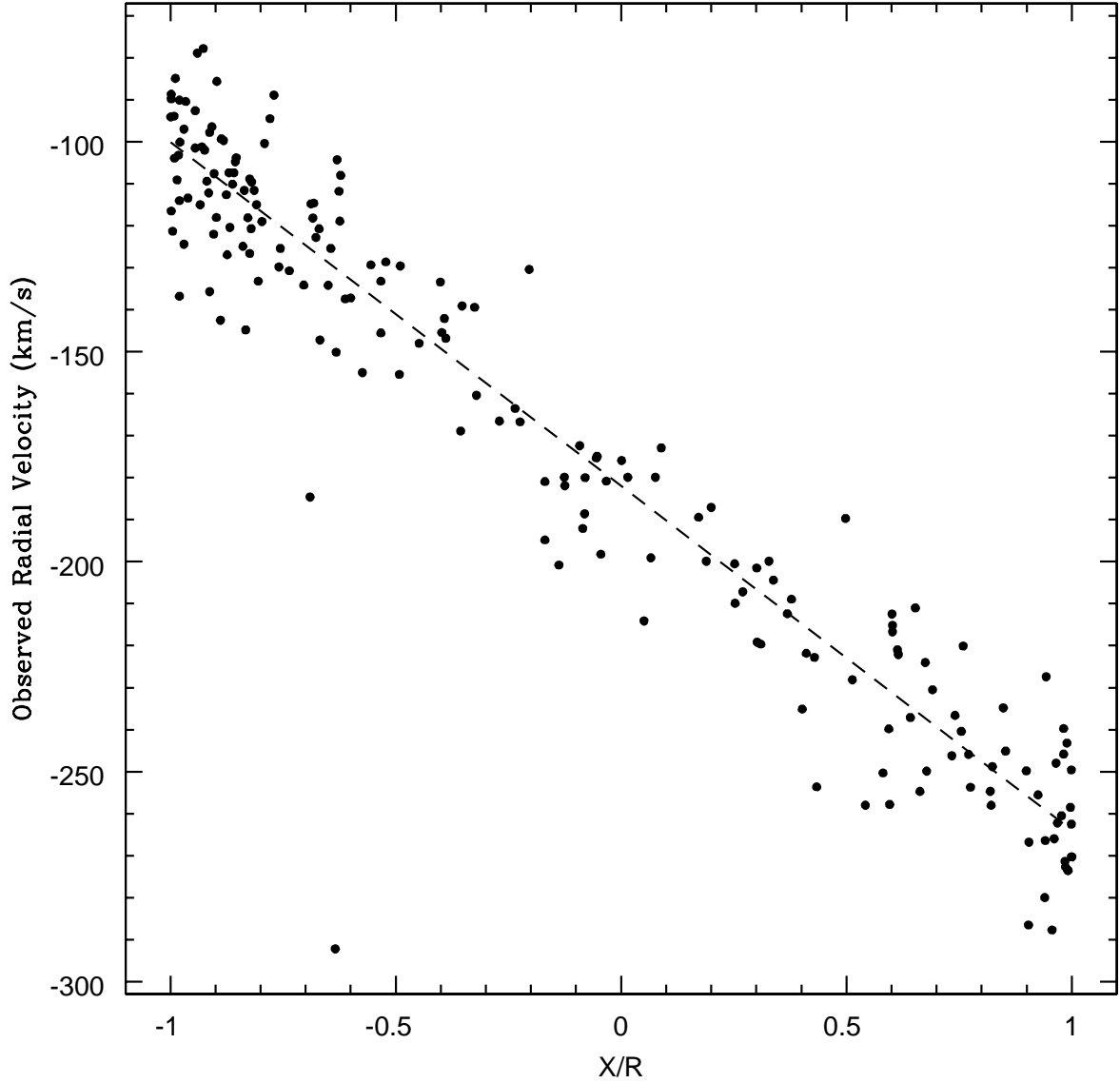}
\caption{\label{fig:VelXRRed} Observed radial velocity versus $X/R$ for our rank 1 red supergiants.  The dashed line represents a linear fit to the data, which we use to construct a more robust formulation of the expected radial velocity for a member of M33.}
\end{figure}

In Figure~\ref{fig:VelXRRed} we plot $X/R$ versus observed radial velocity for our rank 1 red supergiants.  A linear least-square fit to the sample yields a relation (in km s$^{-1}$) of: 
$$V_{\rm expect} = -182 - 81.9 (X/R)$$ 
with a dispersion of 14.3 km s$^{-1}$.  We see that the slope produced by this relation varies slightly from that produced by the HII regions of Zaritsky et al.\ (1989).  We also note that this relation accurately reproduces the systemic velocity of M33 ($-180 \pm 3 $km s$^{-1}$; Kerr \& Lynden-Bell 1986).

\begin{figure}
\plotone{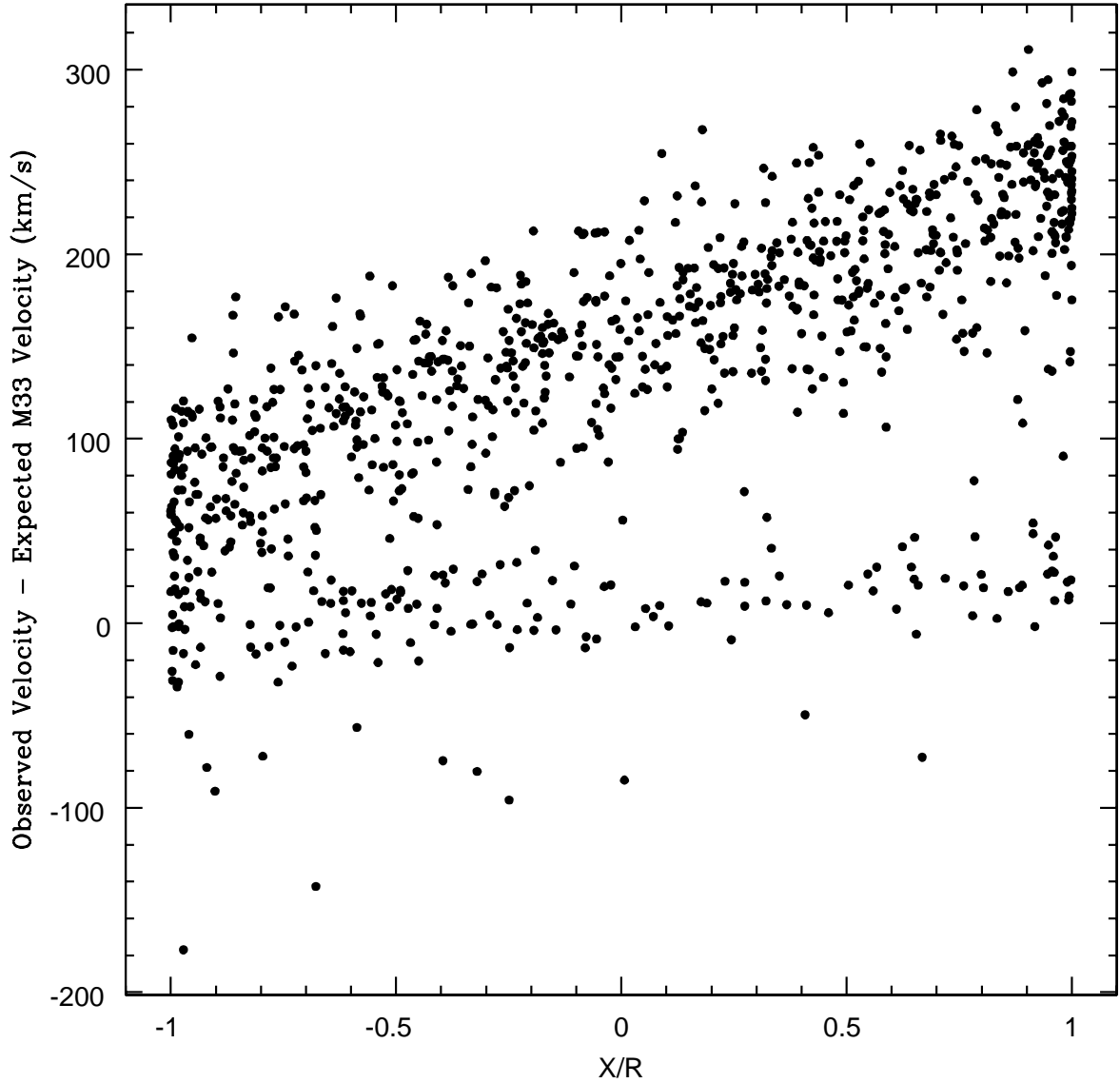}
\caption{\label{fig:XRDif:Yellow} Observed velocity minus expected M33 velocity versus $X/R$ for our sample of yellow stars.  As in Figure~\ref{fig:XRDif:Red} genuine M33 members should fall near the zero point of the y-axis, while we expect a majority of the foreground dwarfs to scatter about $V_{\rm obs} = 0$ km s$^{-1}$ (thus making up the strong upward slanting diagonal band).}
\end{figure}

Using this relation, we calculate the expected radial velocity for each of our yellow stars, were it a genuine M33 member.  In Figure~\ref{fig:XRDif:Yellow} we plot the observed minus expected radial velocities versus $X/R$ for our yellow sample.  As in Figure~\ref{fig:XRDif:Red} for our red sample, we expect the genuine M33 members to fall near the zero point of the y-axis.  We note that although two bands of stars are evident in this figure, the separation is not nearly as distinct as seen in Figure 9 of Drout et al.\ (2009) for the case M31 (where both the systematic and rotational velocities are larger than the modest values of M33).  It is not until the value of $X/R$ approaches zero that the two bands are noticeably distinct.  Thus, if we were to restrict our sample of potential supergiants to those which Figure~\ref{fig:XRDif:Yellow} indicates posses a modest separation from the bulk of Milky Way dwarfs, we would be left only with stars in the NW portion of M33.

\begin{figure}
\plotone{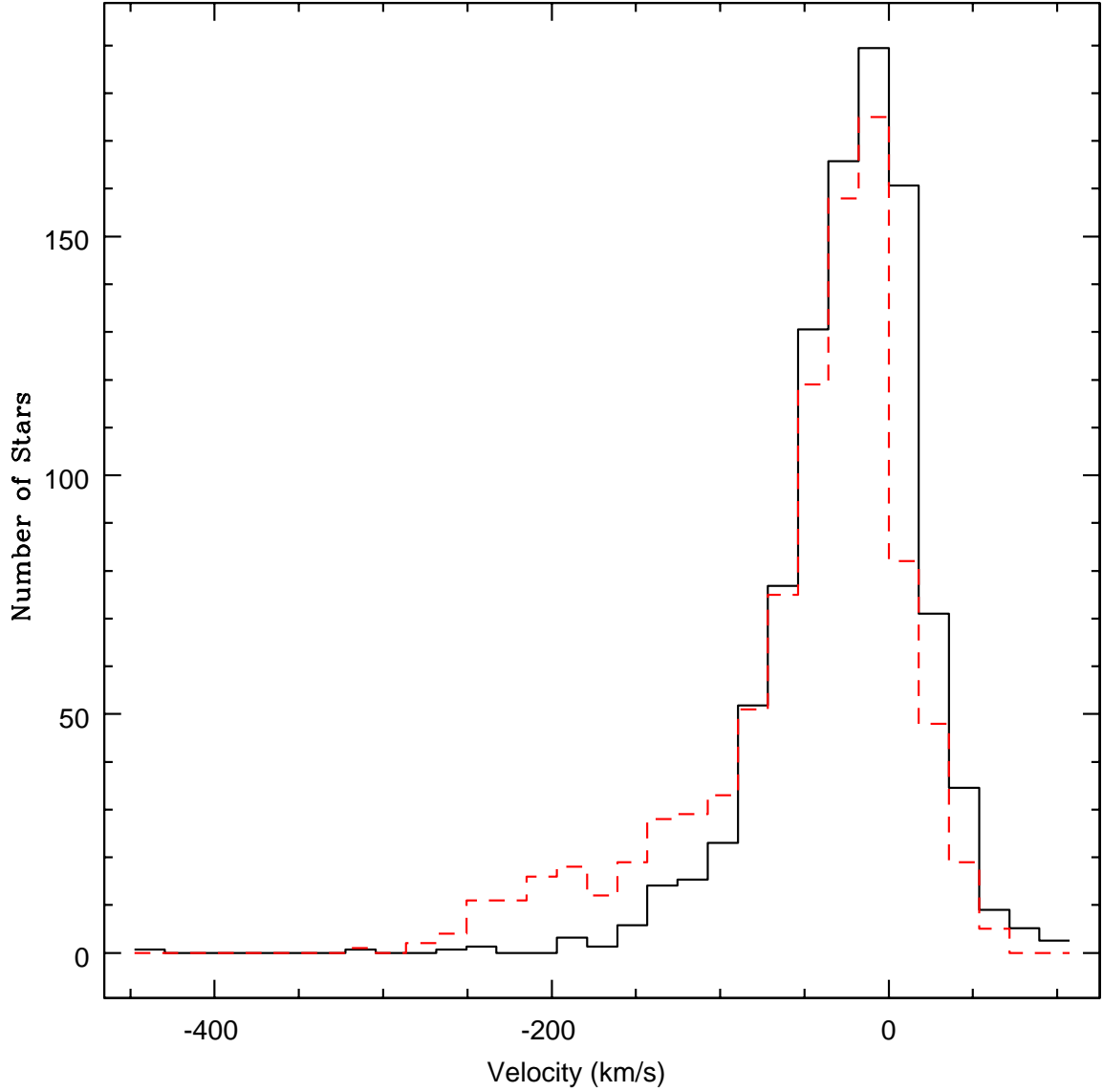}
\caption{\label{fig:BesYellow} A histogram of the radial velocities predicted from the Besan\c{c}on model for galactic dwarfs for stars with the same photometric criteria and within the same area as our yellow sample (black, solid), and the observed radial velocities for our yellow sample (red, dashed).}
\end{figure}

Additionally, the Besan\c{c}on model for galactic dwarfs applied to the color and magnitude range of our yellow sample reveals that we can expect modest contamination from high velocity foreground dwarfs across the entire face of M33.  In Figure~\ref{fig:BesYellow} we plot a histogram of the velocities predicted by the Besan\c{c}on model (solid, black) and the observed velocities of our yellow sample (dashed, red). In this figure, the output of the foreground model has been scaled by the fraction of our initial yellow catalog we actually observed.  We see that at a radial velocity of approximately $-200$ km s$^{-1}$ (corresponding to $X/R = 0.22$ for an typical M33 member in our formulation of M33's rotation curve) a majority of the stars in our sample should be genuine M33 members.  However, at a velocity of $-170$ km s$^{-1}$ ($X/R = -0.15$ for a typical M33 member) a significant portion of our sample is likely composed of foreground dwarfs.  Based on this model, we can predict two foreground stars with velocities beyond $-300$ km s$^{-1}$, 9 foreground stars beyond $-200$ km s$^{-1}$, and 17 foreground stars beyond $-150$ km s$^{-1}$ (the region in which we also expect to find 65\% of our genuine supergiants).  Thus, we conclude that even in M33, which possesses a modest radial velocity separation, it would be prudent to obtain an independent means of membership determination to complement the radial velocity data.   

\subsubsection{OI $\lambda$7774 triplet}

To address this issue, we now examine the possibility of using the OI $\lambda$7774 triplet as a luminosity indicator at M33 metallicities.  Osmer (1972) demonstrated that OI $\lambda$7774 is one of the strongest spectral features in high luminosity A- to F-type supergiants, yet it is relatively weak in dwarfs at galactic metallicities.  Przybilla et al.\ (2000) later showed that this feature is a result of non-LTE atmospheric effects, exacerbated by sphericity.  Although promising, it is not yet understood whether this luminosity dependence extends to the cooler G-type supergiants and what effect, if any, metallicity creates. 

\begin{figure}
\plotone{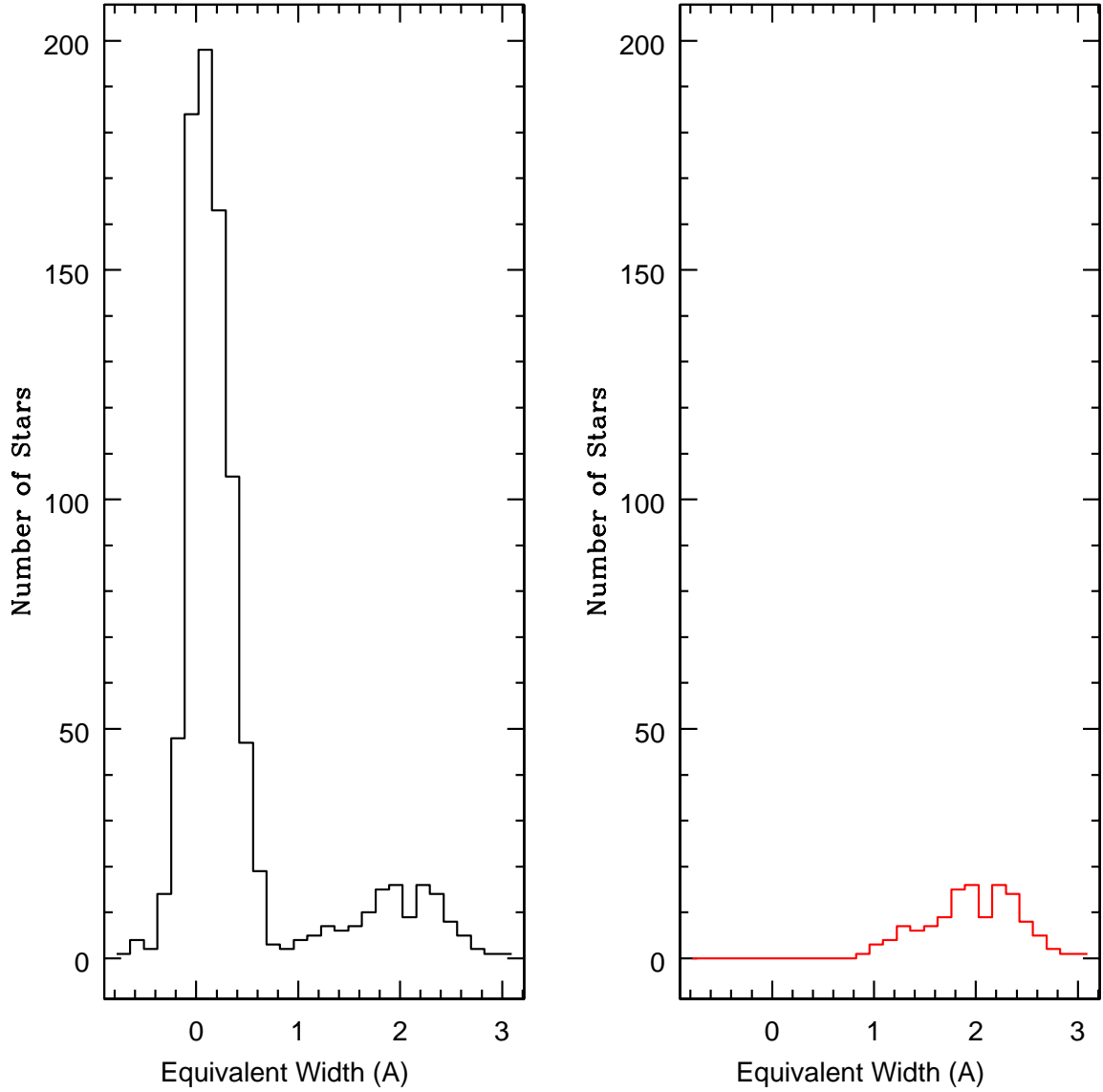}
\caption{\label{fig:ewhist} \emph{left:} A histogram of the equivalent width of the deepest spectral feature in the wavelength range 7760 $-$ 7785~\AA\ for out sample of yellow stars. The distribution is distinctly bimodal. \emph{right:} A histogram of equivalent widths for only stars with a strong OI $\lambda$7774 feature.  In almost every case, an equivalent width above 1~\AA\ represents a strong OI $\lambda$7774 feature.}
\end{figure}

As an initial test we measured the equivalent width of the deepest spectral feature in the wavelength range 7760 - 7785 \AA.  The continuum level was determined by averaging the intensity over a length of the normalized spectra at both longer and shorter wavelengths than the desired line.  These ranges for continuum averaging were taken at least 10~\AA\ away from the line, and the equivalent width was calculated in a wavelength range $\pm 5$~\AA\ from the wavelength of minimum intensity.  For any stars observed in multiple fields, we adopt the average of the measured equivalent widths. Figure~\ref{fig:ewhist} shows a histogram of the results.  We see that the distribution is distinctly bimodal, with a separation at approximately 1~\AA.

However, we caution that a strong spectral feature in the wavelength range given above does not \emph{necessarily} imply a strong OI $\lambda$7774 line.  This is demonstrated by Kirby et al.\ (2008) who generated synthetic spectra in the temperature range $4000 - 8000$ K for a range of surface gravities well matched to our sample.  Using the molecular transitions from Kurucz (1992), they find CN is the most important molecule at red wavelengths for these stellar parameters.  Most notably for our purposes, this includes a CN band head at $\sim$7774~\AA\ along with companion features at $\sim$7865~\AA\ and $\sim$7938~\AA\footnote{We are indebted to Nelson Caldwell for first
noting to us the possible confusion of the OI $\lambda 7774$ feature with the CN band head.}.

\begin{figure}
\epsscale{0.4}
\plotone{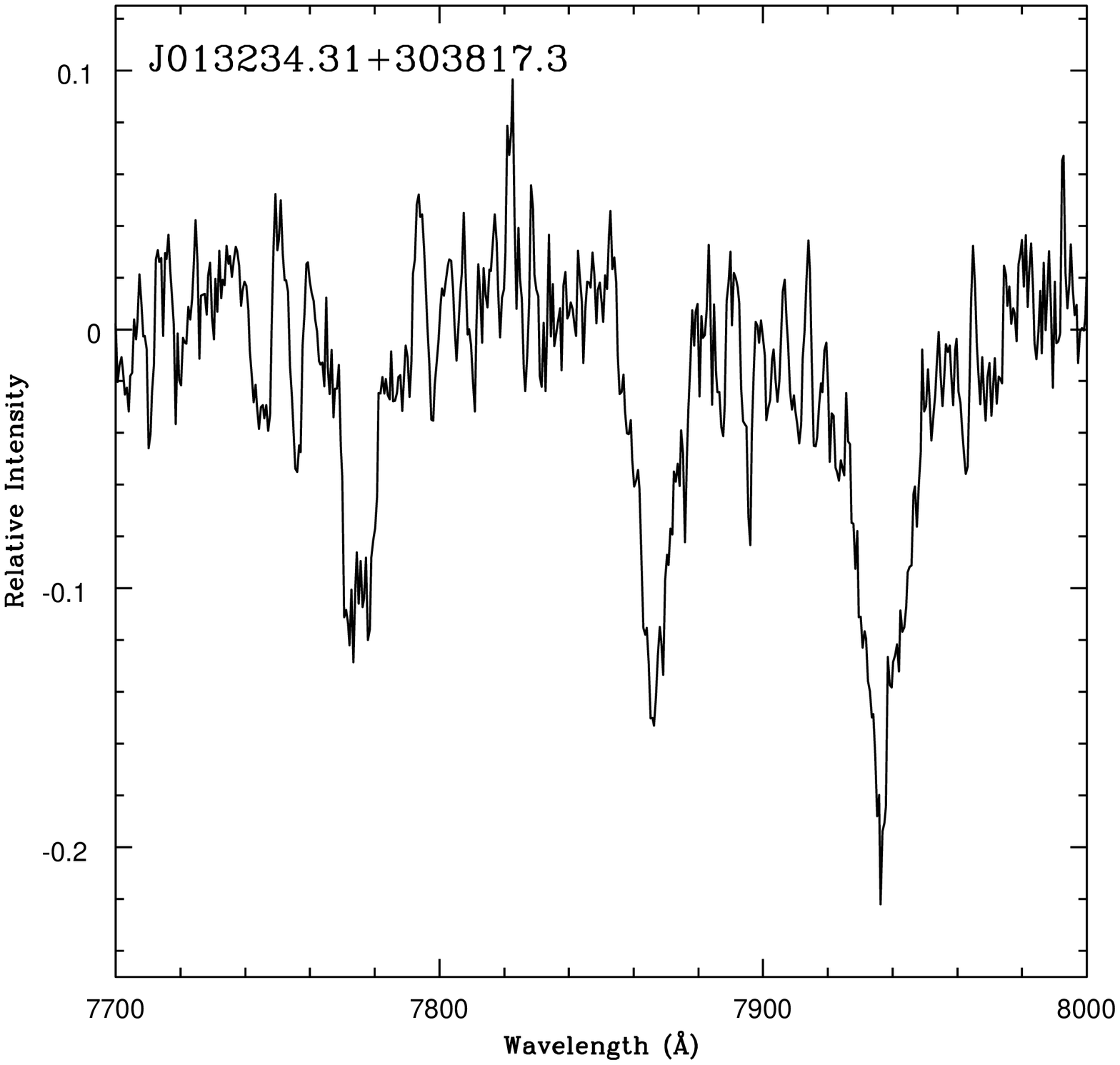}
\plotone{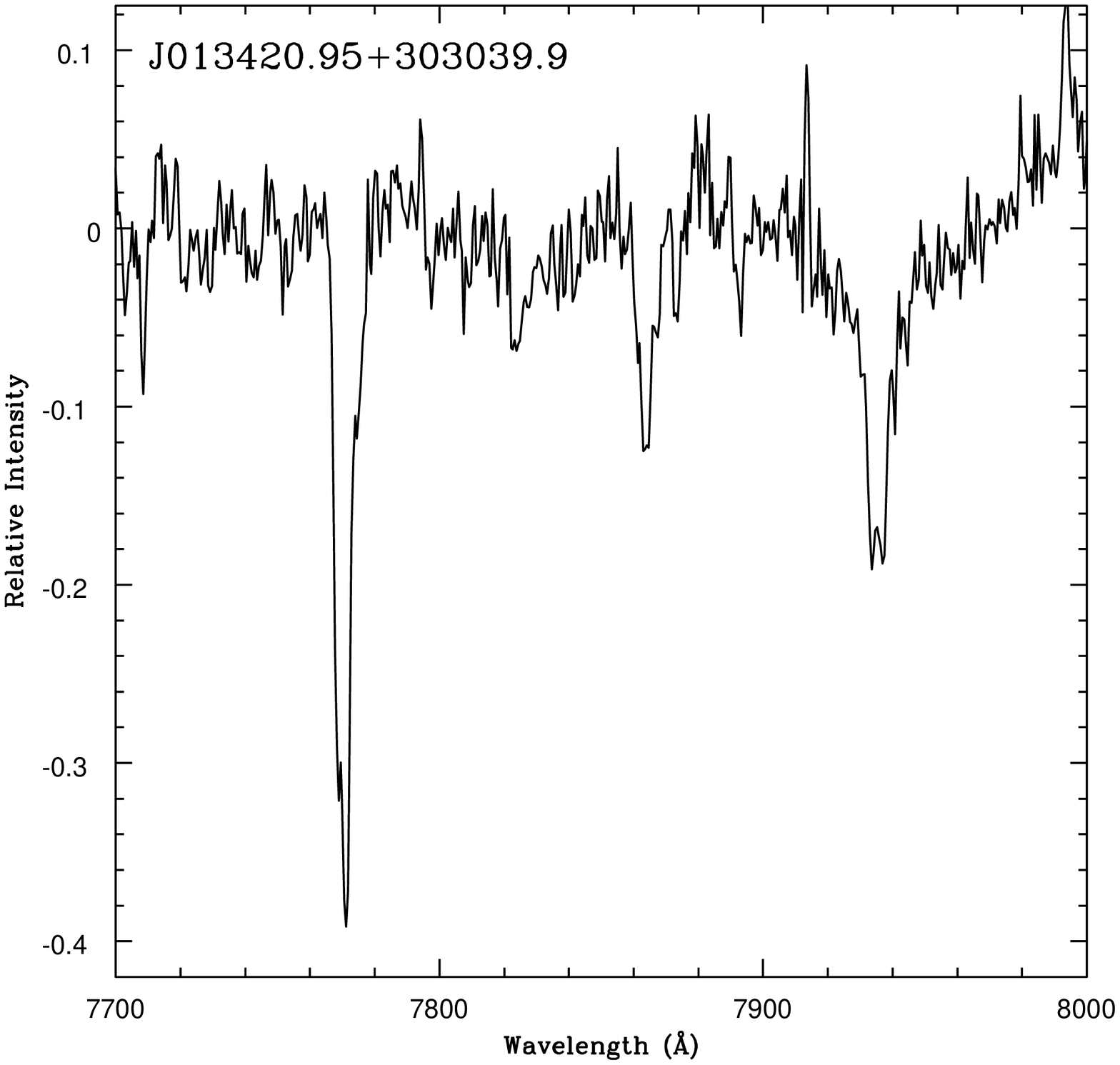}
\plotone{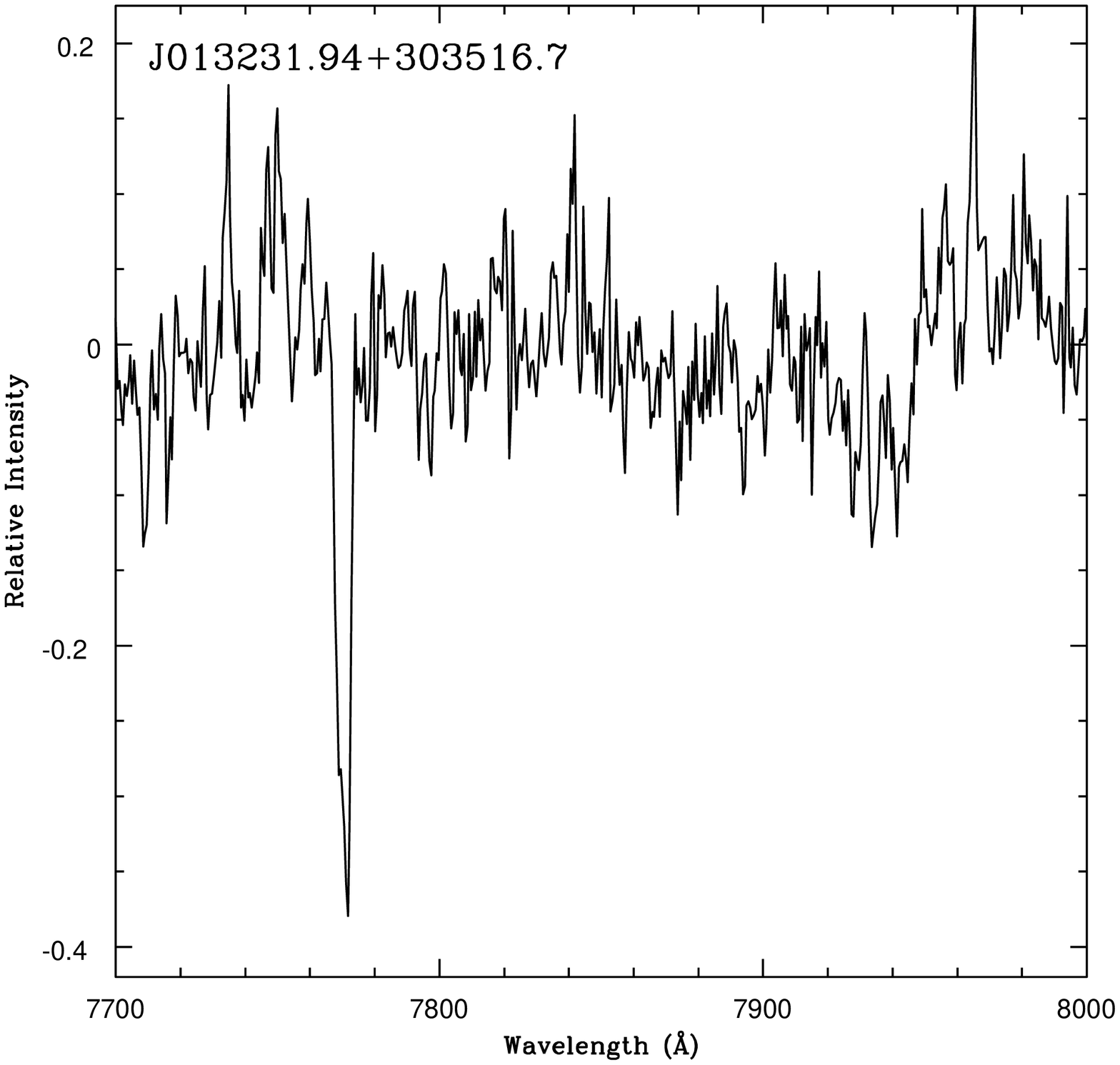}
\caption{\label{fig:OIspec} Sample yellow spectra in the wavelength range 7700 $-$ 8000~\AA. \emph{top:} Three CN band heads with a progression of lines depths. \emph{middle:} Two CN band heads are still evident, but the feature at $\sim$7774~\AA\ has a much larger relative depth, representing the presence of a strong OI $\lambda$7774 feature. \emph{bottom:} An isolated OI $\lambda$7774 feature.} 
\end{figure}

Thus, in order to determine whether the population of stars with high equivalent widths represents a sample of M33 supergiants with a strong OI $\lambda$7774 feature, we examine all stars with an initial equivalent width greater than 0.5~\AA, all stars with a velocity difference less than 100 km s$^{-1}$ in Figure~\ref{fig:XRDif:Yellow}, and all the stars observed in the 2009 fnt-4 field.  We distinguish a CN band head near $\lambda$7774 from a strong OI $\lambda$7774 feature by (1) the presence or absence of the other two band heads and (2) the width of the feature.  We find that our spectra fall roughly into four categories: spectra with no obvious features around $\lambda$7774, spectra with three broad CN features which posses a natural progression of line depths, spectra with features at all three band head wavelengths, but whose feature at $\lambda$7774 is deeper and narrower than the other two, and spectra with a single, narrow feature at $\lambda$7774.  Examples of the latter three are shown in Figure~\ref{fig:OIspec}.

We believe that spectra falling in the latter two categories possess real OI $\lambda$7774 features.  The red histogram in Figure~\ref{fig:ewhist} (right) represents the equivalent widths from these stars.  We see that in almost every case the strongest equivalent widths represent real OI $\lambda$7774 features.  We also note that, with the exception of a handful of spectra (all of which possessed particularly high noise levels) the spectra with strong OI $\lambda$7774 are the same spectra which possess Paschen lines (see Section~\ref{Vels}). This is promising, as the Paschen lines are expected to remain stronger to later spectral types for supergiants than for giants and dwarfs (Andrillat et al.\ 1995).

In Table~\ref{tab:TableMeas:Yel} we list our measured equivalent widths for only the stars we believe have true OI $\lambda$7774 features.  
Propagating the errors in our calculation of the mean continuum level leads to errors in our measured equivalent widths on the order of 0.02~\AA, while a comparison of the two equivalent widths measured for stars observed in both 2009 and 2010 reveal an average difference of 0.02~\AA\ with a standard deviation of 0.36~\AA.

\subsubsection{Radial Velocities and the OI $\lambda$7774 triplet}

\begin{figure}
\epsscale{1}
\plotone{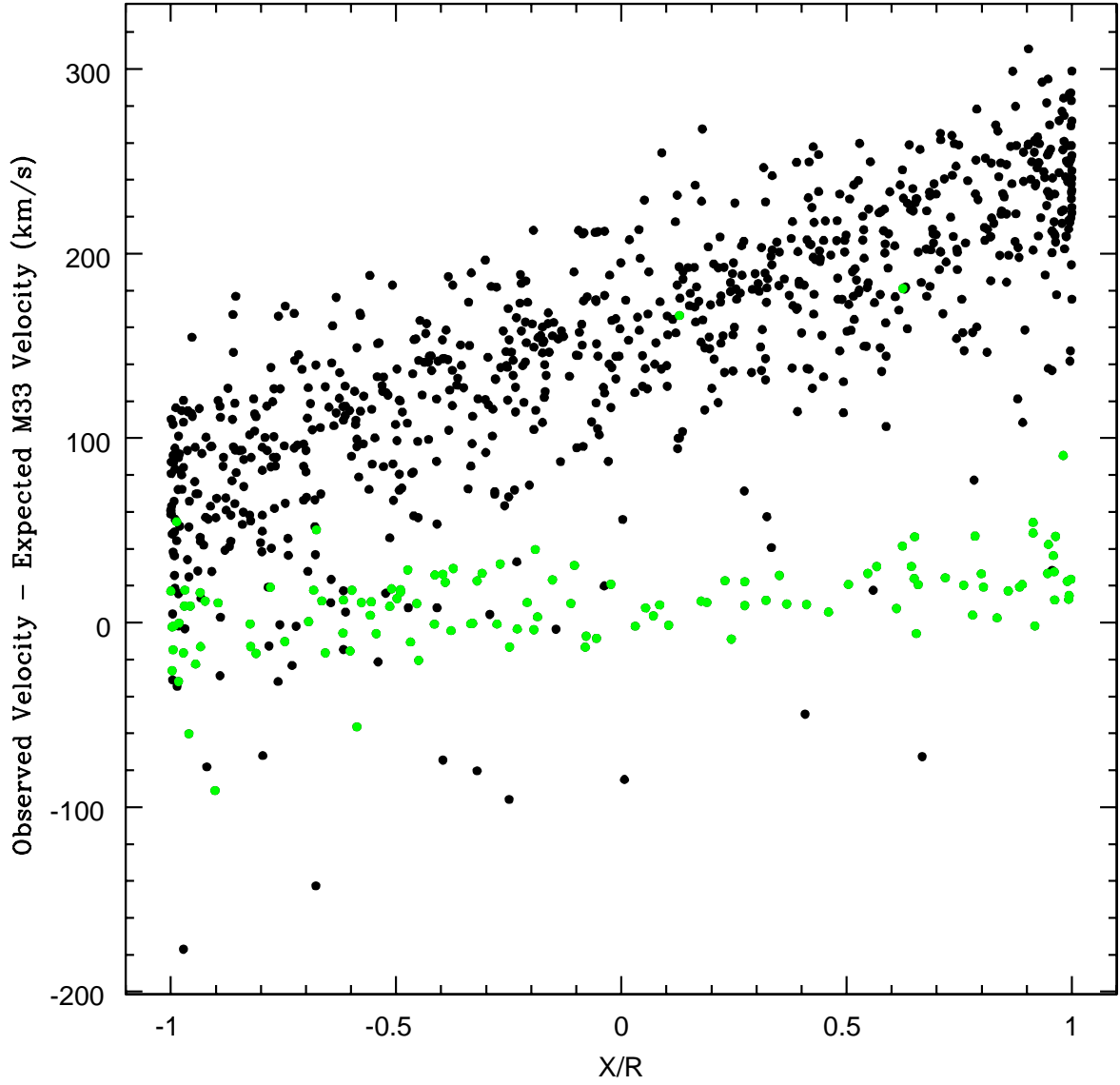}
\caption{\label{fig:XRDif_Ew} Observed minus expected M33 velocity versus $X/R$ for our yellow sample.  Green points represent stars whose spectra show a strong OI $\lambda$7774 feature.}
\end{figure}

In Figure~\ref{fig:XRDif_Ew} we again plot the observed minus expected velocities versus $X/R$ for our yellow sample.  The green points now represent stars with significant amounts of OI $\lambda$7774.  We see that there is excellent agreement between stars with a distinguishable OI $\lambda$7774 feature and stars which have radial velocities consistent with the kinematics of M33.  This indicates that the OI $\lambda$7774 triplet can, indeed, be used as a luminosity indicator for F- and G-type stars at the metallicity of M33.  

We do note that there are (1) two stars who seem to possess significant amounts of OI $\lambda$7774 but whose velocities place them in the middle of the strong band of foreground dwarfs, and (2) a handful of stars who possess velocities consistent with, or more negative than, their expected M33 velocities, but which do not possess significant amounts of OI $\lambda$7774. 

To explain the latter, we first checked our sample for possible non-stellar objects.  Using the R-band images of the LGGS we measured the FWHM of the point-spread function for objects in our sample with either an observed minus expected velocity less than 100 km s$^{-1}$ or significant amounts of OI $\lambda$7774.  This process revealed seven previously known clusters in our sample: J013350.92+303936.9, J013419.85+303613.1, J013456.51+304100.4, J013400.19+303747.3, J013403.30+302756.1, J013256.08+303826.0, J013418.67+303137.7, and one additional extended object on the outskirts of the LGGS images: J013224.07+301242.9.  Despite this fact, our sample still contains eight stars with $-200 > V_{\rm obs} > -300$ km s$^{-1}$, and one star with $V_{\rm obs} < -300$ km s$^{-1}$ which do not possess noticeable amounts of OI $\lambda$7774.  As described above, however, this is consistent with the numbers predicted by the Besan\c{c}on model for foreground stars.  Thus, although it is not impossible for our sample to contain some bona fide M33 members \emph{without} obvious OI  $\lambda$7774, we find our data is not inconsistent with \emph{all} F- and G-type supergiants in M33 possessing significant amounts.  A more full examination of the temperature and metallicity dependence of this line will be addressed in a future paper.

For the purpose of comparing our sample to the current evolutionary models we assign the following ranks to our yellow sample: rank 1 stars possess both velocities consistent with M33 and significant amounts of OI $\lambda$7774, rank 2 stars have \emph{either} a velocity difference less than 60 km s$^{-1}$ and an $V_{\rm expect}$ less than $-150$ km s$^{-1}$ \emph{or} significant amounts of OI $\lambda$7774 and a velocity inconsistent with M33, rank 3 stars have a velocity difference less than 60 km s$^{-1}$ and $V_{\rm expect}$ greater than $-150$ km s$^{-1}$, and the rest of the sample is designated as rank 4.  Even if it is possible for a genuine supergiant in our sample to lack a noticeable OI $\lambda$7774 feature, we expect our rank 3 sample to be \emph{heavily} contaminated by foreground dwarfs (recall from Section~\ref{reds} that M33 only stands out from the galactic clutter at $-150$ km s$^{-1}$).  With these definitions our sample contains 121 rank 1 stars, 14 rank 2 stars, 68 rank 3 stars, 705 rank 4 stars, and 8 clusters.  This corresponds to a foreground contamination of $\sim$83\%.

\begin{figure}
\plotone{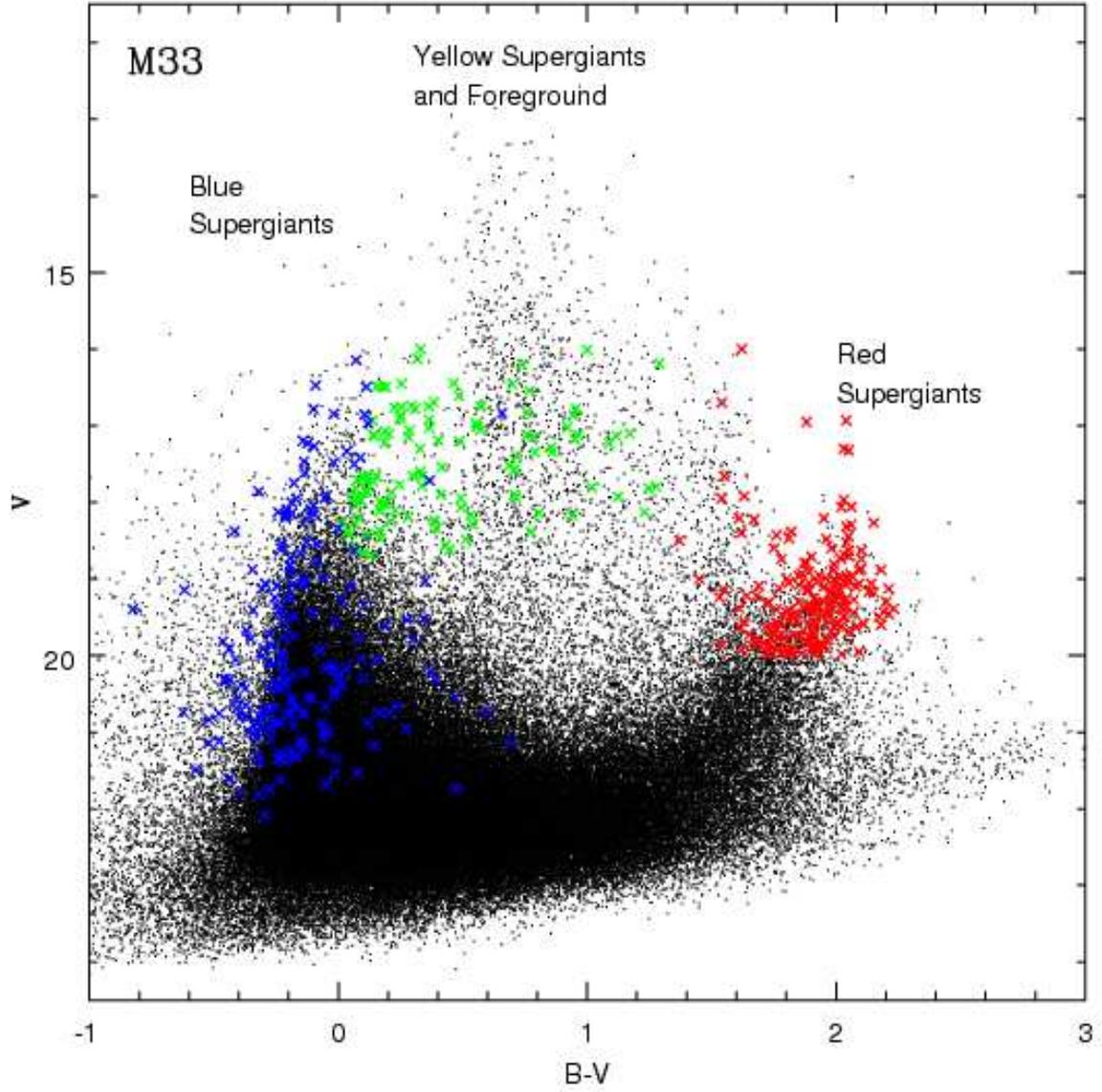}
\caption{\label{fig:CMDpoints} CMD of M33 with evolved massive star populations overlaid.  Same as Figure~\ref{CMD} but with our newly determined RSG and YSG populations (red and green $\times$'s, respectively), as well as the WR population (blue $\times$'s) of Neugent \& Massey (2011) overlaid.}
\end{figure}

\begin{figure}
\plotone{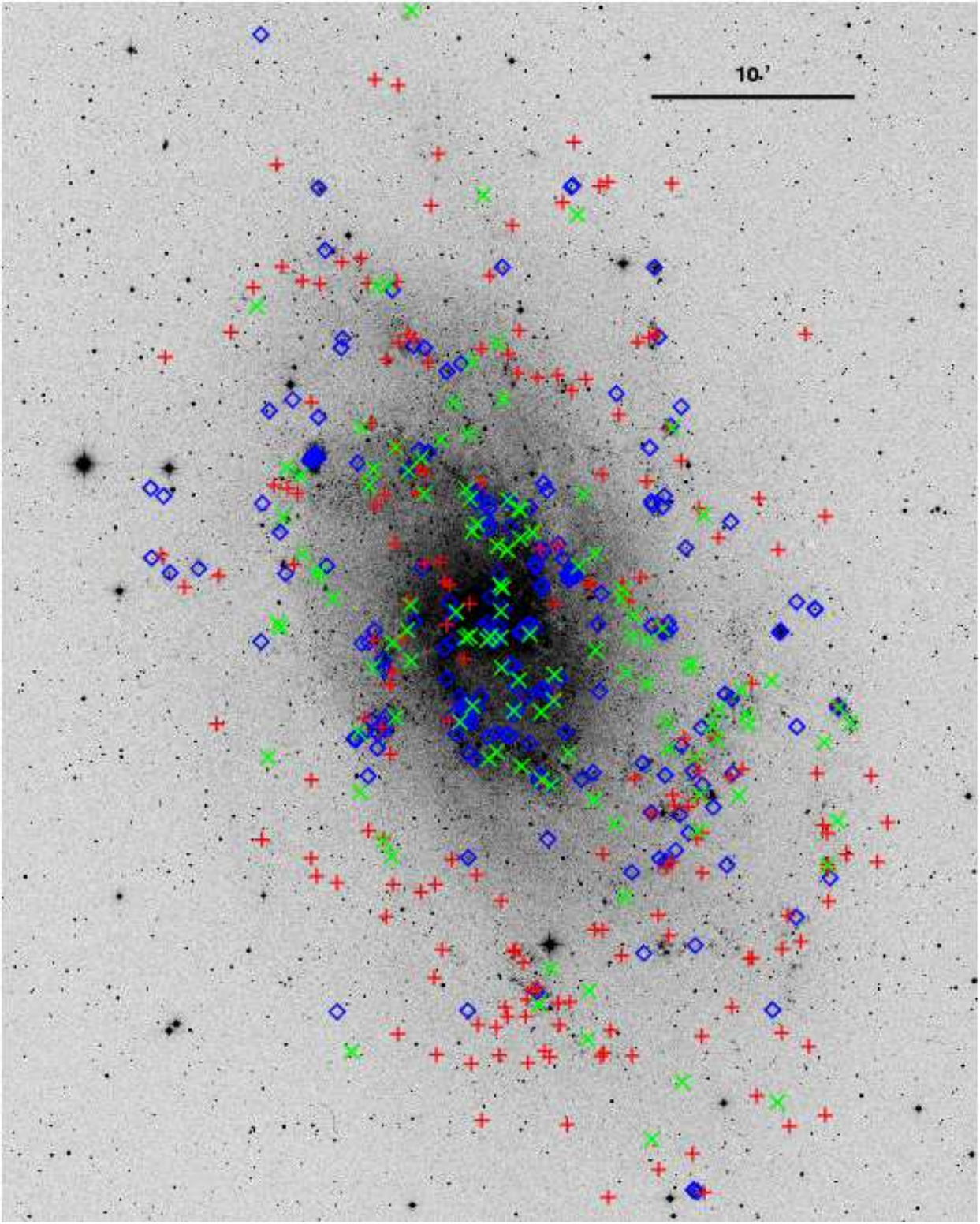}
\caption{\label{fig:locations} Spatial distribution of our newly determined RSG and YSG populations (red $+$'s and green $\times$'s, respectively) , as well as the WR population (blue diamonds) of Neugent \& Massey (2011) in the disc of M33. The lack of RSGs in the centeral region of the galaxy represents an observational bias (see text).}
\end{figure}

In Figure~\ref{fig:CMDpoints} we present the location of our newly determined red and yellow supergiants on the CMD of M33.  Black dots represent the LGGS sample, red $\times$'s our rank 1 RSGs, green $\times$'s our rank 1 YSGs, and blue $\times$'s the M33 WR population, determined by Neugent \& Massey (2011).  In Figure~\ref{fig:locations} we also show the spatial distribution of all three populations in the disc of M33.  We note that the lack of RSGs in the central region of the galaxy is an observational bias, created by our restriction to isolated red stars as described in Section~\ref{Obs}.  

\subsection{Variability \label{Var}}

Although relatively stable in the $K$-band, some RSGs are known to possess variations in the 
$R$-band from 0.2 $-$ 1 mag across a range of metallicities (Josselin et al.\ 2000, Levesque et al.\ 2007, Massey et al.\ 2009).  In some cases this effect is at least partially due to variable dust extinction (Massey et al.\ 2005, Massey et al.\ 2007a), although Levesque et al.\ (2007) also examine a handful of late-type RSGs in the Magellanic Clouds whose variability seem to be due to physical changes in the stars themselves.  These unusual stars may inhabit the Hayashi forbidden region and, hence, be hydrostatically unstable.  Given the potential insight that variability can provide we wish to examine which, if any, of our stars are known to be variable.  

Macri et al.\ (2001), Hartman et al.\ (2006), and Javadi et al.\ (2011) present the most recent catalogs of variable point sources in M33.  Macri et al.\ (2001) identify 544 BVI variables in two fields in the central region of M33 as part of the DIRECT project, Hartman et al.\ (2006) examine a field approximately 1 deg$^2$ (well matched to the LGGS photometry) and identify 36,000 objects variable in g$^\prime$, r$^\prime$, or i$^\prime$, and Javadi et al.\ (2011) examine the central square kpc for JHK variable objects.  

A comparison of our RSG sample with these catalogs reveals that, indeed, 127 of our 189 rank 1
stars ($\sim$67 \%) and 6 of our 12 rank 2 stars are variable in the visible bands.  Thus, although our rank 2 stars are less likely to be genuine M33 members than our rank 1 stars, this data is not inconsistent with many of our rank 2 stars being genuine members.  In contrast, only five of our 207 rank 3 red stars were flagged as visible variables.  Additionally, one RSG was flagged as JHK variable in the catalog of Javadi et al.\ (2011). However, due to the effects of crowding only two of our rank 1 RSGs were located in the central square kpc (the region covered by the Javadi et al.\ 2011 survey).

A similar comparison with our YSG sample revealed only 25 known variables in our 121 rank 1 stars (and zero of our 14 rank 2 stars).  These variables are split between optical (15 stars over the entire disc of M33) and infra-red (10 stars in the central square kpc). Interestingly, our YSG sample contains only 12 stars in the central square kpc, 10 of which are variable in the IR bands, indicating that the IR variability of YSGs warrants more detailed future investigation.  Stars found to be known variables are marked in Tables~\ref{tab:YelProps} and~\ref{tab:RedProps}.

\subsection{Completeness \label{comp}}

\begin{figure}
\plotone{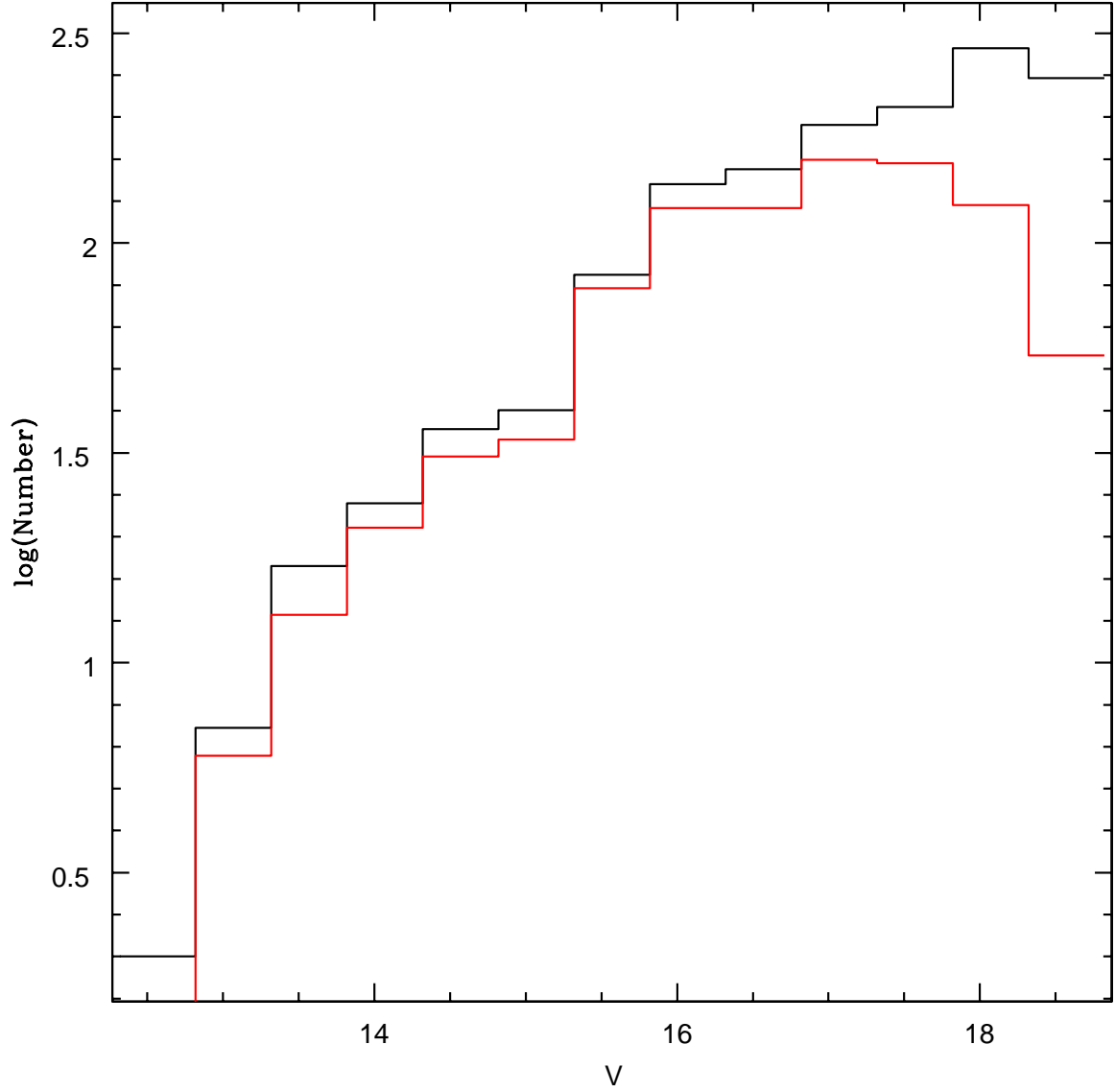}
\caption{\label{fig:Complete}A histogram of the $V$-band magnitudes for our observed yellow sample (red) and the parent population (black).}
\end{figure}

An accurate and meaningful comparison of the relative number of observed YSGs at various luminosities to those predicted by the evolutionary tracks requires that our observed sample be unbiased in luminosity.  To assess what, if any, such biases may be present, we examine the completeness of our observed sample.  In Figure~\ref{fig:Complete} we present the distribution of $V$-band magnitudes for the 1438 yellow LGGS stars which met our original selection criteria (black) and the 916 yellow stars for which we actually obtained radial velocities (red).  We see that there is excellent agreement between the two distributions, with our observed sample falling off only at the highest magnitudes.  If we conservatively set our limiting magnitude to 17.5 (which very roughly corresponds to $\log{L/L_{\odot}} = 4.8$) we find that our sample should be complete down to $\sim$15 M$_\odot$ (see Figure~\ref{fig:toyHR}).

As described in Section~\ref{Intro}, a similar comparison of the relative numbers of our RSG sample at various luminosities to those predicted by the models is made incredibly difficult by the near-vertical nature of the evolutionary tracks.  Thus, we will merely be examining the locations of our RSG populations on the HR diagram, a task which does not depend quite so stringently on the completeness of our sample\footnote{Although, some level of completeness is required in order to accurately estimate the upper luminosity limit for RSGs (i.e., one must observe at least enough stars to have a reasonable bet at catching the most luminous member).}.
 
\section{The HR Diagram \label{HR}}

\subsection{Making the HR Diagram: Effective Temperatures and Luminosities \label{make}}

In order to compare our populations of M33 red and yellow supergiants to the current evolutionary tracks, we must first transform their magnitudes and colors to effective temperatures and luminosities. 

\subsubsection{Yellow Supergiants}

For the YSGs, we follow the same procedure as Drout et al.\ (2009) and Neugent et al.\ (2010), using the Kurucz (1992) Atlas 9 model atmospheres to transform $B - V$ colors into effective temperatures and luminosities.  The $B$- and $V$-band photometry for all of our stars were taken from the LGGS, and we apply a constant reddening correction of $E (B - V) = 0.12$ in keeping with the median value found by Massey et al.\ (2007b).

Before choosing which Atlas 9 model to use for the transformations, we first note that the metallicity of M33 has long been a topic of debate.  As summarized in detail in Neugent \& Massey (2011) there is strong evidence for a metallicity gradient within the plane of M33. Through the years strong gradients (e.g., Kwitter \& Aller 1981, Zaritsky et al.\ 1989, Garnett et al.\ 1997)), weak gradients (e.g., Crockett et al.\ 2006, Rosolowsky \& Simon 2008, Bresolin 2011), and two-component models (Magrini et al.\ 2007) have all been suggested.  By examining the ratio of WC to WN Wolf-Rayet stars in M33, Neugent \& Massey (2011) suggest that the central regions may lie closer to solar metallicity while the outer regions are closer to that of the LMC ($\sim$0.5 solar).

Based on this fact, we initially compute a transformation from $B - V$ to effective temperature using the $Z = 0.6Z_\odot$ Atlas 9 models.  A least square fit to the low surface gravity models over the temperature range $4,000 - 10,000$ K yields the relation:

$$\log{T_{\rm eff}} = 3.936 - 0.5272(B-V)_0 + 0.4752(B-V)_0^2 - 0.1749(B-V)_0^3$$

In order to assess any errors that may result from the metallicity gradient present in M33, we also compute an analogous relation using the $Z = Z_\odot$ Atlas 9 models.  We find that the difference between the two transformations is negligible.  For a star with $(B - V)_0 = 0.6$ the $Z = 0.6Z_\odot$ relation gives $\log{T_{\rm eff}} =$ 3.753 while the $Z = Z_\odot$ relation gives 
$\log{T_{\rm eff}} =$ 3.759.  As there is only a slight, 0.06 dex, difference, we conclude that our original transformation is acceptable over the entire area of M33. Since we restricted our initial fit to the temperature range $4,000 - 10,000$ K the relation for $\log{T_{\rm eff}}$ given above is only valid for the range $-0.11 \leq (B - V)_0 \leq 1.57$ or $0.01 \leq (B - V) \leq 1.69$.  We find, however, that this range encompasses our entire yellow sample.

The bolometric corrections for our yellow sample are modest (a few tenths of a magnitude) and were also computed using the Atlas 9 model atmospheres.  The models yield a relation:
$${\rm BC} = -260.05 + 135.205 \times \log{T_{\rm eff}} - 17.5746 \times (\log{T_{\rm eff}})^2$$
which is also valid over the temperature range $4,000 - 10,000$ K.  Adopting a distance modulus of 24.60 (van den Bergh 2000) we calculate effective temperatures and bolometric luminosities for our rank 1 and rank 2 yellow supergiants.  These derived properties are listed in table ~\ref{tab:YelProps}.

\subsubsection{Red Supergiants}
Deriving a place on the HR diagram for RSGs based on photometry is a more uncertain process than that described above for YSGs.

First, the derivation of the effective temperature is complicated
by the fact that RSGs have so many atomic and molecular lines that
one is not observing the true continuum, but rather regions of relative
transparency between overlapping absorption.  Thus, $B-V$ is no
longer a color indicator, but is instead primarily sensitive to
surface gravity, due to line-blanketing in the $B$-band (see
discussion in Massey 1998 and references therein).  Furthermore,
their surfaces appear to be covered in large spots (Freytag et
al.\ 2002), resulting in slight differences in the effective
temperatures and radii one derives from different bandpasses (Young
et al.\ 2002).  Levesque et al.\ (2006) found, using the
state-of-the-art MARCS models (Gustafsson et al.\ 2003, 2008), that the
effective temperatures derived from $V-K$ were hotter than those
derived by fitting spectral features (i.e., the TiO bands) by 105
K for LMC and by 170 K for the SMC.  They found, however, good
agreement between the temperatures derived from $V-R$ and the
spectral fitting.  Massey et al.\ (2009) attribute the problem with
$V-K$ to the inherent limitations of the one-dimensional models,
given the presence of cooler and warmer regions on the surface.

Second, the bolometric corrections are both large and a strong function
of the assumed effective temperature, while for the YSGs
the bolometric corrections were small and fairly insensitive to the
effective temperature.  This problem can be partially alleviated
by using $K$-band photometry and applying a bolometric correction
to $K$, as discussed by Levesque et al.\ (2006).  There is an
additional advantage to using $K$ to derive the luminosity, namely
that it is less affected by uncertainties in the interstellar (and
circumstellar) reddening.  Massey et al.\ (2009) rederived the
bolometric luminosities of RSGs in the SMC, LMC, the Milky Way, and
M31, and found good agreement on average with the values derived
from $V$ and $K$ when the same effective temperatures were used,
although the individual scatter was large.

All of the RSGs in our sample have $BVR$ data from the LGGS,
and hence we can use the $V-R$ photometry to derive effective
temperatures.  We adopt the conversions given in Section 3.3 of
Levesque et al.\ (2006) for the LMC, since the metallicity of M33
is similar.  Some of the RSGs in our sample also have 2MASS photometry,
and for those stars we compute the bolometric luminosity from $K$,
after converting the 2MASS $K_s$ to standard $K$ following Levesque
et al.\ (2006).  For stars without 2MASS, we compute the bolometric
correction to the LGGS $V$ following the transformations in Levesque
et al.\ (2006), and adopting a reddening of $E(B-V)=0.12$, typical
for OB stars in M33 (Massey et al.\ 2007a).  

For the stars with 2MASS
data, the bolometric luminosity we derive from the $K$-band is, on
average (median), 0.2 mag fainter than that from $V$.  Increasing
the assumed reddening to $E(B-V)=0.5$ would remove most of discrepancy.
This is consistent with our earlier findings that RSGs invariably
have circumstellar reddening, attributable to dust formation (Massey
et al.\ 2005)\footnote{Recall that in our previous comparisons
between the luminosities derived from $V$ and $K$ (Massey et al.\
2009) we explicitly determined the reddening in a star-by-star basis
using spectral fitting.}.  Of course, changing the assumed reddening
would also change the derived effective temperatures.  For now we
have taken the conservative approach of adopting the lower reddening
($E(B-V)=0.12$) but note that this probably underestimates the
effective temperature for some stars.  It also means that the
$V$ band bolometric correction is more positive for $V$ but more
negative for $K$\footnote{We can see this numerically as follows.
Consider a star with a $V-R$ value of $\sim$1.0. If $E(B-V)=0.12$, this star will have an
effective temperature of 3800 K (corresponding to $(V-R)_0\sim0.9$), as $E(V-R)=0.642 E(B-V)$ according to
Schlegel et al.\ (1998).  The bolometric correction will be
$-1.2$~mag at $V$, and $+2.7$~mag at $K$, while the correction for
interstellar reddening will be $-0.37$~mag at $V$, and $-0.04$~mag
at $K$. Thus the apparent magnitude will have to be corrected by
$-1.6$~mag at $V$ and $+2.7$mag at $K$, as well as by the true
distance modulus.  If, however, the reddening had been underestimated
by 0.5~mag in $E(B-V)$, then the star is actually 0.3 less red in
$V-R$ and the effective temperature is more like 4300 K rather than
3800 K.  The bolometric correction will be only $-0.6$~mag in $V$,
and 2.3~mag at $K$, while the interstellar extinction will be
$-$1.9~mag at $V$ and $-0.2$~mag at $K$.  Thus the total corrections
will be $-2.5$~mag at $V$ and $2.1$~mag at $K$.  By assuming too
low a reddening we will have underestimated the bolometric luminosity
of the star by $\sim$1.0 mag at $V$ but by only $\sim$0.6~mag at $K$.}.

Thus, again adopting a distance modulus of 24.60, we calculate effective temperatures and bolometric luminosities for our rank 1 and rank 2 red supergiants.  These derived properties are listed in Table~\ref{tab:RedProps}.

\subsection{Testing the HR diagram: Models and Observations \label{HRdis}}

Having selected our supergiant populations and transformed their photometry into effective temperatures and luminosities, we now compare our results to the Geneva evolutionary tracks.  As mentioned in Section~\ref{make}, the metallicity of M33 is a complex issue and we therefore opt to compare our supergiant populations to evolutionary tracks at multiple metallicities.  Additionally, in the time since our initial studies examining the YSG content of M31 and the SMC, the Geneva group has begun the release a new generation of models (Ekstr\"{o}m et al.\ 2012; Chomienne et al.\ \emph{in prep.}).  Throughout this section we compare our observed populations to both this new set of models and to the previous generation (Meynet \& Maeder 2003, 2005; Schaerer et al.\ 1993).  This will allow us to properly assess what effect the modified physics in the new models creates.  Recall that Drout et al.\ (2009) and Neugent et al (2010) identified a discrepancy between the observed number of YSGs with luminosity and those predicted by the \emph{previous generation} of models.

A thorough description of the differences between the two sets of models is given in Ekstr\"{o}m et al.\ (2012); however, they can be summarized briefly as follows: (1) The initial composition differs, with the new models based on the more recent analysis of solar abundances by Asplund et al.\ (2005).  As a result, the new models adopt $Z_\odot = 0.014$, while the historic value is $Z_\odot = 0.02$. (2) The opacity tables differ as a result of the changed initial composition. (3) The reaction rates have been updated. (4) A new prescription for RSG mass-loss is adopted. An effect of this new prescription is to increase the time-averaged RSG mass-loss by a factor of $\sim$10 for stars above $\sim$15 M$_\odot$. (5) A new prescription for the shear diffusion coefficient, which describes rotationally induced mixing, is utilized. (6) Different initial rotational velocities are considered.  The previous generation of rotating models used an initial velocity of 300 km s$^{-1}$ for all masses.  In contrast, the new models are computed with an initial rotation velocity $v_{\rm rot,i} = 0.4v_{\rm crit}$, where $v_{\rm crit}$ is the critical velocity at which the centripetal force  balances the gravitational acceleration. The value 0.4$v_{\rm crit}$ lies between 200 and 300 km s$^{-1}$ for the mass range examined below\footnote{At Galactic metallicities an initial rotation of 300 km s$^{-1}$ would correspond to an average main sequence rotational velocity of 180 $-$ 240 km s$^{-1}$.  This value is in agreement with observations of Galactic O-type stars (Conti \& Ebbets 1977; Penny 1996) but the appropriate value to use at the metallicities of M33 has yet to be established (e.g., Penny \& Gies 2009).}. We note that both generations of models were also computed with zero initial rotation for comparison. Throughout this section we will refer to non-rotating tracks as S0, the old generation rotating tracks as S3, and the new generation rotating tracks as S4.

In Figures~\ref{BigHR} and~\ref{OldHR} we plot the location of the 121 rank 1 YSGs and 189 rank 1 RSGs, selected in Sections~\ref{reds} and~\ref{yellow}, along with the $Z=0.014$ and $Z=0.006$ (Figure~\ref{BigHR}) new generation models and $Z=0.02$ and $Z=0.008$ old generation models (Figure~\ref{OldHR}).  These models should represent roughly solar and LMC-like metallicities and the values for $Z$ vary between the two generations as a result of the updated compositions described above.  In both panels of Figure~\ref{BigHR} the $Z=0.014$ models are represented by solid lines and the $Z=0.006$ models are represented by dashed lines.  Similarly, in Figure~\ref{OldHR}, the $Z=0.02$ models are represented by solid lines and the $Z=0.008$ models are represented by dashed lines.  In both Figures, the top panel presents models which include the effects of rotation (described above) while the bottom panel presents models which neglect the effects of rotation (shown for comparison). The yellow supergiant region (7500 K $>$ T$_{\rm eff}$ $>$ 4800 K) is designated by vertical lines.

\subsubsection{Locations}  

\begin{figure}
\epsscale{0.5}
\plotone{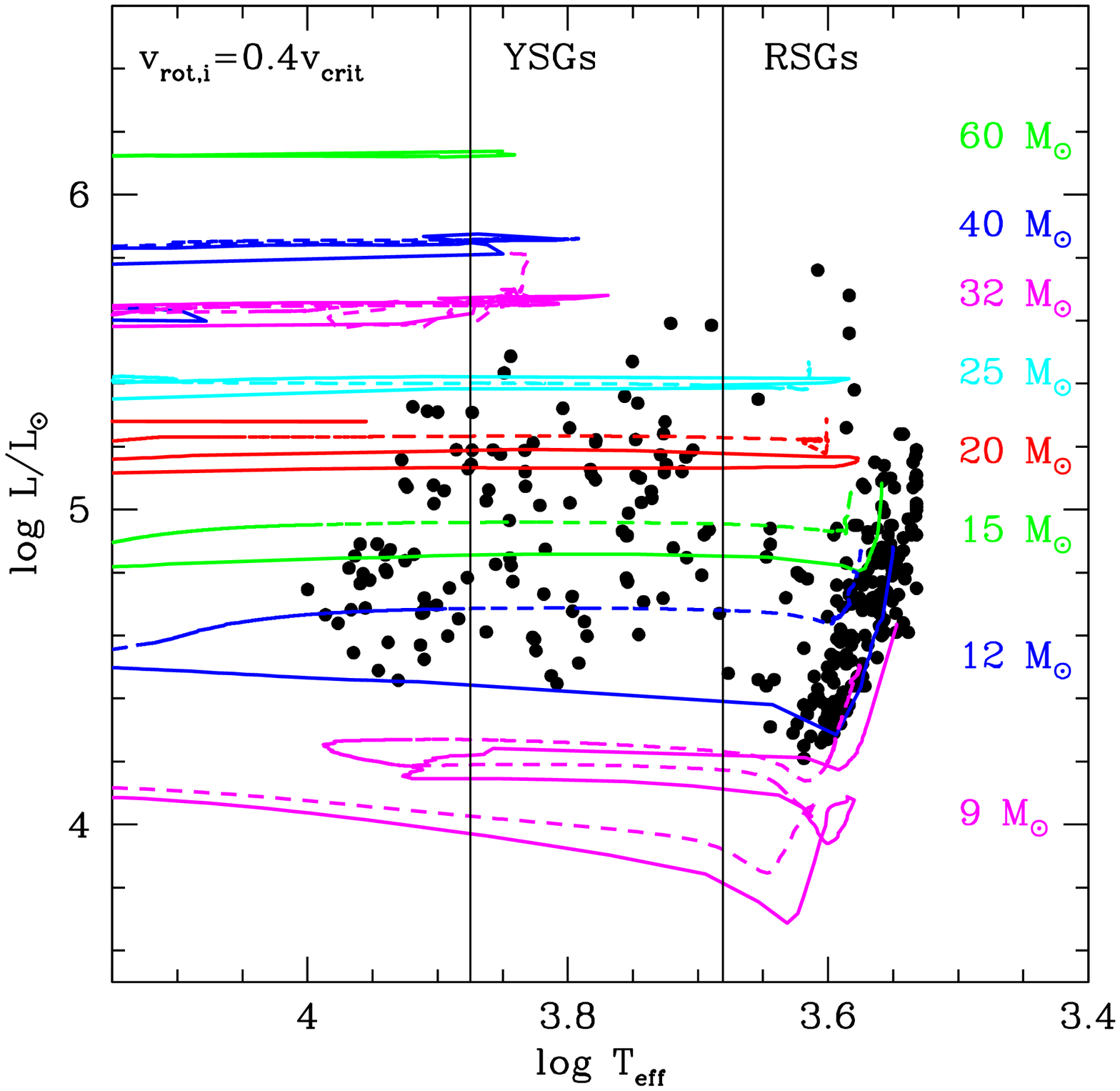}
\plotone{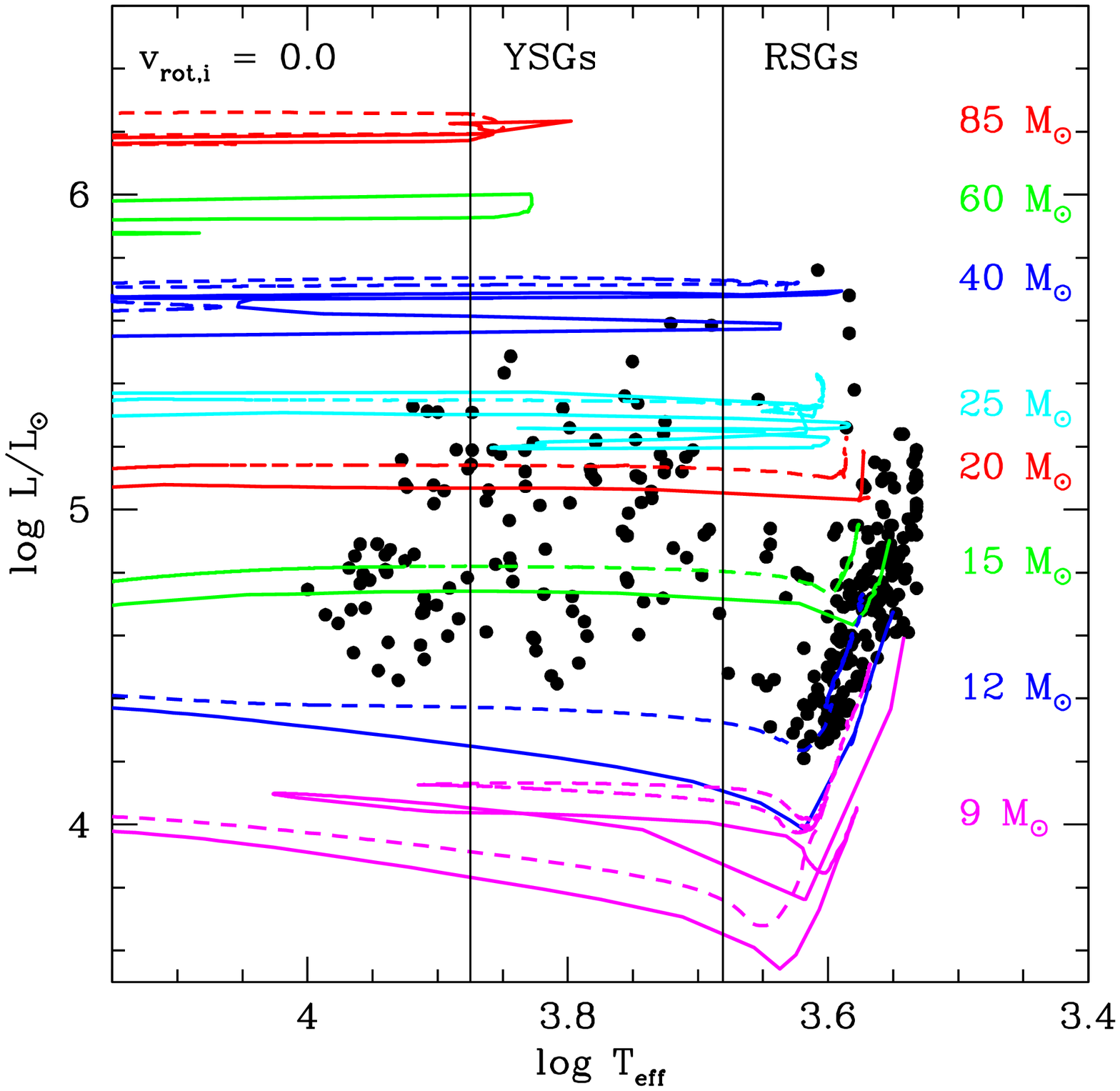}
\caption{\label{BigHR} Our rank 1 yellow and red supergiants plotted with the newest generation of Geneva Group evolutionary models. The vertical lines designate the yellow supergiant region. \emph{top:} Models for $Z=0.014$ (solid lines) and $Z=0.006$ (dashed lines) with an initial rotation of $0.4 v_{\rm crit}$.  \emph{bottom:} Same as top but with zero initial rotation.}
\end{figure}

\begin{figure}
\epsscale{0.5}
\plotone{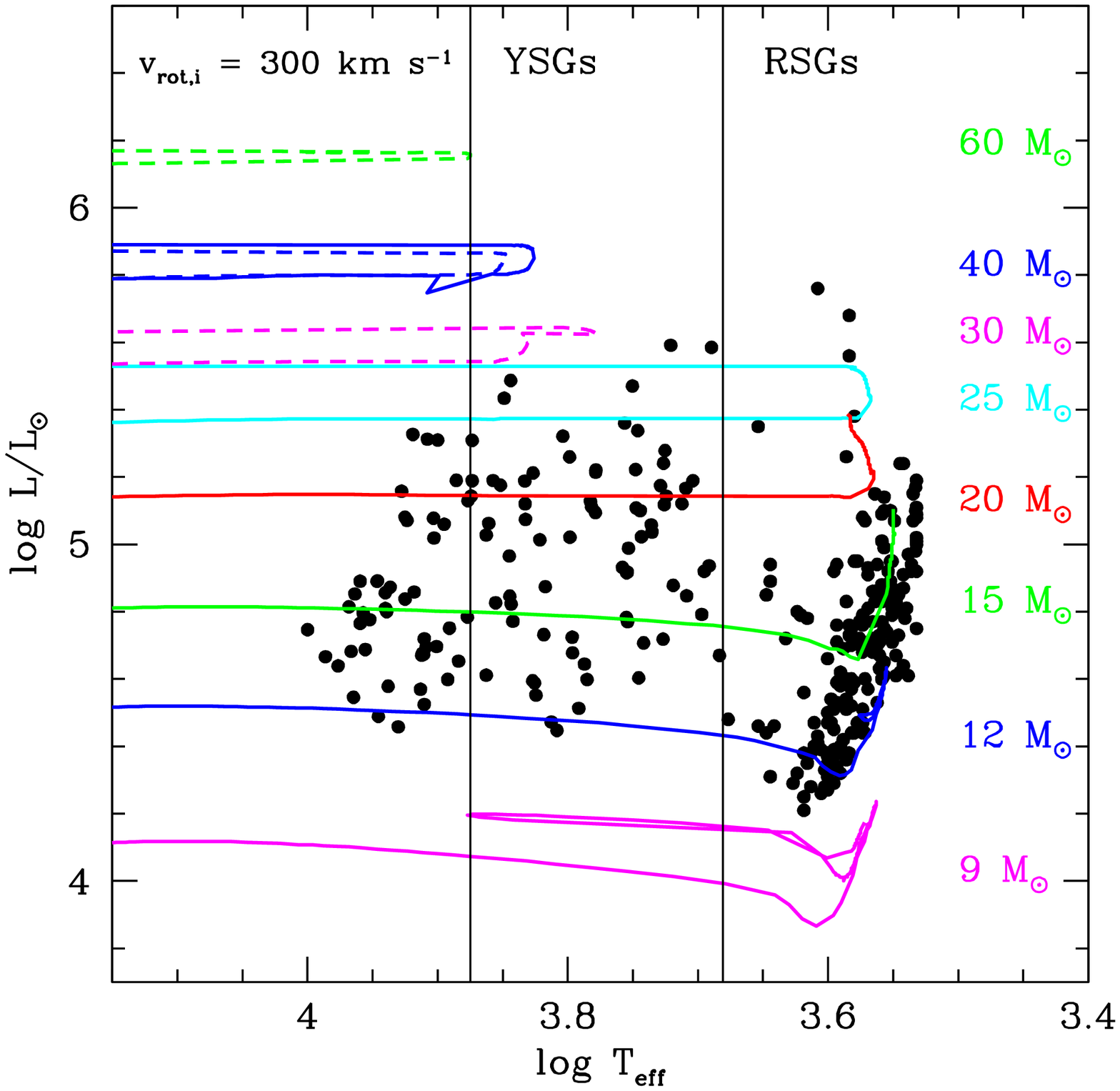}
\plotone{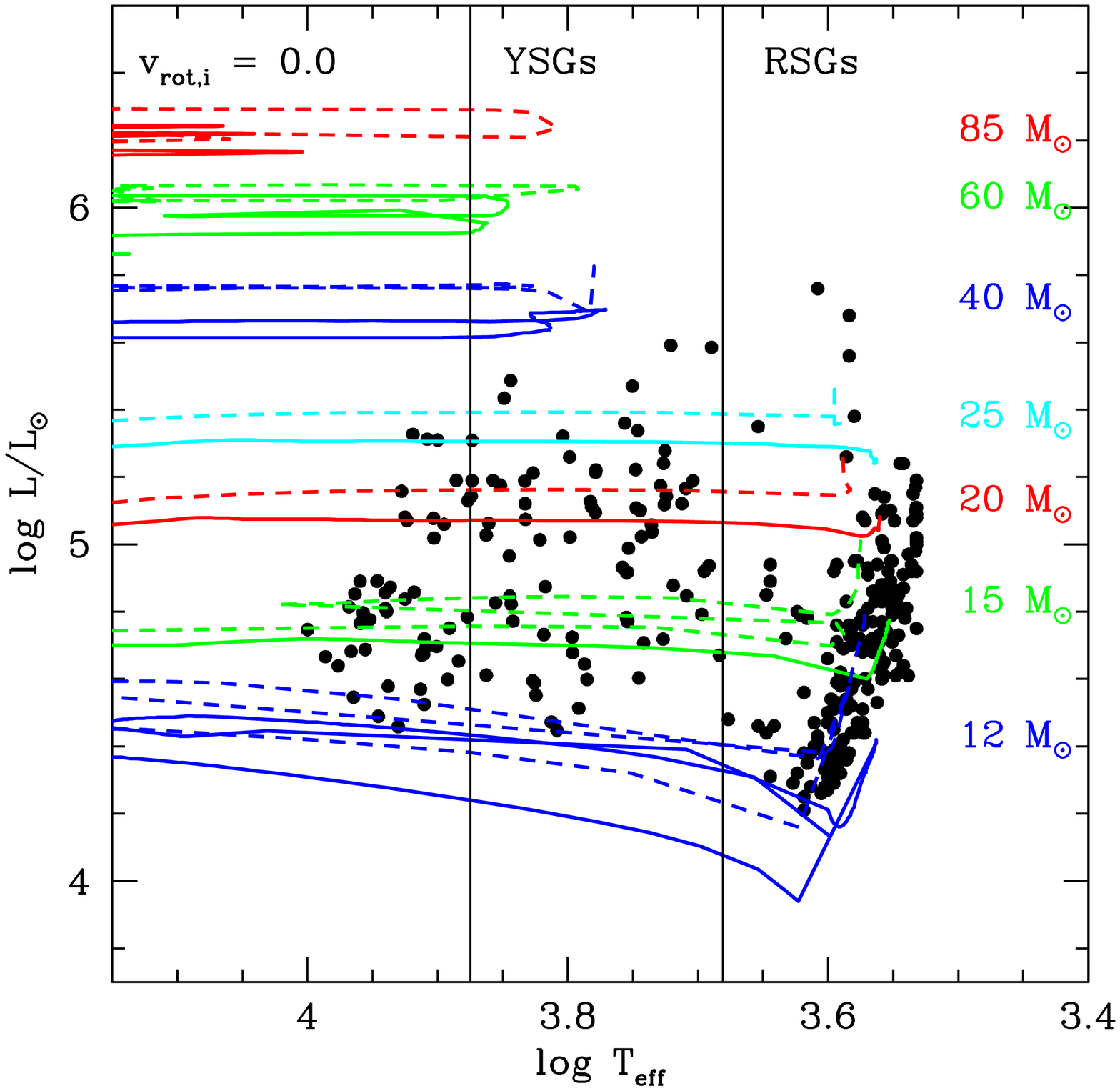}
\caption{\label{OldHR} Our rank 1 yellow and red supergiants plotted with the previous generation of Geneva Group evolutionary models. The vertical lines designate the yellow supergiant region. \emph{top:} Models for $Z=0.02$ (solid lines) and $Z=0.008$ (dashed lines) with an initial rotation of $300$ km s$^{-1}$.  \emph{bottom:} Same as top but with zero initial rotation.}
\end{figure}

In the recent literature, single star evolutionary models have been found to be fairly effective at reproducing the \emph{locations} of red and yellow supergiants on the HR diagram\footnote{At least, once the RSG effective temperatures problem was resolved by Levesque et al.\ (2005) and subsequent papers.}. Despite this historic agreement, we find that much can be learned from the locations of our red and yellow supergiants in Figures~\ref{BigHR} and \ref{OldHR}.

We first note that our RSG population contains a large number of stars below $\log{L/L_\odot} \sim5.2$ and relatively sparse numbers above this point, similar to the upper luminosity limit for RSGs in the Milky Way and Magellanic Clouds found by Levesque et al.\ (2005, 2006).  The presence of a RSG branch should greatly increases the amount of time spent, and, hence, the number of star observed in a given region of the HR-diagram.  Thus, our distribution of RSGs is consistent with the results of the Geneva $Z=0.014$ models (Figure~\ref{BigHR}, solid lines), which predict a RSG branch up to $\log{L/L_\odot} \sim5.2$, while tracks above this point only briefly loop though the RSG portion of the HR diagram.  By contrast, the previous generation of Geneva models predict RSG branches above $\log{L/L_\odot} \sim5.2$.  

We next draw attention to the five most luminous stars in our sample: two YSGs and three RSGs.  These stars appear to be inconsistent with the new generation S4 models and \emph{both} the S0 and S3 old generation models, all of which loop back to the blue portion of the HR diagram without extending to such cool temperatures.  Further examination reveals that two of the three RSGs (J013418.56+303808.6 and J013339.28+303118.8) are moderately crowded in the $B$-band (as described in Section~\ref{reds}), but that the most luminous star (J013312.26+310053.3) is particularly isolated.  Additionally, the two most luminous YSGs (J013358.05+304539.9 and J013349.86+303246.1) both possess particularly strong and isolated OI $\lambda$7774 features.  Thus, for at least three of these five stars, we see no reason to suspect that they were falsely identified as rank 1 supergiants.

As described in detail by Meynet et al.\ (2011) two main factors which affect if, and where, a single star will loop back toward the blue are mass-loss during the RSG phase and rotationally induced mixing (both of which have been modified in the new generation models, see above).  Higher values of either parameter tend to favor blueward evolution.  Indeed, we see that the new generation S0 tracks (Figure~\ref{BigHR}, bottom panel) proceed to cooler temperatures than the S4 tracks (Figure~\ref{BigHR}, top panel) before looping back to the blue.  Although the  distribution of stellar rotational velocities has yet to be fully decoupled from the effect of inclination (since one can only measure $v\sin{i}$) there is increasing evidence that some massive stars are born as genuine slow rotations (e.g., Huang et al.\ 2010).  Thus, although the location of our five most luminous supergiants could likely be explained by a change in the mass-loss prescriptions used by the Geneva models, it is possible that they are simply slow rotators. If this were the case their locations would be fully consistent with the \emph{new generation} models presented here.  

\begin{figure}
\epsscale{1}
\plotone{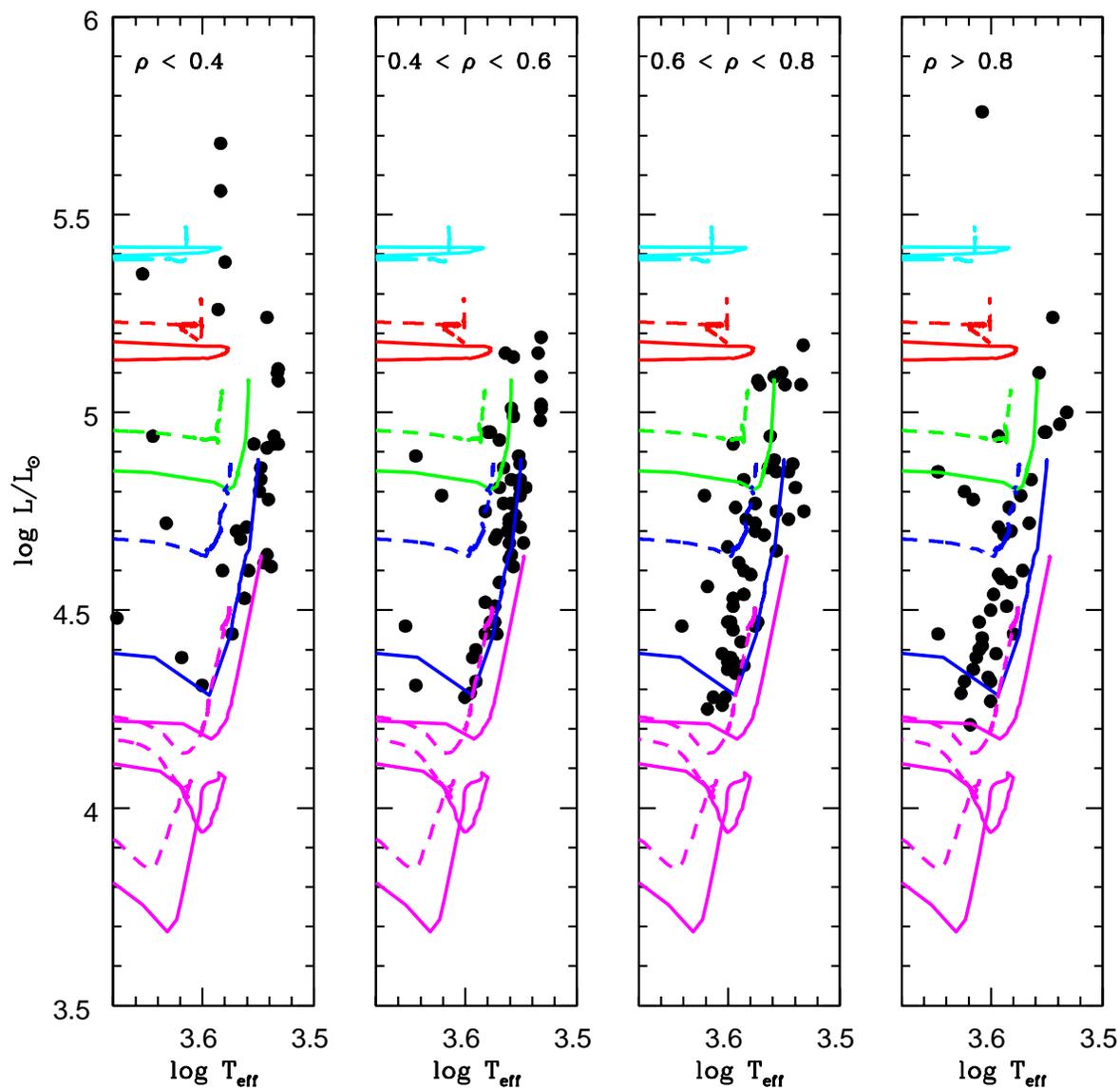}
\caption{\label{HRVar} The location of our rank 1 RSGs binned by radius within the disc of M33.  Moving from left to right, the panels are arranged innermost to outermost.  One can see that RSGs in the outer regions of M33 ($\rho > 0.8$, right panel) are systematically warmer than RSGs in the inner region ($\rho < 0.4$, left panel), as expected if the outer regions lie at lower metallicity.}
\end{figure}

Finally, we call attention to the effective temperatures of a majority of our RSG population.  As can be seen in Figure~\ref{BigHR} the predicted location of the RSG branch moves to warmer temperatures as metallicity decreases.  Recalling that M33 may possess a strong metallicity gradient, in Figure~\ref{HRVar} we again plot the locations of our RSGs on the HR diagram, but now separated into bins based on the value of $\rho$ (the distance from the center within the plane of M33, normalized by the D$_{25}$ isophotal radius of 30\farcm8)\footnote{computed for each star assuming $\alpha_{2000} = 01^{h}33^{m}50^{s}.89$, $\delta_{2000} = 30{^\circ}39{'} 36\farcs8$, an inclination angle of $56^\circ$, and a position angle of the major axis of $23^\circ$, following Kwitter \& Aller (1981) and Zaritsky et al.\ (1989)}.  Moving from left to right, the panels are arranged innermost to outermost.  The evolutionary models present in this figure are the S4 models from the top panel of Figure~\ref{BigHR}.  From this we see that our RSG sample is consistent with a metallicity gradient within the disc of M33.  The supergiants in the outer-most region of M33 are systematically warmer than those in the inner-most region, and the two populations are consistent with the $Z=0.006$ and $Z=0.014$ Geneva models, respectively. The inner two panels appear to show a progression (albeit with some scatter) between the two.

We do note that a handful of our RSGs appear to be inconsistently cool compared to \emph{both} the $Z=0.014$ and $Z=0.006$ Geneva evolutionary models.  Most notable is the group at $\log T_{\rm eff} \sim 3.53$ with $4.9 < \log{L/L_\odot} < 5.2$.  Further examination of these stars reveals that they are all known variables, as discussed in Section~\ref{Var}. Thus, it is possible that our sample contains several of the unusually cool, and potentially unstable, RSGs described by Levesque et al.\ (2007).  Or, it could be that these stars are like the ``too cool'' LMC RSGs discussed by Neugent et al.\ (2011), with high circumstellar reddening.

\subsubsection{Relative Numbers}

Having examined the locations of our red and yellow supergiants on the HR diagram, we now assess the relative numbers of our yellow supergiants at various luminosities.  If we assume the star formation rate \emph{averaged over the entire disc of M33} has been relatively constant over the past 20 Myr (roughly the lifetime of a 12 M$_\odot$ star) then the number of stars expected between masses m$_1$ and m$_2$ will be proportional to:
$$N_{m_1}^{m_2} \propto [m^\Gamma]_{m_1}^{m_2} \times \bar \tau$$
where $\Gamma$ is the slope of the IMF, taken here to be $-1.35$ (Salpeter 1955) and $\bar \tau$ is the average duration of the evolutionary phase for masses $m_1$ and $m_2$.

In Table~\ref{tab:ages} we list the theoretical YSG lifetimes obtained from the new and old generation Geneva models.  Note that (1) we include the lifetime of the 32 M$_\odot$ track for the new non-rotating models, which was omitted from the figure for clarity and, (2) for the old generation of $Z=0.008$ models, there was an insufficient number of rotating tracks available to make a comparison to our data. Recalling that our observed sample should be complete down to $\log{L/L_\odot} \sim 4.8$ (which corresponds to $\sim$15 M$_\odot$), we use the luminosity of the evolutionary tracks with M$\geq$ 15 M$_\odot$ to define mass bins on the HR diagram.  In Table~\ref{tab:num} we list the total number of stars we observe in each mass bin (for each set of tracks) as well as the number relative to the 15 $-$ 20 M$_\odot$ bin.  These relative numbers are then compared to those predicted by the theoretical YSG lifetimes.

For the new $Z=0.006$ models we present results based on a 25 $-$ 40 M$_\odot$ bin in addition to 25 $-$ 32 M$_\odot$ and 32 $-$ 40 M$_\odot$ bins.  Both the rotating and non-rotating $Z=0.006$ 32 M$_\odot$ tracks actually \emph{terminate} in the YSGs regime, and we wish to assess what effect this produces on the predicted number of YSGs with luminosity.

From Table~\ref{tab:num}, we see that both the $Z=0.014$ and $Z=0.006$ new generation S4 models relatively accurately predict our observed ratios of YSGs with luminosity.  In fact, the only main variation we find is the prediction of a few high mass (M $>$ 32 M$_\odot$) YSGs which are not observed. The deviations are only on the order of a handful (2 $-$ 5) of stars, however.  To this end, we emphasize that our observations did not include \emph{every} potential YSG in M33.  As argued in Section~\ref{comp} our sample should be relatively complete, and representative of the relative number of stars at varying luminosities down to $\log{L/L_\odot} \sim 4.8$.  This does not, however, preclude variations on the order of a handful of stars between our observed luminosity distribution and the true luminosity distribution for M33 YSGs.  Indeed, when considering variations of only a few stars small fluxuations in star-formation rate and possible stochastic IMF sampling (see Leitherer \& Ekstr\"{o}m 2011) may become significant.

We note that this effect is very different than that observed by Drout et al.\ (2009) and Neugent et al.\ (2010) when comparing their populations of M31 and SMC YSGs to the old generation models.  In those cases, 10s to 100s of additional high luminosity YSGs would be required to bring their observations into agreement with the evolutionary models, an effect not consistent with the completeness of their samples. 

The non-rotating Geneva models fared slightly worse than the rotating models, over predicting the number of high luminosity (M $>$ 25 M$_\odot$) stars in comparison to low luminosity stars.  They require would on order of 10$-$15 additional stars between 25 and 40 M$_\odot$ to bring our observations inline with the predicted ratios.  We do admit, however, that the mass bins become slightly confused at higher luminosities due to the number of times the tracks seem to loop though the YSG region.  This result may also be indicative of the fact that, although some massive stars are expected to be genuine slow rotators, a much larger percentage of the population likely possesses an initial rotational velocity closer to $0.4v_{\rm crit}$. In fact, the value 0.4$v_{\rm crit}$ was chosen to correspond to the peak in the distribution of initial rotation speeds for young B-type stars in Huang et al.\ (2010). In a more rigorous comparison one would compute an average YSG lifetime, weighted by a velocity distribution for each initial mass.  

We note that for both the S0 and S4 $Z=0.006$ models better results are obtained by ignoring the 32 M$_\odot$ tracks (which, recall, terminated in the YSG regime).  Including the 32 M$_\odot$ track leads to the prediction of $\sim$15 and $\sim$31 stars between 25 M$_\odot$ and 40 M$_\odot$ (for S0 and S4 models, respectively).  On the other hand, ignoring the 32 M$_\odot$ tracks leads to predictions of $\sim$9 and $\sim$11 stars between 25 M$_\odot$ and 40 M$_\odot$ and we observe 5 and 6 (S0 and S4, respectively).  Thus, our observed sample does not \emph{seem} to contain an increase in stars at intermediate luminosities that one would expect as a result of these tracks which terminate in the YSG regime (although we emphasize that even these deviations are \emph{far less} than those observed in our previous works).  This will be discussed briefly in the context of YSG supernova progenitors below.

Given these considerations, we find that the relative number of YSGs at various luminosities predicted by the new generation of Geneva evolutionary tracks are relatively consistent, within our uncertainties, with our sample of observed M33 YSGs.  From Table~\ref{tab:num} we see that this agreement \emph{does} reflect a real improvement over previous models.  At best, the previous generation of Geneva models require an additional 10$-$15 stars at high luminosity ($Z=0.008$, S0 models), and at worst 400+ additional stars at high luminosity ($Z=0.02$, S3 models). Similar results (i.e., a significant improvement in the performance of the new models in comparison to the old) is found by Neugent et al.\ (2011, a companion paper to this manuscript) who compare the YSG population of the LMC to the new Geneva $Z=0.006$ models described here (Chomienne et al.\ \emph{in prep.}).  

One additional concern raised by our initial studies of YSG populations was the actual duration of the YSG phase.  By comparing the (albeit uncertain) number of unevolved OB-type stars to the number of YSGs in M31 and the SMC Drout et al.\ (2009) and Neugent et al.\ (2010) estimated that the duration of the YSG phase should be on the order of 3000 yrs for M $>$ 20 M$_\odot$. As can be seen from Table~\ref{tab:ages}, this value is an order of magnitude below the lifetimes predicted by both the new and old generation Geneva models.  At this time, we do not possess accurate enough information on the number of unevolved OB-type stars in M33 to directly assess whether this discrepancy is due to remaining deficiencies in the models or with the lifetimes predicted from observations (which should be taken as lower limits due to the effects of crowding; see Massey 2003).  However, in a companion paper examining the YSG and RSG populations of the LMC Neugent et al.\ (2011) are able to revise this estimate of the YSG lifetime. By comparing the number of observed LMC YSGs to the number of unevolved LMC OB-type stars from Massey (2010) they estimate that the YSG lifetime is $\sim$17,000 yrs for M $>$ 20 M$_\odot$.  This number, although smaller than those listed in Table~\ref{tab:ages}, is, in our opinion, within the uncertainties of the number of OB-type stars.

\subsubsection{Discussion}

Having demonstrated that the new generation of Geneva evolutionary models offer a dramatic improvement
in predicting both the observed locations and relative numbers of YSGs as a function of luminosity, we now spend a moment discussing the natural question that arises: What is the physical cause of this improvement?  As described above, the new models include several substantial modifications over the previous generation.  However, despite the complex and somewhat degenerate nature of the problem, we are able to offer some modest logical arguments.

One modification which almost certainly impacts the behavior of the evolutionary tracks in this region of the HR diagram is the adoption of a new prescription for RSG mass-loss.  In the new prescription, mass-loss is artificially increased by a factor of three whenever any layer in the envelope of the star reaches a luminosity greater than five times the Eddington luminosity (an effect which is relevant for RSGs with masses above $\sim$15 M$_\odot$). We emphasize that this is an \emph{ad-hoc} prescription, but one which is not without both theoretical and observational precedent (e.g., Heger et al.\ 1997; Vanbeveren et al.\ 2007; Yoon \& Cantiello 2010 ; van Loon et al.\ 2005, 2008).  As discussed in Ekstr\"{o}m et al.\ (2012) this higher mass-loss rate favors blueward evolution, thus lowering the minimum mass required for blueward evolution and increasing the ratio of the blue to red lifetimes for high mass tracks (see section 5.5 of Ekstr\"{o}m et al.\ 2012). In order to investigate whether this new mass-loss prescription can explain the improved behavior of the models in the YSG regime specifically, we again examine the YSG lifetimes listed in Table~\ref{tab:ages}.

From this we see that the improved YSG ratios predicted by the new models is not \emph{simply} a result of a decrease in the YSG lifetimes predicted for the highest mass tracks, nor \emph{simply} an increase in the YSG lifetimes predicted for the lower mass tracks, but rather a complex modification of the predicted lifetimes at all masses.  We do observe significant differences in the lifetimes predicted for some tracks which experience RSG mass-loss before evolving back to the blue portion of the HR diagram (most notably the 20 M$_\odot$ rotating solar mass track which only evolves back to the blue in the new generation models).  However, we also observe significant differences in the predicted lifetimes for tracks which evolve straight across the HR diagram and end their lives as RSGs (e.g., all the 15 M$_\odot$ tracks and the 20 M$_\odot$ non-rotating, low metallicity tracks). These tracks do not possess blue loops, and, hence, their YSG lifetimes should be unaffected by the new prescription for RSG mass-loss. The fact that we observe significant modifications to the predicted YSG lifetimes for even these tracks indicates to us that, although the modified RSG mass-loss undoubtedly affects the behavior of the evolutionary tracks in this region of the HR diagram, other effects must also be behind the modified behavior of the new models.

Thus, we emphasize a few other modifications made to the new set of models.  With respect to rotation: the new models adopt the shear diffusion coefficient proposed by Maeder (1997) as opposed to that of Talon \& Zahn (1997), and the initial rotation velocities included in the models have been modified.  As mentioned above, the effects of rotation can also play a significant role in dictating the behavior of evolutionary tracks in this region. This is evidenced by the 20 M$_\odot$ $Z=0.014$ tracks which only evolve back to the blue in the rotating models.  Additionally, from Table~\ref{tab:num} we see that, although the predictions made by both the rotating and non-rotating tracks are improved in the new models, the rotating tracks now perform \emph{vastly} better than the old, indicating that some of the improved behavior is likely do to the modified descriptions of rotation.  

However, even these modifications cannot explain the contrast between the 15 M$_\odot$ non-rotating $Z=0.008$ and $Z=0.006$ tracks, the  20 M$_\odot$ non-rotating $Z=0.008$ and $Z=0.006$ tracks, and the 15 M$_\odot$ non-rotating $Z=0.014$ and $Z=0.02$ tracks.  All three pairs show a significant change in their predicted YSG lifetimes, and none should be affected by either RSG mass-loss or rotation.  To this end we note that the new models contain modified initial compositions and opacities. We are considering a transitionary phase during which the star is changing shape (inflating when evolving from the blue to the red) and thus changes in the opacities may actually have important implications.

It is therefore our conclusion that no one modification in the new Geneva models can be pointed to as the main cause for their improved overall behavior in the YSG regime.  Rather, when put together, the various modifications produce a set of models which are, at least heuristically, able to reproduce the observed trends.  This emphasizes the role of the YSG regime as a magnifying glass for models of stellar evolution, and that much work can still be done to understand these objects from a theoretical point of view.

\section{Summary and Future Work}

We obtained spectra of 916 yellow stars ($\sim$68\% of those initially selected) and 408 red stars ($\sim$37\% of the isolated stars initially selected) towards M33.  Bona fide RSGs were identified through a combination of radial velocities and a two-color surface gravity discriminant.  After re-characterizing the rotation curve of M33 using our newly identified RSGs, foreground Milky Way contaminants were removed from our YSG sample through a combination of radial velocities and  OI $\lambda$7774 equivalent widths.  In the end we identified a sample of 189 rank 1 RSGs and 121 rank 1 YSGs.
  
After transforming the colors and magnitudes of our rank 1 supergiants into effective temperatures and luminosities, we placed them on the HR diagram.  A comparison of the relative numbers and locations of our supergiant populations to those predicted by the latest generation of Geneva evolutionary tracks yielded excellent results.  We observed a sharp drop-off in the number of RSGs above a luminosity of $\log{L/L_\odot} \sim5.2$, consistent with the upper luminosity limit predicted by the new rotating Geneva models, and an increase in the mean temperature of stars on the RSG branch with radial distance, consistent with metallicity gradient within the disc.  Additionally, our observed relative number of YSGs with luminosity is consistent with the new rotating Geneva models to within a handful of stars. This represents a vast improvement over previous models.

There are several topics pertaining to YSGs, RSGs, and massive star evolution which are worthy of additional study.  One such topic is the effect of binarity on massive star evolution.  All the models examined here are single star tracks.  Many massive stars occur in binary systems, and, for some, this likely has an effect on their evolution through additional mass-loss and interactions (e.g., Roche Lobe overflow, common envelope evolution). The level of this effect on the scale of an entire stellar population depends on many parameters which are, as yet, not well constrained (i.e., binary fraction, distribution of initial separations, distribution of mass-ratios).  Despite these uncertainties, it has been demonstrated that some binary models, such as those calculated with the Binary Population and Spectral Synthesis (BPASS) code as described in Eldridge, Izzard \& Tout (2008) and Eldridge \& Stanway (2009), show similar trends to rotating single-star models in their predicted ratios of supergiant and supernova types with metallicity.  In a private communication J. Eldridge reveals that, at least at the high metallcities of M31, the BPASS models offer a marked improvement over non-rotating, single star tracks with respect to the predicted number of YSGs with luminosity.  Similar to the rotating Geneva models presented here, the largest discrepancy with observations appears to be an over prediction of the highest luminosity stars (on the order of those found here as opposed to our previous work).

With the addition of this paper and that of Neugent et al.\ (2011), we will have characterized the YSG population in four local group galaxies spanning a factor of $\sim$10 in metallicity.  This will allow us to directly compare the predicted trends of the rotating single-star and non-rotating binary tracks in the YSGs region with metallicity and see what constraints, if any, our YSG populations can provide with respect to these two models.

Another topic worthy additional consideration is the increasing number of putative supernova progenitors which fall in the YSG regime (e.g., SN 2008cn, SN 2009kr, and SN2011dh; see Elias-Rosa et al.\ 2009, 2010, and Maund et al.\ 2011, respectively), a regime inconsistent with the end-points of traditional single-star tracks (with the exception of the most recent $Z=0.006$ 32 M$_\odot$ Geneva tracks described above).  Although in most cases the observational data is not inconsistent with (and may even suggest) a binary system (e.g., see Elias-Rosa 2009, Soderberg et al.\ 2011, Arcavi et al.\ 2011), Georgy (2011) also demonstrates that by increasing mass-loss rates in the RSG regime the Geneva code is able to produce tracks for 12 M$_\odot$ and 15 M$_\odot$ single stars which terminate in the YSG region.  As discussed in Georgy (2011) both the physical mechanism and frequency of such enhanced mass-loss is poorly understood, and warrants further investigation.  Any such enhancement would likely alter the predicted ratios of YSGs with luminosity, and thus the populations of this paper, Neugent et al.\ (2011), Neugent et al.\ (2010), and Drout et al.\ (2009) can offer a constraint on such models.

This work was supported by the National Science Foundation through grant AST-1008020 and through a Graduate Research Fellowship provided to MRD. We are grateful to Susan Tokarz for performing the initial reductions on all our spectra. MRD would like to thank John Eldridge for useful comments and for taking the time to investigate the relative number of YSGs predicted with luminosity within his binary models.  Finally, we thank the anonymous referee for comments which led us to examine the physical cause of the improved behavior of the new Geneva models in more detail.

\clearpage

\clearpage

\clearpage

\clearpage

\clearpage

\clearpage

\clearpage

\clearpage

\clearpage

\clearpage

\clearpage

\clearpage

\clearpage

\clearpage

\clearpage

\clearpage

\clearpage

\clearpage

\clearpage

\begin{deluxetable}{lccc}
\tablecaption{\label{tab:obs} Configurations Observed}
\tablewidth{0pt}
\tablehead{\colhead{Configuration} &
\colhead{Exp Time (min)} &
\colhead{UT Date(s)} &
\colhead{N stars observed} \\
\cline{1-4} \\
\multicolumn{4}{c}{2009 Observations}
}
\startdata
Brt-1 & 15 & 2009 Oct 16 & 124 \\
Fnt-1 & 60 & 2009 Oct 15 & 217 \\
Fnt-2 & 60 & 2009 Oct 27 & 210 \\
Fnt-3 & 60 & 2009 Nov 25 & 156 \\
Fnt-4 & 60 & 2009 Dec 15 & 207 \\
\cutinhead{2010 Observations}
Brt-1 & 36 & 2010 Nov 26 & 56 \\
Fnt-1 & 75, 80& 2010 Nov 26, 27 & 211 \\
Fnt-2 & 80 & 2010 Nov 27 & 190 \\
\enddata
\end{deluxetable}

\clearpage

\begin{deluxetable}{lccccccccc}                                                                                                                     
\tabletypesize{\tiny}                                                                                                                                 
\tablecaption{\label{tab:TableMeas:Red} Observed Properties: Red Stars with Radial Velocities}                                                                     
\tablewidth{0pt}                                                                                                                                      
\tablehead{                                                                                                                                           
\colhead{Star Name\tablenotemark{a}} &                                                                                                                                                                                                                \colhead{Vel$_{\rm obs}$} &                                                                                                                               
\colhead{r\tablenotemark{b}} &                                                                                                                        
\colhead{Vel$_{\rm expect}$} &                                                                                                                            
\colhead{Vel$_{\rm obs}$$-$Vel$_{\rm exp}$} &                                                                                                                 
\colhead{$V$} &                                                                                                                                         
\colhead{$B-V$} &                                                                                                                                     
\colhead{$V-R$} &                                                                                                                                     
\colhead{$K$} &                                                                                                                                         
\colhead{Rank\tablenotemark{c}} \\                                                                                                                    
&\colhead{(km s$^{-1}$)} & &\colhead{(km s$^{-1}$)} &\colhead{(km s$^{-1}$)}& & & & &                                                           
}                                                                                                                                                     
\startdata                                                                                                                                            
J013147.87+303648.3 &   -37.2\tablenotemark{d} &  35.4 & -158.1 &  120.9 & 17.38 & 1.17 &  0.69 & \nodata& 3 \\          
J013152.41+302112.2 &    40.5 &  16.8 & -129.7 &  170.2 & 19.00 & 1.07 &  0.67 & \nodata& 3 \\                           
J013153.80+301953.7 &    -8.1 &  37.4 & -127.4 &  119.3 & 17.48 & 1.50 &  0.94 & \nodata& 3 \\                           
J013154.76+301159.1 &   -13.7 &  42.5 & -117.0 &  103.3 & 18.24 & 1.28 &  0.79 & \nodata& 3 \\                           
J013156.14+301123.3 &   -41.0 &  20.6 & -116.1 &   75.1 & 17.96 & 1.48 &  0.94 & \nodata& 3 \\                           
J013158.08+302808.7 &   -19.3 &  32.8 & -140.6 &  121.3 & 17.96 & 1.34 &  0.80 & \nodata& 3 \\                           
J013200.85+302810.0 &   -25.5 &  31.9 & -140.1 &  114.6 & 18.91 & 1.25 &  0.71 & \nodata& 3 \\                           
J013202.95+303157.1 &    -4.4 &  22.0 & -147.4 &  143.0 & 18.32 & 1.43 &  0.87 & \nodata& 3 \\
\enddata    
\tablenotetext{a}{Coordinates may be obtained from object name following the converstion given in Section~\ref{Obs}}                                                                                                                                            
\tablenotetext{b}{Tonry \& Davies (1979) \emph{r} parameter}                                                                                          
\tablenotetext{c}{Rank: 1, highly likely supergiant; 2, possible supergiant; 3, dwarf}                                                                
\tablenotetext{d}{Radial velocity observed on two different nights differed by 10 km s$^{-1}$ or more.}    
\tablenotetext{e}{Star is moderately crowded in the $B$-band.}                                        
\end{deluxetable}                                                                                                                                     
                                                                                                                                                      
\clearpage

\begin{deluxetable}{lcccccccc}                                                                                                                      
\tabletypesize{\tiny}                                                                                                                                 
\tablecaption{\label{tab:TableMeas:Yel} Observed Properties: Yellow Stars with Radial Velocities}                                                                  
\tablewidth{0pt}                                                                                                                                      
\tablehead{                                                                                                                                           
\colhead{Star Name\tablenotemark{a}} &                                                                                                                                                                                                                               \colhead{Vel$_{\rm obs}$} &                                                                                                                               
\colhead{r\tablenotemark{b}} &                                                                                                                        
\colhead{Vel$_{\rm expect}$} &                                                                                                                            
\colhead{Vel$_{\rm obs}$$-$Vel$_{\rm exp}$} &                                                                                                                 
\colhead{$V$} &                                                                                                                                         
\colhead{$B-V$} &                                                                                                                                     
\colhead{OI $\lambda$7774 Equivalent Width} &                                                                                                         
\colhead{Rank\tablenotemark{c}} \\                                                                                                                    
 &\colhead{(km s$^{-1}$)} & &\colhead{(km s$^{-1}$)} &\colhead{(km s$^{-1}$)}& & & \colhead{\AA} &                                               
}                                                                                                                                                     
\startdata                                                                                                                                            
J013146.25+301849.7 &   -39.5 &  47.4 & -124.3 &   84.8 & 16.36 & 0.81 & \nodata &  4 \\                                 
J013147.87+303648.3 &   -39.7\tablenotemark{d} &  51.5 & -157.8 &  118.1 & 17.38 & 1.17 & \nodata &  4 \\                
J013148.14+303206.9 &   -95.1 &  38.8 & -148.5 &   53.4 & 17.47 & 0.73 & \nodata &  3 \\                                 
J013148.74+301941.4 &   -97.2 &  50.3 & -125.0 &   27.8 & 16.06 & 0.62 & \nodata &  3 \\                                 
J013148.89+304020.5 &    -1.7 &  32.3 & -164.6 &  162.9 & 17.72 & 0.64 & \nodata &  4 \\                                 
J013150.25+301822.5 &    45.1 &  30.5 & -122.5 &  167.6 & 16.21 & 1.03 & \nodata &  4 \\                                 
J013150.78+303148.4 &   -10.9 &  37.4 & -147.5 &  136.6 & 16.66 & 0.74 & \nodata &  4 \\                                 
J013151.09+302836.5 &   -33.6 &  51.8 & -141.0 &  107.4 & 16.35 & 0.89 & \nodata &  4 \\
\enddata        
\tablenotetext{a}{Coordinates may be obtained from object name following the converstion given in Section~\ref{Obs}}                                                                                                                                              
\tablenotetext{b}{Tonry \& Davies (1979) \emph{r} parameter}                                                                                          
\tablenotetext{c}{Rank: 1, highly likely supergiant; 2, possible supergiant; 3, possible supergiant, high dwarf contamination; 4, dwarf; Cl, cluster} 
\tablenotetext{d}{Radial velocity observed on two different nights differed by 10 km s$^{-1}$ or more.}                                      
\end{deluxetable}                                                                                                                                     
                                                                                                                                                      
\clearpage                                                                                                                                            
                                                                                                                                                      
\begin{deluxetable}{lcccc}                                                                                                                            
\tabletypesize{\footnotesize}                                                                                                                         
\tablecaption{\label{tab:YelProps} Derrived Properties of Probable Yellow Supergiants}                                                                       
\tablewidth{0pt}                                                                                                                                      
\tablehead{                                                                                                                                           
\colhead{Star Name} &                                                                                                                                 
\colhead{Rank} &                                                                                                                                      
\colhead{$\log{T_{\rm eff}}$} &                                                                                                                                 
\colhead{$\log{L/L_\odot}$} &                                                                                                                          
\colhead{Known Variable Reference\tablenotemark{a}}                                                                                                   
}                                                                                                                                                     
\startdata                                                                                                                                            
J013226.98+303659.7 &  2 &  3.777 &  4.95 &  \\                                                                                                       
J013229.20+303430.2 &  1 &  3.849 &  5.48 &  \\                                                                                                       
J013231.71+302940.6 &  1 &  3.913 &  4.61 &  \\                                                                                                       
J013231.94+303516.7 &  1 &  3.753 &  5.03 &  \\                                                                                                       
J013233.85+302728.9 &  1 &  3.804 &  5.37 &  \\                                                                                                       
J013235.17+303331.6 &  1 &  3.946 &  4.53 &  \\                                                                                                       
J013240.21+304149.6 &  2 &  3.789 &  4.58 &  \\                                                                                                       
J013241.21+304446.4 &  2 &  3.769 &  4.86 &  \\
\enddata                                                                                                                                              
\tablenotetext{a}{1, Hartman et al. (2006); 2, Macri et al. (2001); 3, Javadi et al. (2011)}                                                             
\end{deluxetable}                                                                                                                                     
                                                                                                                                                      
\clearpage                                                                                                                                            
                                                                                                                                                      
\begin{deluxetable}{lcccc}                                                                                                                            
\tabletypesize{\footnotesize}                                                                                                                         
\tablecaption{\label{tab:RedProps} Derrived Properties of Probable Red Supergiants'}                                                                         
\tablewidth{0pt}                                                                                                                                      
\tablehead{                                                                                                                                           
\colhead{Star Name} &                                                                                                                                 
\colhead{Rank} &                                                                                                                                      
\colhead{$\log{T_{\rm eff}}$} &                                                                                                                                 
\colhead{$\log{L/L_\odot}$} &                                                                                                                          
\colhead{Known Variable Reference\tablenotemark{a}}                                                                                                   
}                                                                                                                                                     
\startdata                                                                                                                                            
J013220.36+302938.0 & 1 &  3.627 &  4.29 & 1 \\                                                                                                       
J013222.59+302741.8 & 1 &  3.613 &  4.38 & 1 \\                                                                                                       
J013224.33+303155.5 & 1 &  3.611 &  4.40 & 1 \\                                                                                                       
J013229.64+302802.7 & 1 &  3.603 &  4.33 &  \\                                                                                                        
J013232.49+303521.8 & 1 &  3.564 &  4.83 & 1 \\                                                                                                       
J013233.49+301511.7 & 1 &  3.572 &  4.60 &  \\                                                                                                        
J013233.55+302543.1 & 1 &  3.600 &  4.50 &  \\                                                                                                        
J013233.77+302718.8 & 1 &  3.545 &  5.24 &  \\
\enddata                                                                                                                                              
\tablenotetext{a}{1, Hartman et al. (2006); 2, Macri et al. (2001); 3, Javadi et al. (2011)}                                                             
\end{deluxetable}

\clearpage

\begin{deluxetable}{l c c c c c c c}
\tablecaption{\label{tab:ages}Theoretical Yellow Supergiant Duration (yr)\tablenotemark{a}}
\tablewidth{0pt}
\tablehead{
\colhead{Mass}&\multicolumn{2}{c}{$Z=0.014$} & \multicolumn{2}{c}{$Z=0.006$} & \multicolumn{2}{c}{Old $Z=0.02$} & \colhead{Old $Z=0.008$}\\ 
\cline{2-3} 
\cline{4-5}
\cline{6-8}
\colhead{($M_\odot$)}
&\colhead{S4\tablenotemark{b}}
&\colhead{S0\tablenotemark{b}}
&\colhead{S4\tablenotemark{b}}
&\colhead{S0\tablenotemark{b}}
&\colhead{S3\tablenotemark{b}}
&\colhead{S0\tablenotemark{b}}
&\colhead{S0\tablenotemark{b}}
 }
\startdata
85 & 0 &  5200 & 0 & 3000 & 0 & 0 & 4600 \\
60 & 3600 & 10900 &\nodata & \nodata & 0 & 14100 & 8300 \\
40 & 29700 & 25500 & 29900 & 21100 & 57500 & 403500 & 104500 \\
32 & 69200 & 64200 & 84500 & 173400 & \nodata & \nodata & \nodata \\
25 & 27400 & 293500 & 17900 & 48800 & 5800 & 73000 & 89900 \\
20 & 153100 & 68000  & 68500 & 502000 & 300 & 92600 & 89200 \\
15 & 126900 & 117700 & 71100 & 91400 & 2100 & 51100 & 200000 \\
\enddata
\tablenotetext{a}{For the purposes of this calculation, yellow supergiants are defined as having effective temperatures between 4800~K and 7500~K. Lifetimes are determined from the Geneva group models described in the text.}
\tablenotetext{b}{S4 has initial rotation of 0.4$v_{\rm crit}$, S3 has an initial rotation of 300 km s$^{-1}$, and S0 has an initial rotation of 0 km s$^{-1}$.}
\end{deluxetable}

\clearpage

\begin{deluxetable}{l c c c c c c}
\tabletypesize{\footnotesize}
\tablecaption{\label{tab:num}Numer of Yellow Supergiants: Observed vs.\ Modeled}
\tablecolumns{7}
\tablewidth{0pt}
\tablehead{
\colhead{Mass} &  
\colhead{$\#$ Observed\tablenotemark{a}} & 
\colhead{$\#$ Observed\tablenotemark{a}}& 
\multicolumn{4}{c}{Ratio Relative to $15 - 20 M_\odot$} \\
\cline{4-7}
\colhead{Range} & 
\colhead{S4/S3\tablenotemark{b}} & 
\colhead{S0\tablenotemark{b}} &
\colhead{S4/S3\tablenotemark{b} Observed} & 
\colhead{S4/S3\tablenotemark{b} Modeled}&
\colhead{S0\tablenotemark{b} Observed} & 
\colhead{S0\tablenotemark{b} Modeled} \\
\cline{1-7} \\
\multicolumn{7}{c}{$Z=0.014$ Models}
}
\startdata
$15 - 20 M_\odot$ & 33 & 24 & 1.0 & 1.0 & 1.0 & 1.0\\
$20 - 25 M_\odot$ & 12 & 28 & 0.4 & 0.4 & 1.2 & 1.1\\
$25 - 32 M_\odot$ & 5 & 3 & 0.2 & 0.2 & 0.1 & 0.9\\
$32 - 40 M_\odot$ & 0 & 2 & 0.0 & 0.1  & 0.1 & 0.1\\
$40 - 60 M_\odot$ & 0 & 0 &  0.0 & 0.0 & 0.0 & 0.1\\
$60 - 85 M_\odot$ & 0 & 0 & 0.0 & 0.0 & 0.0 & 0.0 \\
\cutinhead{$Z=0.006$ Models}
$15 - 20 M_\odot$ & 31 & 28 & 1.0 & 1.0 & 1.0 & 1.0\\
$20 - 25 M_\odot$& 7 & 18 & 0.2 & 0.3 & 0.6 & 0.4\\
\enddata
\tablenotetext{a}{The S0 and S4/S3 tracks vary in luminosity for a given set of models (see Figure~\ref{BigHR}).  The number of observed stars in each mass bin must therefore be measured independently.}
\tablenotetext{b}{S4/S3 designates models which include the effects of rotation, S0 designated non-rotating models.  The old generation rotating models are designated S3 because they include an initial rotation of 300 km s$^{-1}$, while the new generation rotating models are designated S4 because the include an initial rotation of 0.4$v_{\rm crit}$ (see text).}
\end{deluxetable}


\begin{references}
\reference {} Andrillat, Y., Jaschek, C., \& Jaschek, M. 1995, A\&AS, 112, 475
\reference {} Arcavi, I., et al. 2011, ApJ, 742, 18
\reference {} Asplund, M., Grevesse, N., \& Sauval, A. J. 2005, in ASP Conf. Ser. 336, Cosmic Abundances as Records of Stellar Evolution and Nucleosynthesis, ed. T. G. Barnes, III \& F. N. Bash (Austin, TX: ASP), 2
\reference {} Bresolin, F. 2011, ApJ, 730, 129
\reference {} Cenarro, A. J., Cardiel, N., Gorgas, J., et al. 2001, MNRAS, 326, 959
\reference {} Conti, P. S., \& Ebbets, D. 1977, ApJ, 213, 438
\reference {} Courteau, S., \& van den Bergh, S. 1999, AJ, 118, 337
\reference {} Crockett, N. R., Garnett, D. R., Massey, P., \& Jacoby, G. 2006, ApJ, 637, 741
\reference {} Drout, M. R., Massey, P., Meynet, G., Tokarz, S., \& Caldwell, N. 2009, ApJ, 703, 441
\reference {} Ekstr\"{o}m, S., et al. 2012, A\&A, \emph{in press}
\reference {} Elias-Rosa, N., et al. 2009, ApJ, 706, 1174
\reference {} Elias-Rosa, N., et al. 2010, ApJ, 714, 254
\reference {} Eldridge, J. J., Izzard, R. G., \& Tout, C. A. 2008, MNRAS, 384, 1109
\reference {} Eldridge, J. J., \& Stanway, E. R. 2009, MNRAS, 400, 1019
\reference {} Fabricant, D., et al. 2005, PASP, 117, 1411
\reference {} Freytag, B., Steffen, M., \& Dorch, B. 2002, Astron. Nachr., 323, 213
\reference {} Garnett, D. R., Shields, G. A., Skillman, E. D., Sagan, S. P., \& Dufour, R. J. 1997, ApJ, 489, 63
\reference {} Georgy, C. 2012, A\&A, \emph{in press}
\reference {} Gustafsson, B., Edvardsson, B., Eriksson, K., Mizuno-Wiedner, M., \& J{\o}rgensen, U. G. 2003, in Stellar Atmosphere Modeling, ASP Conf. Proc. 228, ed. I. Hubeny, D. Mihalas, \& K. Werner (San Francisco, CA: ASP), 331
\reference {} Gustafsson, B., Edvardsson, B., Eriksson, K., et al. 2008, A\&A, 486, 951
\reference {} Hartman, J. D., et al. 2006, MNRAS, 371, 1405
\reference {} Hartmann, D., \& Burton, W. B. 1997, Atlas of Galactic Neutral Hydrogen (Cambridge: Cambridge University Press)
\reference {} Heger, A., Jeannin, L., Langer, N., \& Baraffe, I. 1997, A\&A, 327, 224
\reference {} Huang, W., Gies, D. R., \& McSwain, M. V. 2010, ApJ, 722
\reference {} Javadi, A., van Loon, J. T., \& Mirtorabi, M. T. 2011, MNRAS, 411, 263 
\reference {} Josselin, E., Blommaert, J. A. D. L., Groenewegen, M. A. T., Omont, A., \& Li, F. L. 2000, A\&A, 357, 225
\reference {} Kerr, F., J., \& Lynden-Bell, D. 1986, MNRAS, 221, 1023
\reference {} Kippenhahn, R., \& Weigert, A. 1990, in Stellar Structure and Evolution (Berlin: Springer Verlag)
\reference {} Kirby, E. N., Guhathakurta, P., \& Sneden, C. 2008, ApJ, 682, 1217
\reference {} Kurucz, R. L. 1992, in The Stellar Populations of Galaxies, ed. B. Barbury \& A. Renzini (Dordrecht: Kluwer), 225
\reference {} Kwitter, K. B., \& Aller, L. H. 1981, MNRAS, 195, 939 
\reference {} Leitherer, C. \& Ekstr\"{o}m, S. 2011, in The Spectral Energy Distribution of Galaxies, IAU Symp.\ 284, ed. R. J. Tuffs \& C. C. Popescu (Preston, UK: IAU), in press, arXiv:1111.5204
\reference {} Leitherer, C., et al. 1999, ApJS, 123, 3
\reference {} Levesque, E. M., Massey, P., Olsen, K. A. G., \& Plez, B. 2007, ApJ, 667, 202
\reference {} Levesque, E. M., Massey, P., Olsen, K. A. G., et al. 2005, ApJ, 628, 973
\reference {} Levesque, E. M., Massey, P., Olsen, K. A. G., et al. 2006, ApJ, 645, 1102
\reference {} Maund, J. R., et al. 2011, ApJ, 739, 37
\reference {} Macri, L. M., Stanek, K. Z., Sasselov, D. D., Krockenberger, M., \& Kaluzny, J. 2001, AJ, 121, 870
\reference {} Maeder, A. 1997, A\&A, 321, 134
\reference {} Magrini, L., Stanghellini, L., Corbelli, E., Galli, D., \& Villaver, E. 2010, A\&A, 512, 63
\reference {} Massey, P. 1998, ApJ, 501, 153
\reference {} Massey, P. 2003, ARA\&A, 41, 15
\reference {} Massey, P. 2010, in Hot and Cool: Bridging Gaps in Massive Star Evolution, ASP Conf. 425, ed. C. Leitherer, P. D. Bennett, P. W. Morris, \& J. Th. Van Loon (San Francisco, CA: ASP), 3
\reference {} Massey, P., Olsen, K. A. G., Hodge, P. W., et al. 2006, AJ, 131, 2478
\reference {} Massey, P., Olsen, K. A. G., Hodge, P. W., et al. 2007b, AJ, 133, 2393
\reference {} Massey, P., Olsen, K. A. G., Jacoby, G. H., et al. 2007a, AJ, 133, 2393
\reference {} Massey, P., Plez, B., Levesque, E. M., et al. 2005, ApJ, 634, 1286
\reference {} Massey, P., Silva, D. R., Levesque, E. M., et al. 2009, ApJ, 703, 420
\reference {} Meynet, G., \& Maeder, A. 2003, A\&A, 404, 975
\reference {} Meynet, G., \& Maeder, A. 2005, A\&A, 429, 581
\reference {} Meynet, G., Cyril, G., Hirschi, R., et al. 2011, BSRSL, 80, 266
\reference {} Neugent, K. F., \& Massey, P. 2011, ApJ, 733, 123
\reference {} Neugent, K. F., Massey P., Skiff, B., \& Meynet, G. 2011, ApJ submitted
\reference {} Neugent, K. F., Massey, P., Skiff, B., et al. G. 2010, 719, 1784 
\reference {} Osmer, P. S. 1972, ApJS, 24, 247
\reference {} Penny, L. R. 1996, ApJ, 463, 737
\reference {} Penny, L. R., \& Gies, J. D. 2009, ApJ, 700, 844
\reference {} Plez, B. 2003, in ASP Conf. Proc. 298, GAIA Spectroscopy: Science and Technology, ed. U. Munari (San Francisco, CA: ASP), 189
\reference {} Plez, B., Brett, J. M., \& Nordlund, \r{A}. 1992, A\&A, 256, 551
\reference {} Przybilla, N., Butler, K., Becker, S. R., Kudritzki, R. P., \& Venn, K. A. 2000, A\&A, 359, 1085
\reference {} Robin, A. C., Reyl\'{e}, S., Derri\`{e}re, S., \& Picaud, S. 2003, A\&A, 409, 523
\reference {} Rosolowsky, E., \& Simon, J. D. 2008, ApJ, 675, 1213
\reference {} Rubin, V. C., \& Ford, W. K., Jr. 1970, ApJ, 159, 379
\reference {} Salpeter, E. E. 1955, ApJ, 121, 161
\reference {} Schaerer, D., Meynet, G., Maeder, A., \& Schaller, G. 1993, A\&AS, 98, 523
\reference {} Schlegel, D. J., Finkbeiner, D. P., \& Davis, M. 1998, ApJ, 500, 525
\reference {} Skrutskie, M. F., et al. 2006, AJ, 131, 1163
\reference {} Soderberg, A. M., et al. 2011, ApJ submitted (arXiv:1107.1876)
\reference {} Talon, S.,\& Zahn, J.-P. 1997, A\&A, 317, 749
\reference {} Tonry, J., \& Davis, M. 1979, AJ, 84, 1511
\reference {} Vanbeveren, D., Van Bever, J., \& Belkus, H. 2007, ApJ, 662, 107
\reference {} van den Bergh, S. 2000, The Galaxies of the Local Group (Cambridge: Cambridge Univ. Press)
\reference {} van Loon, J. Th., Cioni, M.-R. L., Zijlstra, A. A., \& Loup, C. 2005, A\&A, 438, 273
\reference {} van Loon, J. Th., et al. 2008, A\&A, 487, 1055
\reference {} Yoon, S., \& Cantiello, M. 2010, ApJ, 717, 62
\reference {} Young, J. S., et al. 2000, MNRAS, 315, 635
\reference {} Zacharias, N. et al. 2010, AJ, 139, 2184
\reference {} Zaritsky, D., Elston, R., \& Hill, J. M. 1989, AJ, 97, 1
\end{references}
\end{document}